\documentclass[twocolumn,twocolappendix,trackchanges]{aastex631}	
\usepackage[mathlines]{lineno}	

\shorttitle{Modeling APOKASC-3 red giants I}
\shortauthors{K. Cao \& Pinsonneault}
\usepackage{CJKutf8}

\usepackage{graphicx}	
\usepackage{amsmath}	
\usepackage{amssymb}	
\allowdisplaybreaks

\begin{document}

\title{Modeling APOKASC-3 red giants: I. The first dredge-up and red giant branch bump}

\correspondingauthor{Kaili Cao}
\email{cao.1191@osu.edu}

\author[0000-0002-1699-6944]{Kaili Cao (\begin{CJK*}{UTF8}{gbsn}曹开力\end{CJK*}$\!\!$)}
\affiliation{Center for Cosmology and AstroParticle Physics (CCAPP), The Ohio State University, 191 West Woodruff Ave, Columbus, OH 43210, USA}
\affiliation{Department of Physics, The Ohio State University, 191 West Woodruff Ave, Columbus, OH 43210, USA}

\author[0000-0002-7549-7766]{Marc H. Pinsonneault}
\affiliation{Center for Cosmology and AstroParticle Physics (CCAPP), The Ohio State University, 191 West Woodruff Ave, Columbus, OH 43210, USA}
\affiliation{Department of Astronomy, The Ohio State University, 140 West 18th Avenue, Columbus, OH 43210, USA}

\begin{abstract}

We focus on two key diagnostics of stellar physics in red giant branch (RGB) stars: the first dredge-up (FDU) of nuclear processed material and the location of the red giant branch bump (RGBB). We compare asteroseismic and spectroscopic APOKASC-3 data with theoretical MESA models. Our FDU predictions have similar mass and metallicity trends to the data, but the observed magnitude of the change in $[{\rm C}/{\rm N}]$ in data is smaller than theoretical predictions by $[0.1615 \pm 0.0760 \,({\rm obs}) \pm 0.0108 \,({\rm sys})] \,{\rm dex}$. These results are insensitive to the input physics, but they are at a level consistent with systematic uncertainties in the abundance measurements. When we include observed trends in birth $[{\rm C}/{\rm Fe}]$ and $[{\rm N}/{\rm Fe}]$ in our models, it modestly increases the metallicity dependent difference relative to the data. We find a well-defined empirical RGBB locus: $\log g = 2.6604 - 0.1832 (M/{\rm M}_\odot-1) + 0.2824 \,[{\rm Fe}/{\rm H}]$. Our model RGBB loci have mass and composition trends that mirror the data, but we find that the observed RGBB is $[0.1509 \pm 0.0017 \,({\rm obs}) \pm 0.0182 \,({\rm sys})] \,{\rm dex}$ higher than predicted across the board, similar to prior literature results. We find that envelope undershooting, proposed solution to reconcile theory with data, increases ${\rm Li}$ destruction during the FDU at higher metallicities, creating tension with depletion observed in GALAH data. We propose ${\rm Li}$ in the FDU as a sensitive test of the RGBB and FDU, and discuss other potential solutions.

\end{abstract}


\section{Introduction} \label{sec:intro}

The theory of stellar structure and evolution naturally predicts fundamental observed properties of stars. For example, the luminosity-temperature trends on the main sequence (MS) are naturally explained by the dependence of stellar structure on mass. However, stellar theory goes deeper, and makes many detailed predictions about how evolutionary changes in stellar interiors can manifest themselves. With the advent of large surveys, the degree of data about stars has grown enormously. As a result, we can now test stellar theory in unprecedented detail. In this paper, we study two key aspects of the red giant branch (RGB): the first dredge-up (FDU) and the red giant branch bump (RGBB). Our goal is to test the predictive power of stellar models.

\subsection{Theoretical understanding}

During the core ${\rm H}$ burning main sequence, the ${\rm CNO}$ cycle processes $^{12}{\rm C}$ into $^{13}{\rm C}$ and ${\rm C}$ into ${\rm N}$ in stellar cores. The light element ${\rm Li}$ is destroyed at temperatures of order $2.5 \times 10^6 \,{\rm K}$, so ${\rm Li}$ can survive only in the cool outer layers of stars. After core ${\rm H}$ exhaustion, stars enter the RGB stage and develop a ${\rm H}$ burning shell surrounding an inert ${\rm He}$ core and expand in radius, becoming large and cool. A deep surface convection zone develops during this transition and reaches its maximum extent on the lower RGB. The ${\rm He}$ core then grows in mass, which pushes the lower boundary of the convective envelope out in mass fraction. During the FDU, the incorporation of ${\rm CNO}$-processed and ${\rm Li}$-depleted material into the convection zone causes observable changes in the abundances of ${\rm C}$, ${\rm N}$, and ${\rm Li}$. The carbon to nitrogen abundance ratio $[{\rm C}/{\rm N}]$, the $^{12}{\rm C}/^{13}{\rm C}$ ratio, and the surface ${\rm Li}$ abundance all decrease. Of these three, the $[{\rm C}/{\rm N}]$ ratio is particularly important because it can now be measured for hundreds of thousands of evolved red giant stars. More massive stars have more extensive nuclear processing in their cores, so the post-FDU surface $[{\rm C}/{\rm N}]$ depends on mass and birth composition in theoretical models \citep{Iben1965ApJb, Salaris2015A&A} and is therefore a potent age indicator \citep{Martig2016MNRAS, Ness2016ApJ}. Strong correlations between post-dredge-up $[{\rm C}/{\rm N}]$, mass, and composition are seen in stars with masses inferred directly from asteroseismology. For a recent discussion, see \citet{Roberts2024MNRAS}.

The FDU also imprints a distinctive signal in the luminosity evolution of stars. This feature was recognized in the first generation of stellar models of giants; see \citet{Refsdal1970A&A} for an insightful physical discussion. \citet{Christensen-Dalsgaard2015MNRAS} also examined this question, with updated models. As stars ascend the RGB, the hydrogen-burning shell moves outwards in terms of enclosed mass as the helium core grows, and the surface convection zones extend inwards, sinking unprocessed material to deeper layers. When the advancing hydrogen-burning shell approaches the basis of the chemically uniform and hydrogen-rich envelope, the lower mean molecular weight $\mu$ impacts the pressure at fixed temperature, causing a drop in nuclear reaction rates and leading to a short period where the luminosity decreases and the surface temperature slightly increases. This short period  of ``retrograde motion'' on the Hertzsprung-Russell diagram is recognized as the RGBB \citep[e.g.,][]{Hekker2020MNRAS}, another over-density region of stars near the red clump (RC), which corresponds to the more advanced core helium-burning stage of the post-MS evolution, but at a lower temperature. A related concept, the asymptotic giant branch bump, has been investigated by \citet{Bossini2015MNRAS}. The RGBB has been revealed by many modern photometric observations. The amplitude of the bump reflects the initial helium abundance of stars \citep[e.g.,][]{Nataf2013ApJ}; however, the RGBB is not in the place predicted by models \citep{Lattanzio2015MNRAS, Christensen-Dalsgaard2015MNRAS}.

To address this problem, prior works have invoked envelope undershooting \citep[e.g.,][]{Alongi1991A&A, Khan2018ApJ} to effectively deepen the convective zone, so that it meets with the growing ${\rm He}$ core earlier. This is a reasonable hypothesis in principle, as some degree of convective penetration must occur into the radiatively stable layers below a turbulent convection zone. Furthermore, convective core overshooting is almost universally used in stellar models to reproduce the observed width of the the main sequence in massive stars \citep[see for example][]{Demarque2004ApJS, Pietrinferni2004ApJ, VandenBerg2006ApJS, Choi2016ApJ}. However, undershooting below a surface convection zone\footnote{We consider boundary layer mixing above the core and below the envelope to be different cases. We therefore prefer to use different words (undershooting for envelope, overshooting for core) to distinguish between them.} is different physically and creates some known problems in other astrophysical contexts. Convective core overshooting occurs in a domain with high radiative flux, and it tends to reduce $\mu$ gradients that would formally inhibit convection. Envelope undershooting, however, must maintain a sharp lower boundary with a steep $\mu$ gradient, to permit the RGBB to exist, while deepening the convection zone significantly in a regime with a much lower flux \citep[see][for a discussion]{Christensen-Dalsgaard2021LRSP}.

Additionally, pre-MS stars are located in a similar part of the HR diagram as RGB stars. During their early evolution, they burn ${\rm Li}$ in deep convective envelopes \citep{Bodenheimer1965ApJ}, and this ${\rm Li}$ depletion is known to be highly sensitive to envelope undershooting \citep{Pinsonneault1997ARA&A}. Envelope undershooting sufficient to change the location of the RGBB induces severe pre-MS ${\rm Li}$ depletion in severe contradiction with the data \citep{Fu2015MNRAS, Chaname2022ApJ, Nguyen2025A&A}. This is evident as well in the MESA Isochrones and Stellar Tracks \citep[MIST;][]{Choi2016ApJ}. No physically compelling proposal to explain the difference between the severe ${\rm Li}$ depletion predicted and the modest depletion that is observed. We do note high stellar activity can alter stellar structure and reduce pre-MS ${\rm Li}$ depletion significantly \citep{Somers2020ApJ}. As we demonstrate in this paper, ${\rm Li}$ destruction during the FDU is another important test of envelope undershooting.

\subsection{New observational regime}

Fortunately, the advent of large, modern surveys --- spectroscopic, astrometric, and asteroseismic --- has enabled astronomers to study RGB stars in greater detail. Spectroscopic surveys yield detailed information about stellar abundances, as well as $\log g$ and $T_{\rm eff}$. Major recent examples include the Apache Point Observatory Galactic Evolution Experiment \citep[APOGEE;][]{Majewski2010IAUS, Majewski2017AJ}, the LAMOST Experiment for Galactic Understanding and Exploration \citep[LEGUE;][]{Zhao2012RAA, Deng2012RAA}, and the GALactic Archaeology with HERMES \citep[GALAH;][]{DeSilva2015MNRAS, Buder2018MNRAS, Buder2021MNRAS}. Astrometry supplements spectroscopy by providing parallax and apparent brightness, from which we can derive intrinsic luminosities and radii of stars of known $T_{\rm eff}$ \citep{Gaia2016A&A, Gaia2023A&A}.

Time domain surveys are also powerful. Turbulence in the envelopes of cool stars induces waves. For discrete frequencies a standing wave pattern can be created, causing periodic brightness variations. The study of these oscillations, asteroseismology, is now a flourishing field \citep[see][for a review]{Chaplin2013ARA&A}. Asteroseismic oscillations are highly visible in photometry obtained in space-based time domain surveys, such as COnvection, ROtation and planetary Transits \citep[COROT;][]{Baglin2003AdSpR}, Kepler \citep{Borucki2010Sci, Koch2010ApJ} and K2 \citep{Howell2014PASP}, and Transiting Exoplanet Survey Satellite \citep[TESS;][]{Ricker2015JATIS}. There are also upcoming missions, such as PLAnetary Transits and Oscillation of stars \citep[PLATO;][]{Rauer2014ExA} and the Nancy Grace Roman Space Telescope \citep[Roman;][]{Huber2023arXiv, Weiss2025ApJ}. All of these observational projects were designed to study exoplanets, yet they are also valuable sources of asteroseismic data.

APOKASC \citep{Pinsonneault2014ApJS, Pinsonneault2018ApJS, Pinsonneault2025ApJS} is an effort to compile APOGEE spectroscopic data and Kepler asteroseismic data\footnote{KASC stands for the Kepler Asteroseismic Science Consortium.} into large data sets, enabling detailed population studies. The first APOKASC catalog \citep{Pinsonneault2014ApJS} includes spectroscopic and asteroseismic properties of $1916$ red giants observed in the Kepler fields. As an upgraded version, the sample size was enlarged to $6676$ evolved stars in the second APOKASC catalog \citep{Pinsonneault2018ApJS}. Using the APOKASC data, astronomers have been able to derive other important quantities of the stars, including but not limited to age \citep{Martig2015MNRAS}, distance and extinction \citep{Rodrigues2014MNRAS}, and chemical abundance \citep{Hawkins2016A&A}. Various other interesting studies can also be conducted, for instance researching the population of rapidly rotating, low-mass giant stars \citep{Tayar2015ApJ}, calibrating the mixing length parameter in theoretical models \citep[e.g.,][]{Tayar2017ApJ, Salaris2018A&A}, and confirming the Gaia Data Release 2 \citep{Gaia2018A&A} parallax zero-point offset \citep{Zinn2019ApJ}. This work uses the third APOKASC catalog \citep{Pinsonneault2025ApJS}, which contains $15, 808$ targets, including $12, 573$ high-quality detections with mass and radius measurements. We choose this source because it empirically combines $10$ independent asteroseismic pipelines, and is calibrated to an absolute system by using Gaia radius (from luminosity and spectroscopic temperature) as reference.

\subsection{Testing theory with data}

The FDU and the RGBB are both measurable in data and predicted by theoretical models. This paper conducts systematic studies on the FDU modeling and the RGBB location, using the state-of-the-art APOKASC-3 measurements of red giant properties and MESA \citep{Paxton2011, Paxton2013, Paxton2015, Paxton2018, Paxton2019, Jermyn2023ApJS} numerical simulations. As we demonstrate below, these two diagnostics provide contradictory information: The observed FDU indicates less mixing while the RGBB location necessitates more mixing to explain. This combination may require some mechanism beyond our current understanding of stellar structure and evolution. This paper is structured as follows. We describe our observational data and theoretical models in Sections~\ref{sec:data} and \ref{sec:models}, respectively, focusing on how the stellar properties are measured or derived, and what physical assumptions and numerical schemes are adopted. In Sections~\ref{sec:fdu} and \ref{sec:rgbb}, we study first dredge-up (especially surface abundances of carbon and nitrogen) and red giant branch bump location as a function of surface gravity, respectively. We contrast our benchmark predictions to observed samples, and explore how different physical and numerical choices change the answer. Finally, we summarize and discuss our main results in Section~\ref{sec:discuss}.

\section{Observational data} \label{sec:data}

Precisely mapping out the dependence of the FDU and RGBB properties on mass and birth composition requires large, homogeneously analyzed data sets. For this purpose, we use the APOKASC-3 catalog \citep{Pinsonneault2025ApJS}, a compilation of effective temperatures and surface abundance ratios from APOGEE spectroscopy with masses and surface gravities from Kepler asteroseismology for $15, 808$ targets. We describe how the stellar properties are measured or derived from these observations in Section~\ref{ss:apokasc}, and discuss our sample selection process in Section~\ref{ss:select}. In addition, in Section~\ref{ss:apok2}, we introduce a sample of $398$ stars in the APO-K2 catalog with ${\rm Li}$ data from GALAH, which will be used in Section~\ref{ss:envos_li} to test FDU models.

\subsection{The APOKASC-3 catalog} \label{ss:apokasc}

The APOKASC-3 data represent the full overlap between data from the spectroscopic APOGEE survey \citep{Majewski2010IAUS, Majewski2017AJ} and time domain data from the Kepler survey \citep{Kjeldsen2010AN, Gilliland2010PASP}. APOGEE, an important portion of the Sloan Digital Sky Survey \citep[SDSS;][]{York2000AJ, Eisenstein2011AJ, Blanton2017AJ}, is a high-resolution ($R \simeq 22,000$) H-band spectroscopic survey optimized to study all components of the Milky Way galaxy \citep[e.g.,][]{Hayden2014AJ, Weinberg2022ApJS}. The APOGEE Stellar Parameter and Chemical Abundances Pipeline \citep[ASPCAP;][]{GarciaPerez2016AJ} derives (via a multi-dimensional least squares fit) stellar properties including the effective temperature $T_{\rm eff}$, the (logarithmic) spectroscopic surface gravity $\log g_{\rm spec}$, and a detailed heavy element mixture.

For our purposes, the key data are contained in four surface abundance ratios: $[{\rm M}/{\rm H}]$ (``metallicity''), $[\alpha/{\rm M}]$ (``alpha enhancement''), $[{\rm C}/{\rm Fe}]$, and $[{\rm N}/{\rm Fe}]$, where ${\rm M}$ denotes all the metals (elements heavier than ${\rm He}$) and $\alpha$ denotes the $\alpha$-capture elements (defined as ${\rm O}$, ${\rm Mg}$, ${\rm Si}$, ${\rm S}$, ${\rm Ca}$, and ${\rm Ti}$). $[{\rm M}/{\rm H}]$ is closely tied to $[{\rm Fe}/{\rm H}]$ in practice, and we therefore use it as a $[{\rm Fe}/{\rm H}]$ proxy. The SDSS survey publishes regular data releases that incorporate new data and improvements in analysis techniques. APOKASC-3 adopts Data Release 17 \citep[DR17;][]{Abdurrouf2022ApJS} data where available. For a minority of stars, APOKASC-3 uses Data Release 16 \citep[DR16;][]{Ahumada2020ApJS}, which also serves as a measure of systematic measurement uncertainties. Since DR17 random errors are likely underestimated, APOKASC-3 has recommended errors in $T_{\rm eff}$, $[{\rm Fe}/{\rm H}]$, and $[\alpha/{\rm Fe}]$ in the APOKASC-3 catalog \citep[see Section~3.4 of][]{Pinsonneault2025ApJS}; in this work, we rescale $[{\rm C}/{\rm Fe}]$ and $[{\rm N}/{\rm Fe}]$ errors in Appendix~\ref{app:CN_errors}.

The Kepler mission obtained $4$ years of almost continuous data, sampled at $30$-minute intervals. Non-radial oscillations have been detected in thousands of evolved red giants, including almost $16,000$ stars with APOGEE data. The frequency pattern is collapsed down to two figures of merit, the frequency of the maximum power $\nu_{\max}$ and the large frequency spacing $\Delta\nu$ \citep{Chaplin2013ARA&A, Miglio2021A&A}, which can be combined with spectra and stellar models to infer mass, radius, and age \citep[see][for details]{Pinsonneault2025ApJS}. The $15, 808$ targets are classified into several samples based on the number of pipelines which produce consistent results for each star: ``gold'' (five or more $\Delta\nu$ detections), ``silver'' (two to four $\Delta\nu$ detections), ``detect'' (at least two $\nu_{\max}$ detections), and ``non-detect'' (others).

\subsection{Sample selection} \label{ss:select}

For this work, we select APOKASC-3 ``gold'' sample RGB data. These stars have consistent measurements from $5$ or more methods, and they have the highest precision and accuracy. We use asteroseismic and spectroscopic data to remove core He-burning stars and stars on the upper RGB, which are a mix of first ascent red giants and double shell source asymptotic giant branch (AGB) stars. Core ${\rm He}$ burning stars have much lower central densities than shell ${\rm H}$ burning stars. This difference can clearly be seen in the asteroseismic frequency patterns \citep{Bedding2011Natur}. As a result, evolutionary states can be inferred from asteroseismology \citep{Elsworth2019MNRAS, Vrard2024A&A}, and we refer to these as asteroseismic states.

It is now recognized that stars in general do not have the same mixture of heavy elements as that of the Sun. Most of this variation can be measured by the ratio of $\alpha$-capture elements, primarily from core-collapse supernovae, to iron, which comes from a mix of Type Ia supernovae and core-collapse source, $[\alpha/{\rm Fe}]$. The APOGEE stars can be roughly grouped into an $\alpha$-poor population and an $\alpha$-rich one \citep[see][]{Hayden2015ApJ}, which can be thought of as the chemical thin and thick disks. This work focuses on the $\alpha$-poor population, or the chemical thin disk, which samples a wide range of ages and therefore masses. The $\alpha$-rich stars, by comparison, are mostly old, with a minority of either truly young stars or merger products. Furthermore, many of these stars are metal-poor, and extra mixing on the RGB is known to modify the $[{\rm C}/{\rm N}]$ measurements in evolved RGB stars \citep[see][for a discussion in the APOGEE context]{Shetrone2019ApJ}. The extra mixing occurs in luminous giants and in RC stars, so the FDU can be measured in $\alpha$-rich stars. However, bias from RC and AGB contamination of the sample is a larger issue in this case than for the $\alpha$-poor sample, where there is little evidence for extra mixing impacting the $[{\rm C}/{\rm N}]$ ratios. We therefore exclude $\alpha$-rich stars from our sample. Following \citet{Pinsonneault2025ApJS}, we apply the $[\alpha/{\rm Fe}]$ cut as
\begin{equation}
    \left\{\begin{aligned}
    &[\alpha/{\rm Fe}] < 0.14, &&[{\rm Fe}/{\rm H}] \leq -0.4, \\
    &[\alpha/{\rm Fe}] < 0.08 - 0.15 \,[{\rm Fe}/{\rm H}], &&-0.4 < [{\rm Fe}/{\rm H}] < 0.2, \\
    &[\alpha/{\rm Fe}] < 0.05, &&[{\rm Fe}/{\rm H}] \geq 0.2.
    \end{aligned}\right.
\end{equation}
Further treatment for $\alpha$-enhancement \citep[e.g.,][]{Salaris1993ApJ} is outside the scope of the current paper.

\begin{figure}
    \centering
    \includegraphics[width=\columnwidth]{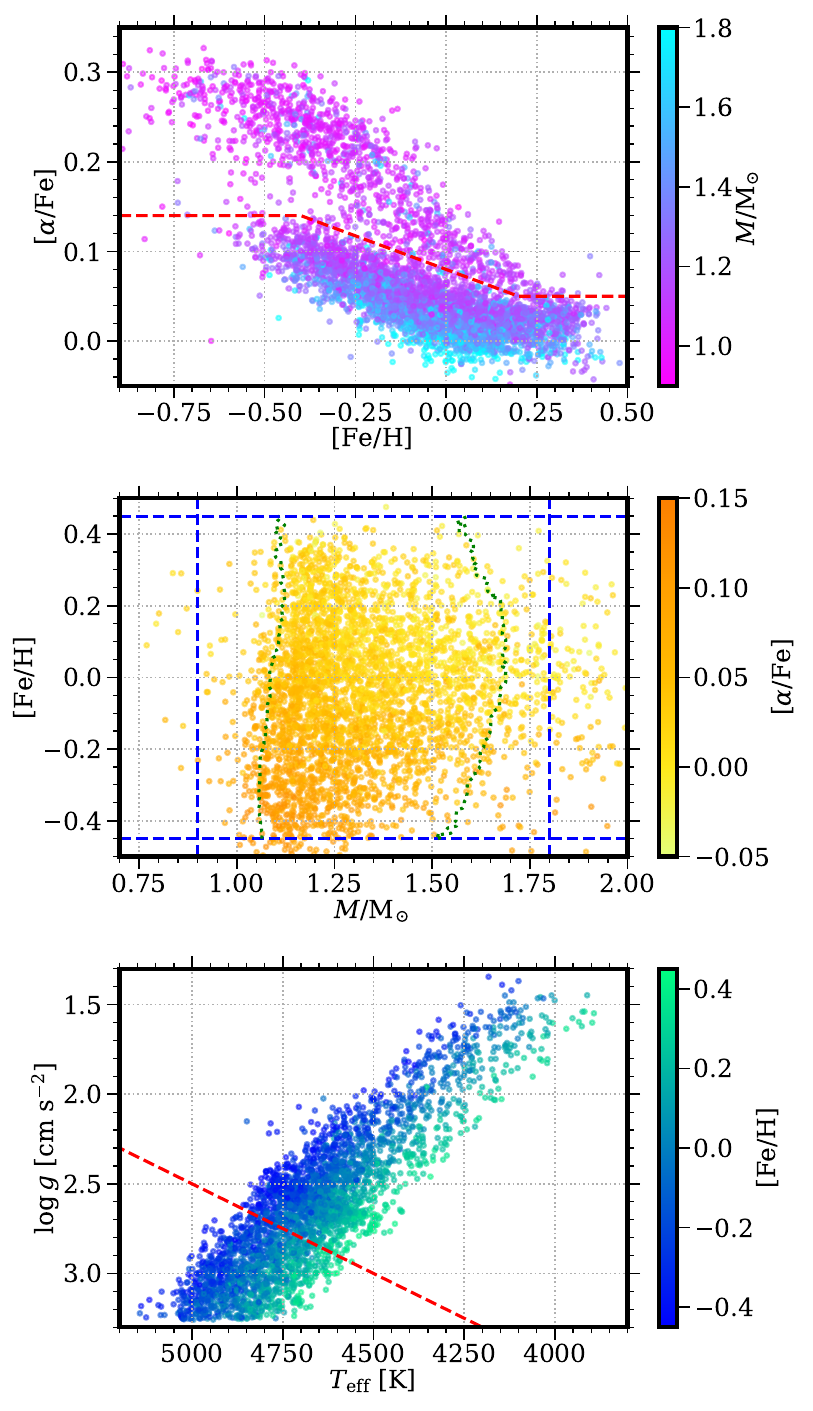}
    \caption{\label{fig:selection}Observational sample of RGB stars used in this work. The cuts we apply, shown as red or blue dashed lines, are explained in the text. Top panel: $[\alpha/{\rm Fe}]$ versus $[{\rm Fe}/{\rm H}]$, color-coded by mass; we select the $3801$ $\alpha$-poor stars from the $4997$ ``gold'' RGB stars. Middle panel: $[{\rm Fe}/{\rm H}]$ versus mass, color-coded by $[\alpha/{\rm Fe}]$; we use the $3513$ stars within the metallicity range $-0.45 < [{\rm Fe}/{\rm H}] < 0.45$ and the mass range $0.9 \,{\rm M}_\odot < M < 1.8 \,{\rm M}_\odot$. The green dotted lines show the $5^{\rm th}$ and $95^{\rm th}$ percentiles in mass at each metallicity within the $3513$ selected stars. Bottom panel: Kiel diagram ($\log g$ versus $T_{\rm eff}$) of selected stars, color-coded by metallicity; we select the $1261$ lower RGB stars (well below the RGBB) for RGB mixing length calibration purposes and FDU comparisons.}
\end{figure}

Since our main interest is the precise and accurate characterization of the FDU and RGBB, we prioritize the domain where we have a sufficient sample to explore mass and metallicity trends. Specifically, we select the stars in the metallicity range $-0.45 < [{\rm Fe}/{\rm H}] < 0.45$ and the mass range $0.9 \,{\rm M}_\odot < M < 1.8 \,{\rm M}_\odot$. These ranges are justified as follows:
\begin{itemize}
    \item Our sample below $[{\rm Fe}/{\rm H}] = -0.45$ is small ($77$, or $2.03\%$, out of $3801$ $\alpha$-poor ``gold'' RGB stars), with a narrow age range, and has the same issues as the $\alpha$-rich population discussed above.
    \item Only $2$ (i.e., $0.05\%$) APOKASC-3 ``gold'' RGB stars have $[{\rm Fe}/{\rm H}] > 0.45$. Domains that contain few or no stars would only yield weak statistical significance, and are therefore excluded from our comparisons.
    \item For our metallicity range, the $15$ (i.e., $0.39\%$) stars below $0.9 \,{\rm M}_\odot$ are unlikely to be the products of single star evolution, and our models therefore do not apply to them. Instead, they are very likely to have experienced a binary interaction, which makes their current mass an unreliable measure of their age.
    \item Relatively high-mass ($M > 1.8 \,{\rm M}_\odot$) RGB stars are less common ($199$, or $5.24\%$), as the lifetime of the RGB phase drops rapidly as mass increases.
    Since they span a wide range of masses (up to $3.2 \,{\rm M}_\odot$), their distribution is sparse and insufficient to provide reliable statistics.
\end{itemize}
For our core sample, prior studies have found that there is little to no evidence for extra mixing. However, these studies were not performed at our level of detail. To avoid any issues with theoretically possible extra mixing on the upper RGB, we only use stars below the RGBB for RGB mixing length calibration purposes (see Section~\ref{ss:cal_synth}) and FDU comparisons (see Section~\ref{sec:fdu}). Specifically, we apply a cut at $T_{\rm eff} / {\rm K} + 1000 \log g > 7500$. Figure~\ref{fig:selection} visualizes our sample selection procedure.

\subsection{APO-K2 sample with GALAH ${\rm Li}$ data} \label{ss:apok2}

The APO-K2 catalog \citep{Zinn2022ApJ, Schonhut-Stasik2024AJ, Warfield2024AJ} is a recent effort similar to APOKASC, but with asteroseismic data from K2 \citep{Howell2014PASP} instead of Kepler \citep{Borucki2010Sci, Koch2010ApJ}. In addition to spectroscopic data from APOGEE, among $4661$ RGB stars, $662$ have lithium abundance measurements from GALAH \citep{DeSilva2015MNRAS, Buder2018MNRAS, Buder2021MNRAS}. We perform a sample selection process similar to that described in Section~\ref{ss:select}:  For the $[\alpha/{\rm Fe}]$ cut, we exclude $\alpha$-rich stars indicated by the APO-K2 {\tt \textquotesingle ALPHA\_RICH\_FLAG\textquotesingle}; otherwise the sample selection process is exactly the same. Thus we obtain $398$ $\alpha$-poor stars on the RGB, ready for ${\rm Li}$ comparisons (see Section~\ref{ss:envos_li}). Since APO-K2 has larger uncertainties in asteroseismic measurements, we mainly use this catalog in the context of ${\rm Li}$ depletion during the FDU. Because GALAH is in the southern hemisphere, there is no overalap between APOKASC-3 and GALAH.

\section{Theoretical models} \label{sec:models}

We use the Modules for Experiments in Stellar Astrophysics \citep[MESA;][]{Paxton2011, Paxton2013, Paxton2015, Paxton2018, Paxton2019, Jermyn2023ApJS}, specifically MESA r22.11.1,\footnote{There has been a serious bug in the reverse triple-alpha process rate in MESA r22.05.1 and r22.11.1. However, this does not affect our simulations, which end before core helium ignition.} to calculate a suite of stellar models, focusing on their predictions for the FDU and the location of the RGBB. We start by setting a benchmark grid of simulations and then explore the sensitivity of our results to the choices for input physics. Our simulations are modified versions of the MESA test case {\tt 1M\_pre\_ms\_to\_wd}.\footnote{\url{https://docs.mesastar.org/en/release-r22.11.1/test_suite/1M_pre_ms_to_wd.html}}

\begin{table}
    \caption{\label{tab:mesa_params}Summary of initial condition and input physics variations explored in this work. See the text for further explanations and references.}
    \centering
    \begin{tabular}{ccc}
    \hline
        Parameter & Benchmark & Variation(s) \\
    \hline
        Solar mixture & GS98 & $\left\{ \begin{aligned} &{\rm AGSS09} \\ &{\rm varying\ C\ and\ N} \end{aligned} \right.$ \\
        Reaction rates & NACRE II & MESA default \\
        High-$T$ opacities & OP & OPAL \\
        Low-$T$ opacities & {\AE}SOPUS 1.0 & Ferguson \\
        Opacity interp. & cubic & linear \\
        Atmosphere & {\tt Eddington} & {\tt photosphere} \\
        Mixing & core overshooting & $\left\{ \begin{aligned} &{\rm no\ overshooting} \\ &{\rm envelope\ undershooting} \end{aligned} \right.$ \\
    \hline
    \end{tabular}
\end{table}

Table~\ref{tab:mesa_params} summarizes different options explored in this work. Each variation grid is a perturbation of the benchmark grid with one major input physics ingredient varied. Below we briefly introduce the key ingredients of our grids; for further explanations, numerical choices, calibration of models, and synthesis of tracks, see Appendix~\ref{app:models}. To validate our models, we will compare our results to literature data from \citet{SilvaAguirre2020A&A}.

\paragraph{Chemical composition} To characterize the stellar birth mixture, we adopt a reference solar mixture and heavy element to hydrogen ratio $Z/X$, and use $[{\rm Fe}/{\rm H}]$ as a global scaling factor for all the metals. We further assume that $Y$ scales linearly as
\begin{equation}
    Y = Y_{\rm BBN} + \frac{Y_{\odot, {\rm birth}} - Y_{\rm BBN}}{Z_{\odot, {\rm birth}}} \cdot Z,
    \label{eq:helium}
\end{equation}
where $Y_{\rm BBN} = 0.24709$ is the primordial helium abundance taken from \citet{Cyburt2016RvMP}, while $Y_{\odot, {\rm birth}}$ and $Z_{\odot, {\rm birth}}$ are the solar values at birth (see Section~\ref{ss:cal_synth} for how we calibrate these). In this paper, we explore two standard solar compositions: GS98 \citep{Grevesse1998SSRv}, which is adopted for our {\tt benchmark} grid and most variations, and AGSS09 \citep{Asplund2009ARA&A}, which is only used for the corresponding variation grid, {\tt var\_chem\_agss09}.

The change in the ${\rm C}$ to ${\rm N}$ ratio is known to depend strongly on mass and birth composition \citep{Iben1965ApJb, Salaris2015A&A}, and we use it as our primary diagnostic of the FDU. \citet{Roberts2024MNRAS} found that ${\rm C}$ and ${\rm N}$ abundances of APOGEE stars are different from those of the Sun, and characterized median trends of $\alpha$-poor subgiants as
\begin{equation}
    \left\{\begin{aligned}
    [{\rm C}/{\rm N}] &= -0.118 \times [{\rm Fe}/{\rm H}]^2 -0.344 \times [{\rm Fe}/{\rm H}] -0.0343, \\
    [{\rm C}/{\rm Fe}] &= 0.268 \times [{\rm Fe}/{\rm H}]^2 + 0.0258 \times [{\rm Fe}/{\rm H}] -0.00983, \\
    [{\rm N}/{\rm Fe}] &= 0.373 \times [{\rm Fe}/{\rm H}]^2 + 0.373 \times [{\rm Fe}/{\rm H}] + 0.0260.
    \end{aligned}\right.
    \label{eq:Roberts24}
\end{equation}
In a dedicated variation grid, {\tt var\_chem\_r24cn}, we postulate that these are the same as the ``birth'' trends of RGB stars, and superimpose them onto standard GS98. This grid is of particular interest for FDU studies (see Section~\ref{sec:fdu}). Note that mixture variation requires customized opacity tables to be made (see Section~\ref{ss:comp_nuc}).

\paragraph{Equation of state (EOS)} We adopt the MESA EOS, which is a blend of the OPAL \citep{Rogers2002ApJ}, SCVH \citep{Saumon1995ApJS}, FreeEOS \citep{Irwin2004}, HELM \citep{Timmes2000ApJS}, PC \citep{Potekhin2010CoPP}, and Skye \citep{Jermyn2021ApJ} EOSes, for all grids.

\paragraph{Nuclear reactions} To keep track of the major ${\rm pp}$ chain and ${\rm CNO}$ nuclear burning reactions, we adopt {\tt \textquotesingle pp\_and\_cno\_extras.net\textquotesingle} for all our simulations. By default, MESA nuclear reaction rates are from JINA REACLIB \citep{Cyburt2010ApJS} and NACRE \citep{Angulo1999NuPhA}. These do not include the more recent, and lower $^{14}{\rm N}$ proton capture rate, which is crucial for setting the nuclear equilibrium $[{\rm C}/{\rm N}]$ in stellar interiors and which can impact FDU predictions. Ergo we use Nuclear Astrophysics Compilation of REactions II \citep[NACRE II;][]{Xu2013NuPhA} rates for our {\tt benchmark} grid and most variations; for comparison purposes, we also study MESA default rates in a variation grid, {\tt var\_rates\_default}.

\paragraph{Radiative opacities} For high-$T$ opacities, we use tables from the Opacity Project \citep[OP;][]{Badnell2005MNRAS} for our {\tt benchmark} grid and most variations, and usage of the default OPAL \citep{Iglesias1993ApJ, Iglesias1996ApJ} tables is one of the variations, {\tt var\_kap\_opal}. In the low-$T$ regime, this work adopts {\AE}SOPUS 1.0 opacities \citep{Marigo2009A&A}, which include ${\rm CNO}$ variations as additional dimensions,\footnote{Customized {\AE}SOPUS 1.0 tables can be made at \url{http://stev.oapd.inaf.it/cgi-bin/aesopus_1.0}. {\AE}SOPUS 2.0 \citep{Marigo2022ApJ} has become available during the preparation of this work and is not explored here.} for the {\tt benchmark} grid and most variations, while the default \citet{Ferguson2005ApJ} tables are used as a variation, {\tt var\_kap\_fa05}. We adopt cubic interpolations \citep{Farag2024ApJ} for our {\tt benchmark} models and most variations, and study linear interpolations (MESA default) in a specific variation grid, {\tt var\_kap\_linear}, for comparison purposes.

\paragraph{Surface boundary conditions} Our {\tt benchmark} grid and most variations adopt the MESA default {\tt \textquotesingle Eddington\textquotesingle} grey atmosphere. A dedicated grid using the {\tt \textquotesingle photosphere\textquotesingle} boundary condition \citep{Hauschildt1999ApJa, Hauschildt1999ApJb, Castelli2003IAUS}, {\tt var\_atm\_phot}, is included to test the impact of the choice of surface boundary conditions on model predictions.

\paragraph{Convection and convective zones} To model convective energy transport, the mixing length theory \citep[MLT;][]{Henyey1965ApJ, Cox1968pss} is useful in 1D stellar evolution codes. By default, MESA implements MLT as time-dependent convection based on the \citet{Kuhfuss1986A&A} model. We adopt core overshooting for our {\tt benchmark} models and most variation grids; we turn this off in the no-overshooting grid, {\tt var\_mix\_noos}. Envelope undershooting has been considered as a viable explanation for previously found disagreement between observed and predicted RGBB locations \citep{Christensen-Dalsgaard2015MNRAS, Lattanzio2015MNRAS}. This possibility is also explored in a dedicated grid, {\tt var\_mix\_envos}. When included, both of core overshooting and envelope undershooting are set following MIST \citep{Choi2016ApJ}.

\section{First dredge-up} \label{sec:fdu}

In this section, we examine the first dredge-up modeling by comparing our theoretical predictions to observed surface composition changes. Our models include mass loss and gravitational settling. As a result, both the current mass and the current surface metallicity do not match the birth values in the tracks. We therefore interpolate between tracks to obtain a grid of fixed observed metallicity and current mass, as described in Section~\ref{ss:cal_synth}.

\subsection{FDU: benchmark results} \label{ss:fdu_base}

\begin{figure*}
    \centering
    \includegraphics[width=0.95\textwidth]{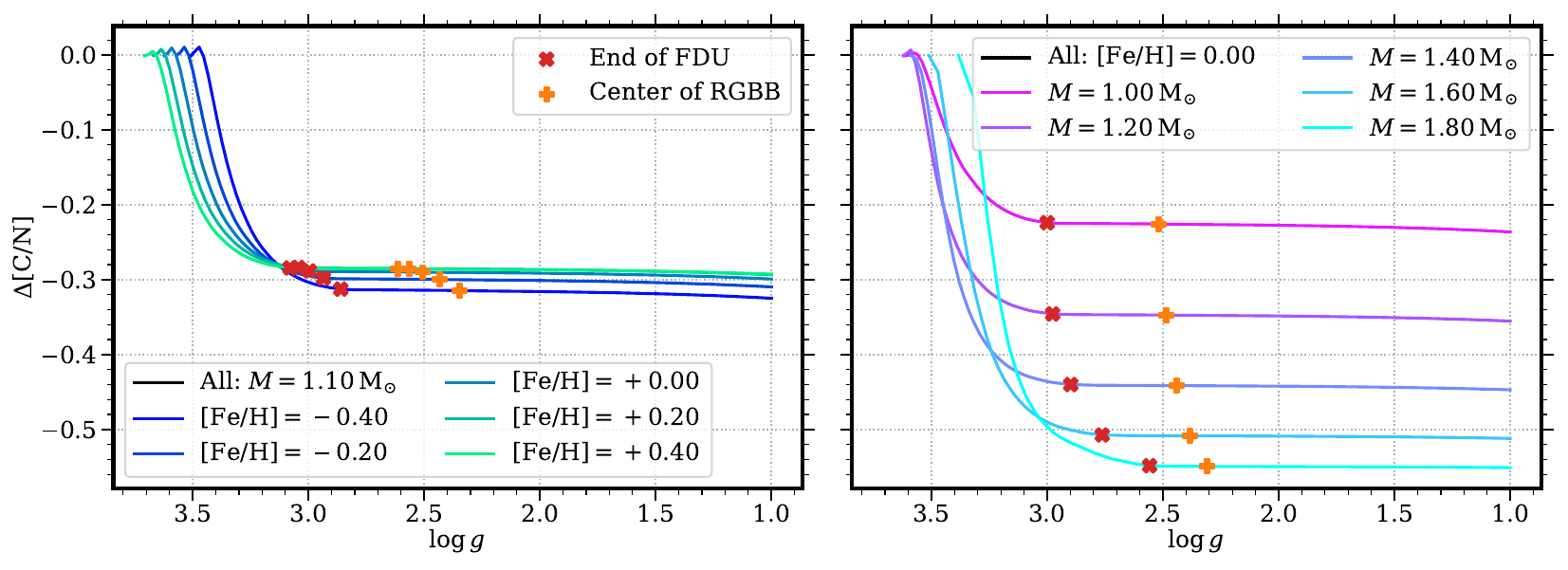}
    \caption{\label{fig:FDU_gravity_base}Predicted $\Delta[{\rm C}/{\rm N}]$ (change from birth value) as a function of $\log g$ according to our {\tt benchmark} models. Left panel: tracks at and stars near $M = 1.10 \,{\rm M}_\odot$, color-coded by metallicity; right panel: tracks at and stars near $[{\rm Fe}/{\rm H}] = 0.00$, color-coded by mass. End of FDU and center of RGBB are marked with red crosses and orange plus signs, respectively. Note that the curves begin where $\Delta[{\rm C}/{\rm N}]$ values start to change; in other words, $\Delta[{\rm C}/{\rm N}]$ values are constantly zero at higher $\log g$.}
\end{figure*}

\begin{figure}
    \centering
    \includegraphics[width=\columnwidth]{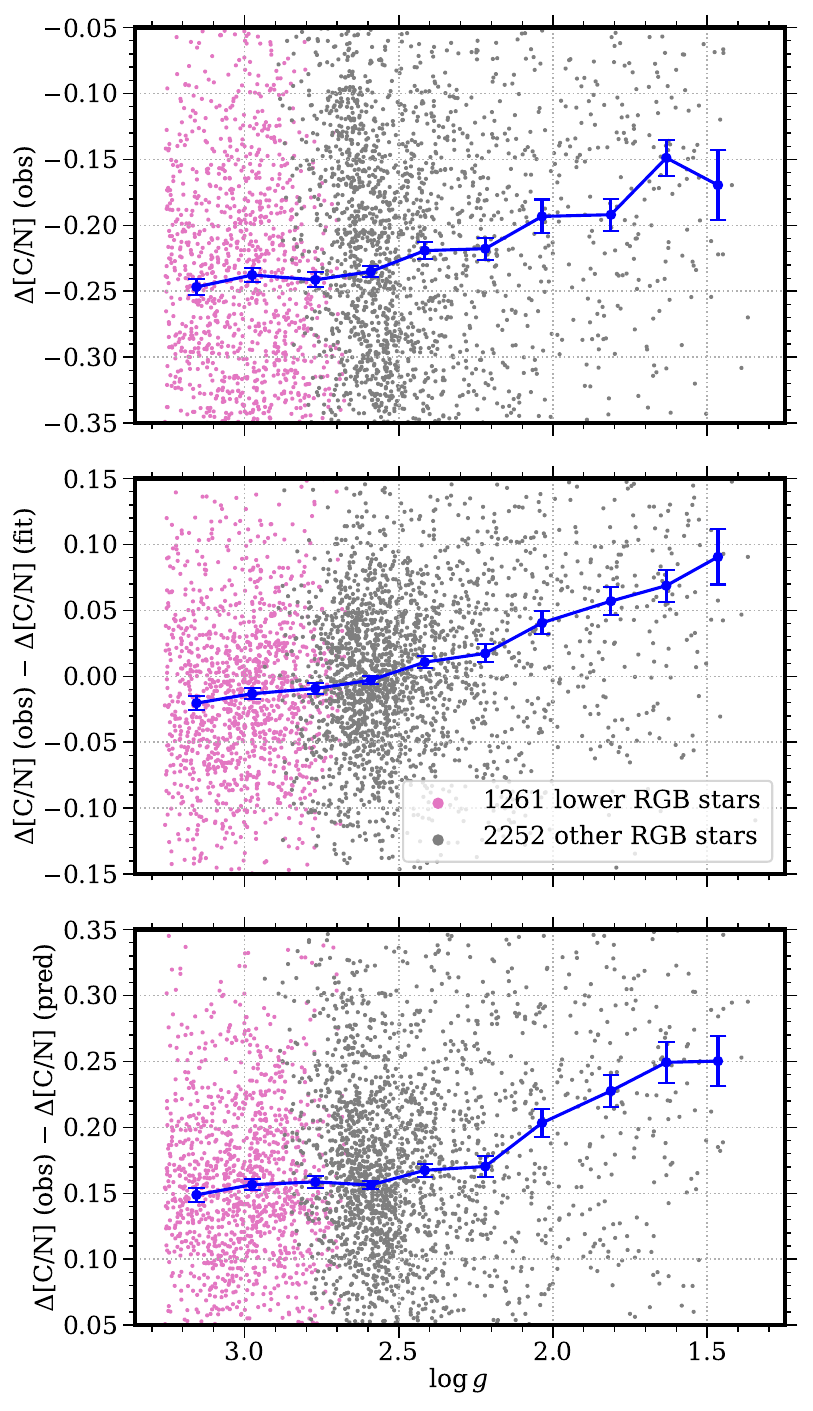}
    \caption{\label{fig:FDU_gravity_obs}Gravity dependence of observed $\Delta[{\rm C}/{\rm N}]$ (RGB value minus birth value, which is assumed to be Equation~(\ref{eq:Roberts24})). In each panel, lower RGB stars used in FDU comparisons are shown in pink, while other RGB stars are shown in gray; blue points and error bars are medians and dispersions in $10$ bins of equal widths. Upper panel: observed $\Delta[{\rm C}/{\rm N}]$ values as a function of $\log g$; middle panel: we fit observed $\Delta[{\rm C}/{\rm N}]$ as a function of mass and metallicity, but not gravity, and plot the residuals as a function of $\log g$; lower panel: we subtract {\tt benchmark} predictions in the middle of RGBB at given masses and metallicities from observed values, and again plot the residuals as a function of $\log g$.}
\end{figure}

Figure~\ref{fig:FDU_gravity_base} presents predicted change in surface ${\rm C}$ to ${\rm N}$ ratio $\Delta[{\rm C}/{\rm N}]$ as a function of $\log g$ for selected models. The degree of dredge-up depends sensitively on mass (right panel), but weakly on metallicity (left panel). The FDU is nearly complete by $\log g = 3.2$, the lower edge (on a Kiel diagram) of the APOKASC-3 sample, but there is some small dependence of the surface ${\rm C}$ and ${\rm N}$ abundances on surface gravity predicted by the models at lower surface gravity. Motivated by these features, we then examine the observational data to check for evidence of a $\log g$ dependence within the APOKASC-3 sample, which could be evidence for an incomplete dredge-up. However, as illustrated in Figure~\ref{fig:FDU_gravity_obs}, our observational sample does not show a statistically significant trend at high gravity. We therefore treat the lower RGB sample as all being post-dredge-up, and average the measurements accordingly. Our conclusion is consistent with \citet{Liagre2025A&A}, who obtained new data for less evolved giants in the Kepler sample. We measure the theoretically predicted FDU abundances and abundance ratios at the middle of the RGBB. This ensures that the FDU is complete for all models and that it is being measured at a consistent location.

\begin{figure}
    \centering
    \includegraphics[width=\columnwidth]{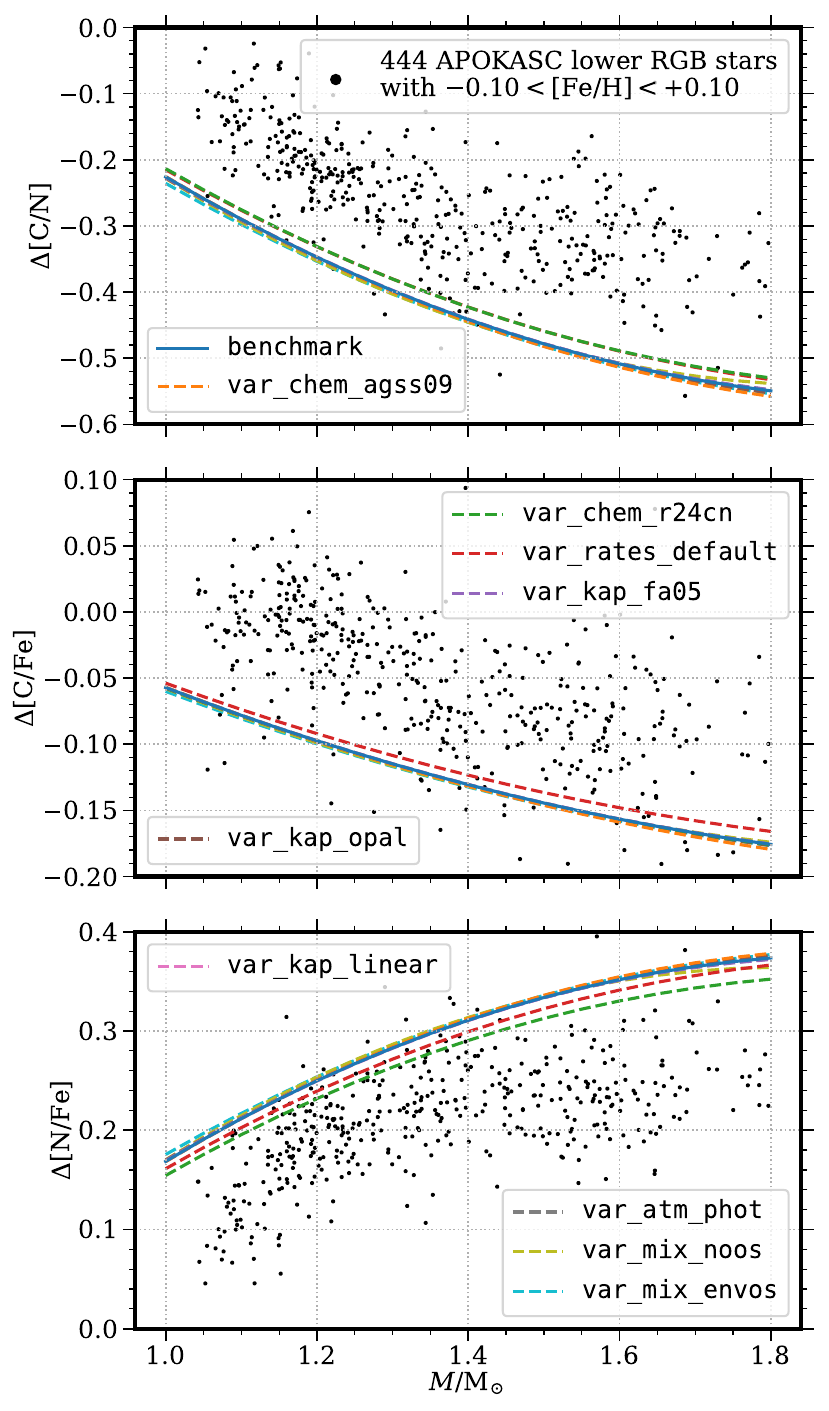}
    \caption{\label{fig:FDU_solar_base}Comparisons between the APOKASC-3 lower RGB sample (black dots) and theoretical predictions (curves) regarding FDU at solar metallicity. Changes in three abundance ratios from birth values (Equation~(\ref{eq:Roberts24}) assumed for observational data) to RGB, $\Delta[{\rm C}/{\rm N}]$, $\Delta[{\rm C}/{\rm Fe}]$, and $\Delta[{\rm N}/{\rm Fe}]$ are shown in the upper, middle, and lower panels, respectively. Solid blue curves present our {\tt benchmark} predictions, and dashed curves are variations (see Section~\ref{ss:fdu_var}).}
\end{figure}

\begin{figure}
    \centering
    \includegraphics[width=\columnwidth]{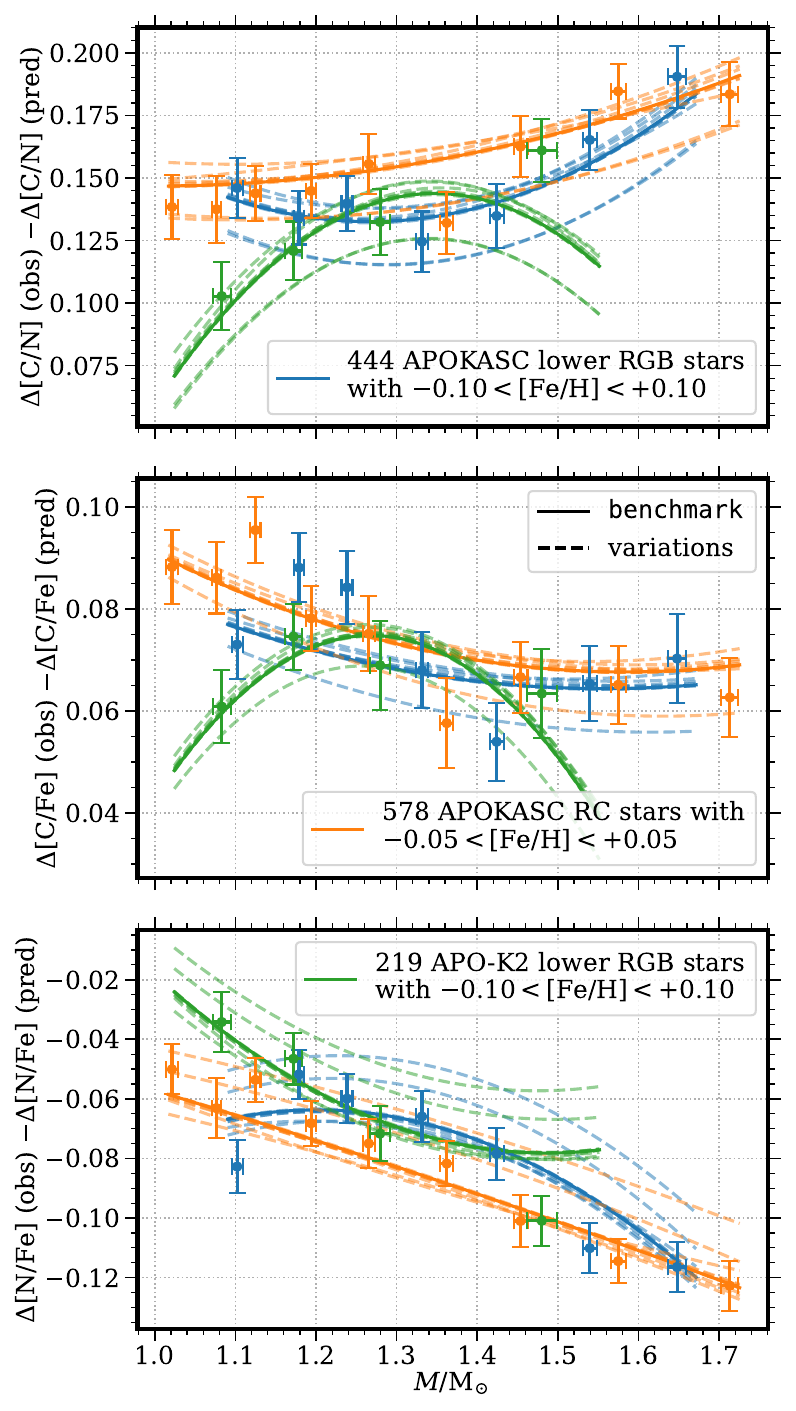}
    \caption{\label{fig:FDU_solar_obs}Comparisons between different observational samples regarding FDU at solar metallicity. APOKASC-3 lower RGB, APOKASC-3 RC, and APO-K2 lower RGB samples are shown in blue, orange, and green, respectively. We subtract predicted $\Delta[{\rm C}/{\rm N}]$, $\Delta[{\rm C}/{\rm Fe}]$, and $\Delta[{\rm N}/{\rm Fe}]$ values as functions of mass and metallicity from individual observations, fit the residuals, and visualize the results based on our {\tt benchmark} grid (other grids) as solid (dashed) curves. The curves start (end) at the $5^{\rm th}$ ($95^{\rm th}$) mass percentile of each sample. In addition, we plot medians and dispersions in several (depending on the number of stars) bins of equal sizes as points and error bars.}
\end{figure}

Figure~\ref{fig:FDU_solar_base} is a summary of our FDU comparisons at solar metallicity. Note that our {\tt benchmark} predictions for $\Delta[{\rm C}/{\rm N}]$ values at $M = 1.0$ and $1.5 \,{\rm M}_\odot$ are $-0.2258$ and $-0.4783$, respectively, in reasonable agreement with those reported in \citet{SilvaAguirre2020A&A}: about\footnote{These ranges are extracted from figures via visual inspection.} $[-0.095, -0.235]$ (see their Figure~4) and $[-0.35, -0.46]$ (Figure~10). Theoretical predictions largely replicate the relative mass and metallicity trends, yet there are significant zero-point shifts between theory and data. To test observational systematics, we compare three different observational samples in Figure~\ref{fig:FDU_solar_obs}, still at solar metallicity. In general, the APOKASC-3 RGB and RC samples agree with each other at the $0.01 \,{\rm dex}$ level (within $1\sigma$ in most cases), implying a potential lack of extra mixing. However, the APO-K2 RGB sample only agrees with APOKASC-3 around $1.4 \,{\rm M}_\odot$, and differs at the $0.1 \,{\rm dex}$ level at $0.9$ or $1.8 \,{\rm M}_\odot$, which we attribute to small number statistics in the APO-K2 sample for these domains.

\begin{figure*}
    \centering
    \includegraphics[width=0.95\textwidth]{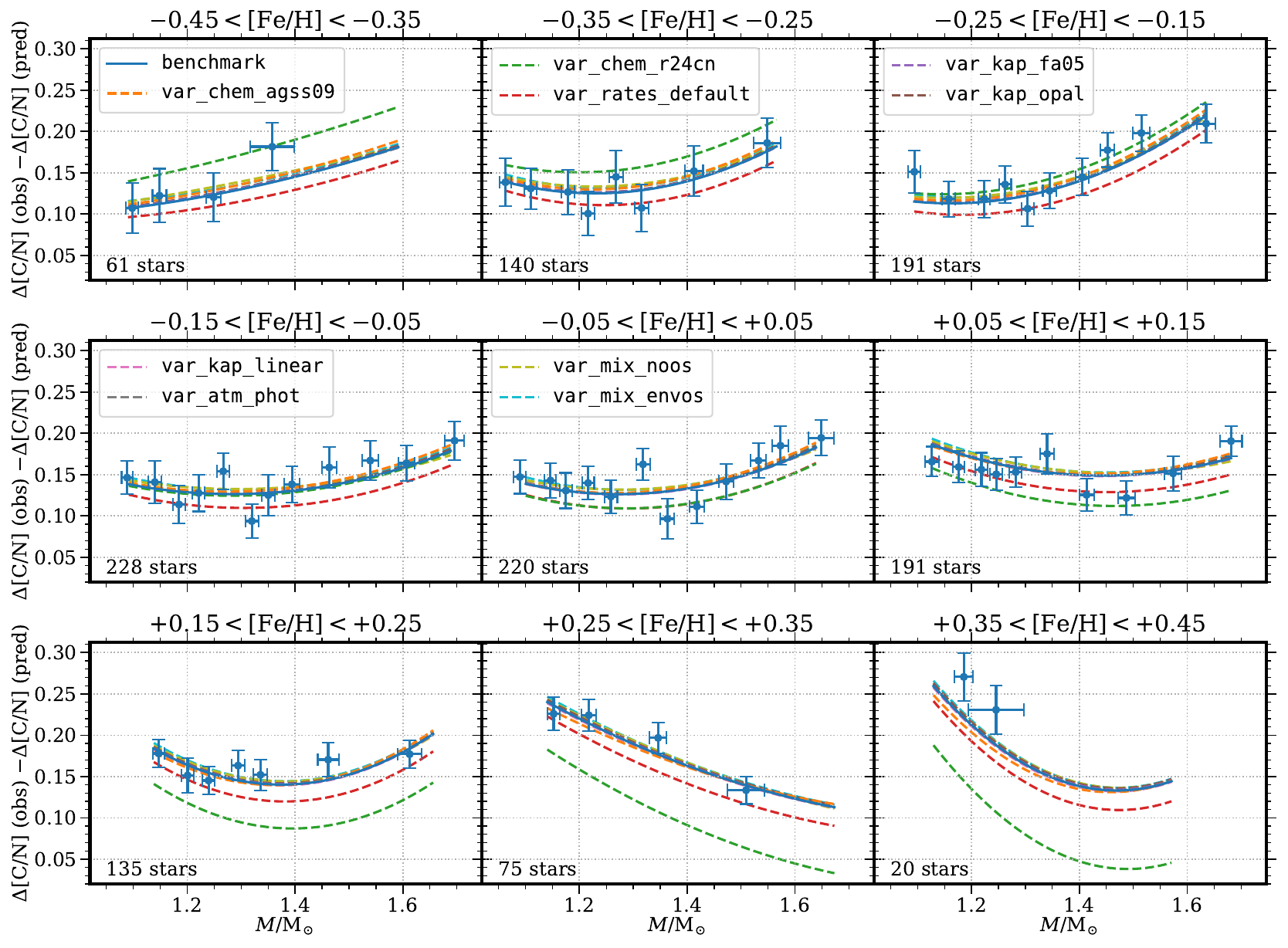}
    \caption{\label{fig:FDU_DCN}Discrepancies between observed and predicted $\Delta[{\rm C}/{\rm N}]$ values (post-FDU minus birth). Each panel represents a metallicity bin of width $0.1$; we plot $\Delta[{\rm C}/{\rm N}]$ versus mass. Like in Figure~\ref{fig:FDU_solar_base}, solid blue curves present fitting results based on our {\tt benchmark} predictions, and dashed curves correspond to the variations. Like in Figure~\ref{fig:FDU_solar_obs}, the curves start (end) at the $5^{\rm th}$ ($95^{\rm th}$) mass percentile of each sample, and we plot medians and dispersions based on our {\tt benchmark} models in several bins of equal sizes as points and error bars.}
\end{figure*}

\begin{table}
    \caption{\label{tab:FDU_all_base}Some discrepancies extracted from the right column of Figure~\ref{fig:FDU_all_base} (at cyan crosses). All $\Delta[*/*]$ values are post-FDU minus birth; ``disc.'' means fitting to observations minus predictions by the {\tt benchmark} grid. Note that $1.085$ and $1.685 \,{\rm M}_\odot$ are the $5^{\rm th}$ and $95^{\rm th}$ mass percentiles near solar metallicity, respectively, while $1.295 \,{\rm M}_\odot$ is the median mass of our entire APOKASC-3 lower RGB sample.}
    \centering
    \begin{tabular}{ccccc}
    \hline
        $M/{\rm M}_\odot$ & $1.085$ & $1.685$ & $1.295$ & $1.295$ \\
        $[{\rm Fe}/{\rm H}]$ & $0.00$ & $0.00$ & $-0.30$ & $+0.30$ \\
    \hline
        $\Delta[{\rm C}/{\rm N}]$ disc. & $0.1555$ & $0.1851$ & $0.1292$ & $0.1841$ \\
        $\Delta[{\rm C}/{\rm Fe}]$ disc. & $0.0823$ & $0.0706$ & $0.0779$ & $0.0910$ \\
        $\Delta[{\rm N}/{\rm Fe}]$ disc. & $-0.0747$ & $-0.1161$ & $-0.0507$ & $-0.0944$ \\
    \hline
    \end{tabular}
\end{table}

Figure~\ref{fig:FDU_DCN} explores $\Delta[{\rm C}/{\rm N}]$ discrepancies at all metallicities, using only the APOKASC-3 lower RGB sample. In most regions of the mass-metallicity space, all our grids predict more negative $\Delta[{\rm C}/{\rm N}]$ values, indicating more mixing during first dredge-up. Roughly speaking, the discrepancies are between $0.1$ and $0.2 \,{\rm dex}$; specifically, some typical values are extracted and tabulated in Table~\ref{tab:FDU_all_base}. The mass dependence is relatively weak, except in the highest-metallicity bin; the latter is understandable, as the non-uniformity of mass distribution is the most serious in that bin (see also the middle panel of Figure~\ref{fig:selection}). The metallicity trend is also weak but noticeable: APOKASC-3 data manifest larger $\Delta[{\rm C}/{\rm N}]$ values in metal-rich stars, but theoretical models do not replicate this trend. Varying the birth ${\rm C}$ and ${\rm N}$ abundances following \citet{Roberts2024MNRAS}, the {\tt var\_chem\_r24cn} grid (green dashed curves) has similar mass dependence, but introduces an interesting stretch in metallicity space. Compared to the {\tt benchmark} results, the agreement with observational data is not as good at low $[{\rm Fe}/{\rm H}]$, but is better at high $[{\rm Fe}/{\rm H}]$. Such differences suggest that birth mixture is crucial for surface abundance studies.

\begin{figure*}
    \centering
    \includegraphics[width=0.95\textwidth]{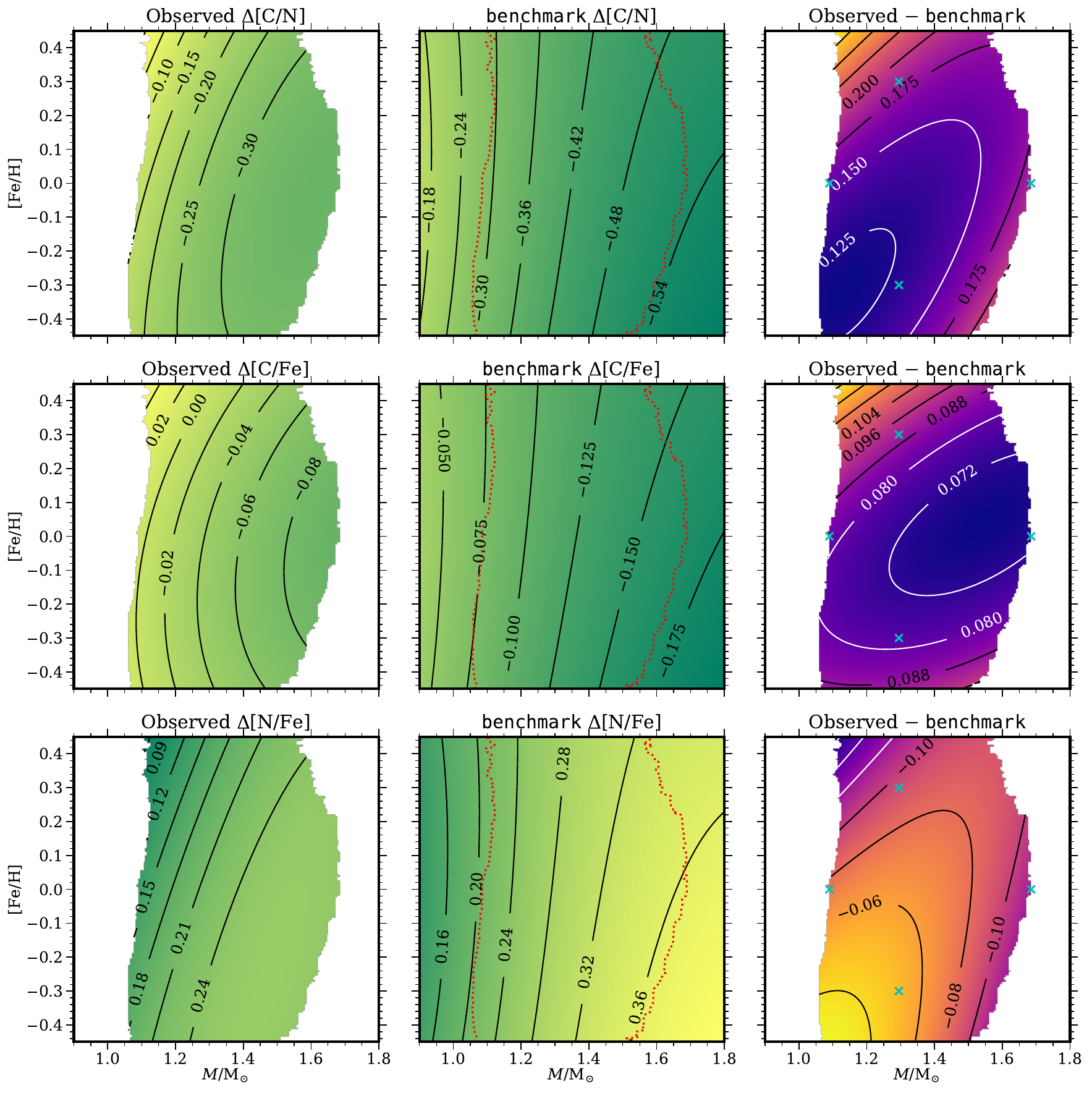}
    \caption{\label{fig:FDU_all_base}Comparisons between observed and predicted $\Delta[{\rm C}/{\rm N}]$ (top row), $\Delta[{\rm C}/{\rm Fe}]$ (middle row), and $\Delta[{\rm N}/{\rm Fe}]$ (bottom row); $\Delta$ means post-FDU minus birth. Each panel visualizes a bivariate function of mass and metallicity as a map with contours; panels involving observations only show the function within the contour derived in the middle panel of Figure~\ref{fig:selection}, which is also displayed as red dotted line in purely theoretical panels. Bivariate quadratic fits to APOKASC-3 data are shown in the left column, MESA predictions ({\tt benchmark} grid) are shown in the middle column, and the discrepancies (observed minus predicted) are shown in the right column. The coloring scheme is unified for the first two panels in each row, but is not unified across rows. Values at cyan crosses are extracted and tabulated in Table~\ref{tab:FDU_all_base}.}
\end{figure*}

To further compare our {\tt benchmark} FDU predictions to APOKASC-3 data, Figure~\ref{fig:FDU_all_base} visualizes observed and predicted $\Delta[{\rm C}/{\rm N}]$ as bivariate functions, and decomposes it into $\Delta[{\rm C}/{\rm Fe}]$ and $\Delta[{\rm N}/{\rm Fe}]$. From the left column (observational results), we see clearly that low-mass, high-metallicity stars have $\Delta[{\rm C}/{\rm Fe}] > 0$. This is implausible, as the net effect of the  ${\rm CNO}$ cycle is to convert carbon into nitrogen (see Section~\ref{sec:intro}). We tentatively conclude that there are some zero-point shifts in the data, either in APOGEE abundance measurements or in the assumption that the ${\rm C}$ and ${\rm N}$ trends observed in subgiants \citep{Roberts2024MNRAS} are the same as the birth trends of RGB stars. Comparing the three rows (three abundance ratios), $\Delta[{\rm N}/{\rm Fe}]$ has a larger contribution to $\Delta[{\rm C}/{\rm N}]$ than $\Delta[{\rm C}/{\rm Fe}]$ in both observational and theoretical results --- note that by definition, $\Delta[{\rm C}/{\rm N}] = \Delta[{\rm C}/{\rm Fe}] - \Delta[{\rm N}/{\rm Fe}]$. Another interesting aspect of this figure is that, contour lines are mostly vertical in the middle column ({\tt benchmark} predictions), but significantly tilted in the left column (APOKASC-3 results). This supplements the lack of predicted metallicity dependence in Figure~\ref{fig:FDU_DCN}. Such an effect could be induced by metallicity-dependent systematics in the abundance measurements.

\begin{figure*}
    \centering
    \includegraphics[width=0.95\textwidth]{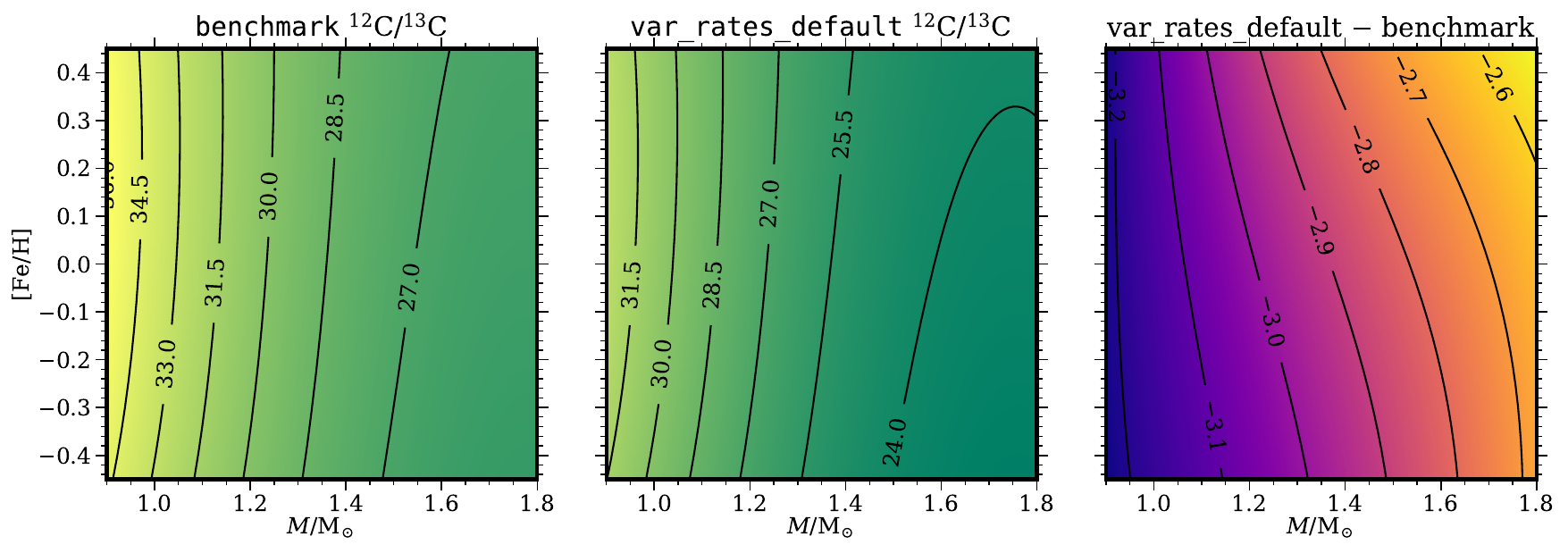}
    \caption{\label{fig:FDU_Ciso}Predicted post-FDU $^{12}{\rm C}/^{13}{\rm C}$. Each panel visualizes a bivariate function of mass and metallicity as a map with contours. Our {\tt benchmark} ({\tt var\_rates\_default}) predictions are shown on the left (middle) panel, while the discrepancy is shown on the right panel. Note that the coloring scheme is unified for the first two panels.}
\end{figure*}

When we examine the tracks in Figure~\ref{fig:FDU_gravity_base} in more detail, at low mass, $\Delta[{\rm C}/{\rm N}]$ rises by a small amount at the beginning of FDU before dropping to the final value. This is because $^{13}{\rm C}$ is dredged up before ${\rm N}$. We therefore include a brief discussion about the predicted post-FDU surface $^{12}{\rm C}/^{13}{\rm C}$ ratio, which is shown in Figure~\ref{fig:FDU_Ciso}. Theoretically, $^{12}{\rm C}$ is processed into $^{13}{\rm C}$ before $^{13}{\rm C}$ is processed into $^{14}{\rm N}$, so the $^{12}{\rm C}/^{13}{\rm C}$ ratio changes at lower temperatures than the $[{\rm C}/{\rm N}]$ ratio does. Therefore, $^{13}{\rm C}$ is dredged-up before $^{14}{\rm N}$ \citep{Pinsonneault1989ApJ}. In addition to our {\tt benchmark} predictions, we also include {\tt var\_rates\_default} results and the discrepancies between these two grids. Among all the variation grids, this is the one which differs from {\tt benchmark} at the $1 \,{\rm dex}$ level; other variations only differ at the $0.1$ or $0.01 \,{\rm dex}$ level. Fitting coefficients for all the grids are tabulated in Table~\ref{tab:Ciso_coeffs}. Since the APOKASC-3 data catalog does not include isotopic ratios, we stop this discussion here, but warn the readers that nuclear reaction rates (see Section~\ref{ss:comp_nuc}) are important when isotopic ratios are of interest. For example, \citet{Karakas2014PASA} reported much lower $^{12}{\rm C}/^{13}{\rm C}$ ratios ($\sim 20$--$25$) as they used different rates.

\subsection{FDU: variations} \label{ss:fdu_var}

\begin{figure*}
    \centering
    \includegraphics[width=0.95\textwidth]{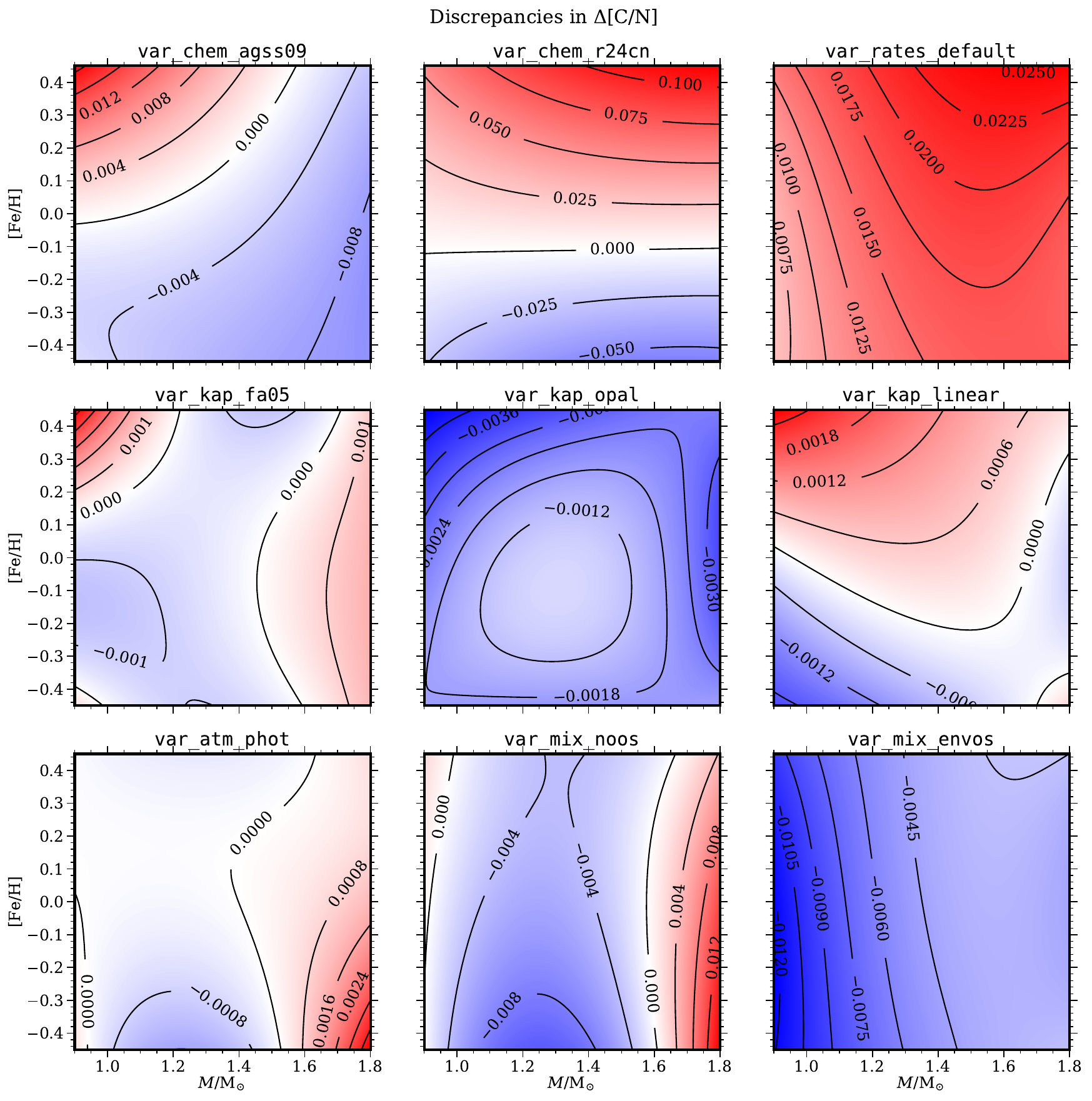}
    \caption{\label{fig:FDU_DCN_var}Discrepancies in predicted $\Delta [{\rm C}/{\rm N}]$ (variations minus {\tt benchmark}). See Section~\ref{sec:models} for the definition of grid names. Each panel visualizes a bivariate function of mass and metallicity as a map with contours. Note that different panels have different coloring schemes.}
\end{figure*}

We then compare FDU predictions yielded by our variation grids to the {\tt benchmark} predictions. Figure~\ref{fig:FDU_DCN_var} shows discrepancies in predicted $\Delta[{\rm C}/{\rm N}]$. While the maps display interesting patterns, most variations only introduce differences at the $0.01$ or $0.001 \,{\rm dex}$ level, evidently insufficient to account for the zero-point shift between theoretical predictions and observational data (see Section~\ref{ss:fdu_base}). Fitting coefficients for observed and predicted $\Delta[{\rm C}/{\rm N}]$ trends are tabulated in Tables~\ref{tab:CN_coeffs}. We omit similar maps and tables for $\Delta[{\rm C}/{\rm Fe}]$ and $\Delta[{\rm N}/{\rm Fe}]$, but note that the latter is still the dominating component of $\Delta[{\rm C}/{\rm N}]$. Below we comment on some individual variation grids.

\paragraph{{\tt var\_chem\_agss09}} Although the \citet{Asplund2009ARA&A} solar composition leads to a substantially lower calibrated $Z_{\odot, {\rm birth}}$ than \citet{Grevesse1998SSRv} (see Table~\ref{tab:SunCalibr}), this variation grid only differs from the {\tt benchmark} case at the $0.01 \,{\rm dex}$ level in terms of $\Delta[{\rm C}/{\rm N}]$ at fixed mass and $[{\rm Fe}/{\rm H}]$.

\paragraph{{\tt var\_chem\_r24cn}} In subgiant data there is a strong observed trend of birth ${\rm C}$ and ${\rm N}$ abundances as a function of $[{\rm Fe}/{\rm H}]$, which is not accounted for in a fixed mixture model. It is therefore not surprising that accounting for this effect causes the largest change in FDU predictions, at the $0.1 \,{\rm dex}$ level. As seen in Section~\ref{ss:fdu_base}, there is a consistent metallicity stretch (positive discrepancy at high metallicity and negative discrepancy at low metallicity) at all masses, although the stretch is more significant at low mass. Note that the best agreement ($0.0$ discrepancy) between {\tt var\_chem\_r24cn} and {\tt benchmark} grids occurs at $[{\rm Fe}/{\rm H}] \approx -0.1$, where the birth mixture variation Equation~(\ref{eq:Roberts24}) is the smallest. This stretch indicates that post-FDU $[{\rm C}/{\rm N}]$ depends on birth ${\rm C}$ and ${\rm N}$ abundances in a non-linear manner, and there is an important degeneracy between mass and birth mixture that needs to be accounted for to interpret observed $[{\rm C}/{\rm N}]$ values in red giants.

\paragraph{{\tt var\_rates\_default}} This grid uses MESA default nuclear reaction rates instead of NACRE II; the most important difference for our purpose is that the $^{14}{\rm N}$ proton capture rate is lower by a factor of $\sim 2$ compared to modern measurement. This only yields positive discrepancy in $\Delta[{\rm C}/{\rm N}]$ (and negative discrepancy in $\Delta[{\rm N}/{\rm Fe}]$), again manifesting the importance of nuclear reaction rates for FDU predictions.

\paragraph{{\tt var\_mix\_noos}} Core overshoot has a large effect on the main sequence lifetime of massive stars, but neglecting it induces only small changes, mostly at the $0.001 \,{\rm dex}$ level, in the FDU. This effect is somewhat larger at higher mass, reaching $\sim 0.01 \,{\rm dex}$ at $M = 1.8 \,{\rm M}_\odot$, solar metallicity; however, it is still small overall.

Other variation grids only differ from {\tt benchmark} at the $0.001 \,{\rm dex}$ level.

\begin{table*}
    \caption{\label{tab:fdu_shifts}Zero-point shifts of FDU results. Reported are medians and dispersions of discrepancies between observed values and theoretical predictions (observed minus predicted) based on mass and metallicity.}
    \centering
    \begin{tabular}{lccc}
    \hline
        Grid name & $\Delta[{\rm C}/{\rm N}]$ shift & $\Delta[{\rm C}/{\rm Fe}]$ shift & $\Delta[{\rm N}/{\rm Fe}]$ shift \\
    \hline
        {\tt benchmark} & $0.161464 \pm 0.076008$ & $0.089125 \pm 0.042050$ & $-0.074765 \pm 0.053136$ \\
        {\tt var\_chem\_agss09} & $0.163929 \pm 0.075197$ & $0.090217 \pm 0.041607$ & $-0.076193 \pm 0.052882$ \\
        {\tt var\_chem\_r24cn} & $0.173292 \pm 0.097525$ & $0.090583 \pm 0.042146$ & $-0.080630 \pm 0.078020$ \\
        {\tt var\_rates\_default} & $0.145031 \pm 0.075250$ & $0.083394 \pm 0.042224$ & $-0.064280 \pm 0.052852$ \\
        {\tt var\_kap\_fa05} & $0.161939 \pm 0.075745$ & $0.089265 \pm 0.042020$ & $-0.075218 \pm 0.052839$ \\
        {\tt var\_kap\_opal} & $0.162589 \pm 0.076241$ & $0.089730 \pm 0.042058$ & $-0.075450 \pm 0.053457$ \\
        {\tt var\_kap\_linear} & $0.161208 \pm 0.075852$ & $0.089026 \pm 0.042048$ & $-0.074579 \pm 0.053019$ \\
        {\tt var\_atm\_phot} & $0.161688 \pm 0.075797$ & $0.089138 \pm 0.042016$ & $-0.074854 \pm 0.052963$ \\
        {\tt var\_mix\_noos} & $0.165650 \pm 0.074896$ & $0.090684 \pm 0.042132$ & $-0.077627 \pm 0.052163$ \\
        {\tt var\_mix\_envos} & $0.166965 \pm 0.075913$ & $0.091392 \pm 0.042085$ & $-0.078270 \pm 0.053080$ \\
    \hline
    \end{tabular}
\end{table*}

To conclude this section, we report measured zero-point shifts between data and theory. For each star in our sample (see Section~\ref{ss:select}), we use its mass and metallicity to derive its expected changes in abundance ratios, and study the distribution of discrepencies between observed values and theoretical predictions (observed minus theoretical). The results are tabulated in Table~\ref{tab:fdu_shifts}. To prevent these numbers from being biased by outliers, the central values are the medians, and the dispersions are half the differences between upper and lower $1\sigma$ percentiles. In addition, we estimate theoretical systematics by summing up squares of half the discrepancies between variations and {\tt benchmark}, and then taking the square roots. In summary, the zero-point shifts are:
\begin{itemize}
    \item $\Delta[{\rm C}/{\rm N}]$: $0.1615 \pm 0.0760 \,({\rm obs}) \pm 0.010787 \,({\rm sys})$;
    \item $\Delta[{\rm C}/{\rm Fe}]$: $0.0891 \pm 0.0420 \,({\rm obs}) \pm 0.003322 \,({\rm sys})$;
    \item $\Delta[{\rm N}/{\rm Fe}]$: $-0.0748 \pm 0.0531 \,({\rm obs}) \pm 0.006472 \,({\rm sys})$.
\end{itemize}
By comparing several high-resolution spectroscopic sky surveys, \citet{Hegedus2023A&A} found that observational systematics can be at the $0.1 \,{\rm dex}$ level. Therefore, the zero-point shifts listed above could be caused by observational systematics. That said, an alternative explanation on the theoretical side would be that the maximum extent of convective envelopes in current models extend too deep into the stellar interiors.

\section{Red giant branch bump} \label{sec:rgbb}

In this section, we examine the RGBB location by comparing our model predictions to observations. Following \citet{Nataf2013ApJ}, we use an exponential background plus Gaussian bump model to fit the distribution
\begin{align}
    \mathcal{P} (\log g) = \mathcal{N} \bigg\{ &\exp (k \log g) \nonumber \\
    + &\frac{h_{\rm b}}{w_{\rm b} \sqrt{2 \pi}} \exp \bigg[ - \frac{1}{2} \left( \frac{\log g - c_{\rm b}}{w_{\rm b}} \right)^2 \bigg] \bigg\},
    \label{eq:bump_fit}
\end{align}
where $\mathcal{N}$ is the normalization factor so that $\mathcal{P} (3.0 > \log g > 2.0) = 1$, $k$ is the slope of the exponential background, $h_{\rm b}$, $w_{\rm b}$, and $c_{\rm b}$ are the relative height, width, and center location of the RGBB, respectively. The range $3.0 > \log g > 2.0$ is chosen because it includes the entire RGBB but excludes the edges of our sample (see the lower panel of Figure~\ref{fig:selection}) so that our statistical results below are not affected by selection bias. We note that since the APOKASC selection function is complex \citep[see][]{Pinsonneault2014ApJS}, the exponential background (i.e., the slope parameter $k$ in Equation~(\ref{eq:bump_fit})) of the APOKASC-3 may not be a perfect match for the overall population. However, the asteroseismic sample is nearly complete in the RGBB regime \citep{Pinsonneault2025ApJS}, and the selection effects become more significant in giants well above the RGBB, which were selected against for planet transit studies. As a result, we believe that selection effects do not bias this measurement. The relative height $h_{\rm b}$ is principally a function of the helium abundance $Y$ of the model, but because we are working with a field sample that has a wide metallicity range, it is difficult to quantify this diagnostic at a precise enough level to be an interesting measurement. We therefore do not infer it here.

\begin{figure*}
    \centering
    \includegraphics[width=\textwidth]{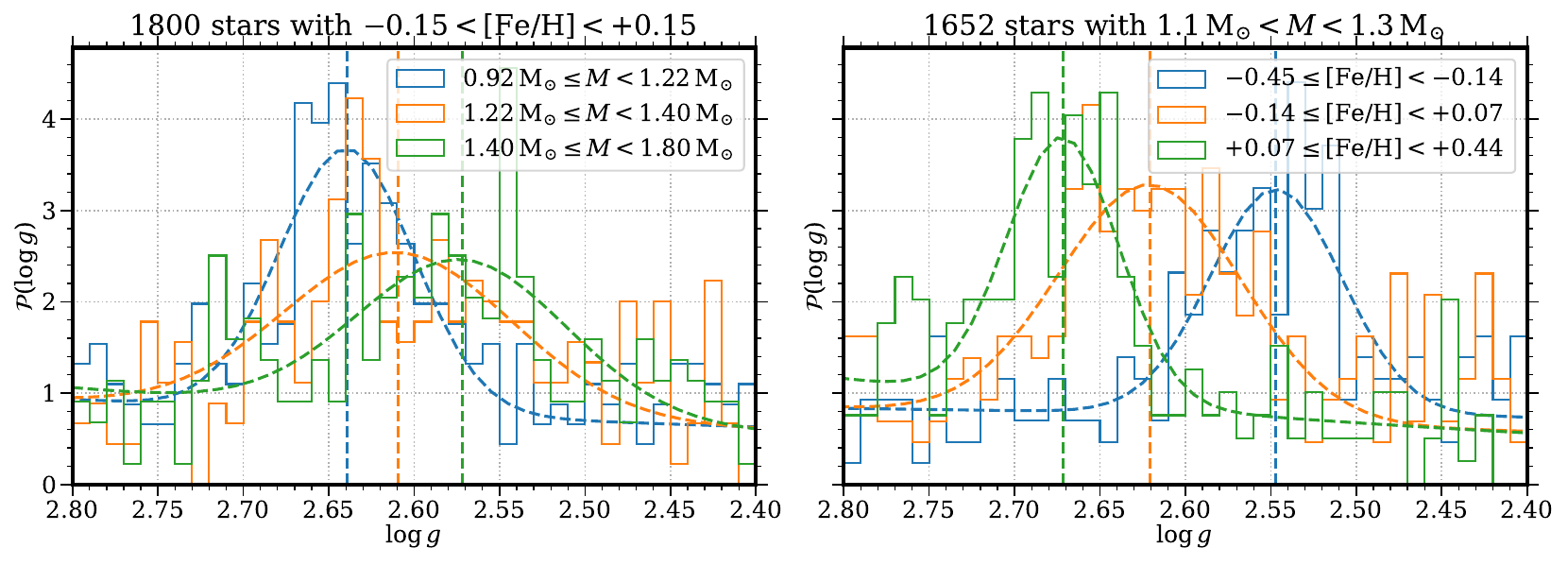}
    \caption{\label{fig:RGBB_recasting}Trends in observed RGBB location. On the left (right) panel, we select APOKASC-3 RGB stars in a relatively narrow metallicity (mass) range, subdivide the sample into three mass (metallicity) bins with equal number of stars, and fit Equation~(\ref{eq:bump_fit}) to stars in each bin.}
\end{figure*}

To start with, we explore the mass and metallicity trends of the observed RGBB location in the APOKASC-3 sample (see Section~\ref{ss:select}) in Figure~\ref{fig:RGBB_recasting}. For each subsample, we select APOKASC-3 stars in the surface gravity range $3.0 > \log g > 2.0$, fit Equation~(\ref{eq:bump_fit}) to them, and plot the probability density function. It is clear that the RGBB location depends on both mass and metallicity; however, due to our limited sample size, fitting results in each bin are noticeably noisy. Therefore, to investigate how the RGBB location shifts with mass and metallicity given our relatively limited sample size (a few thousand stars in $100$ bins), we implement a holistic method.

\begin{figure*}
    \centering
    \includegraphics[width=\textwidth]{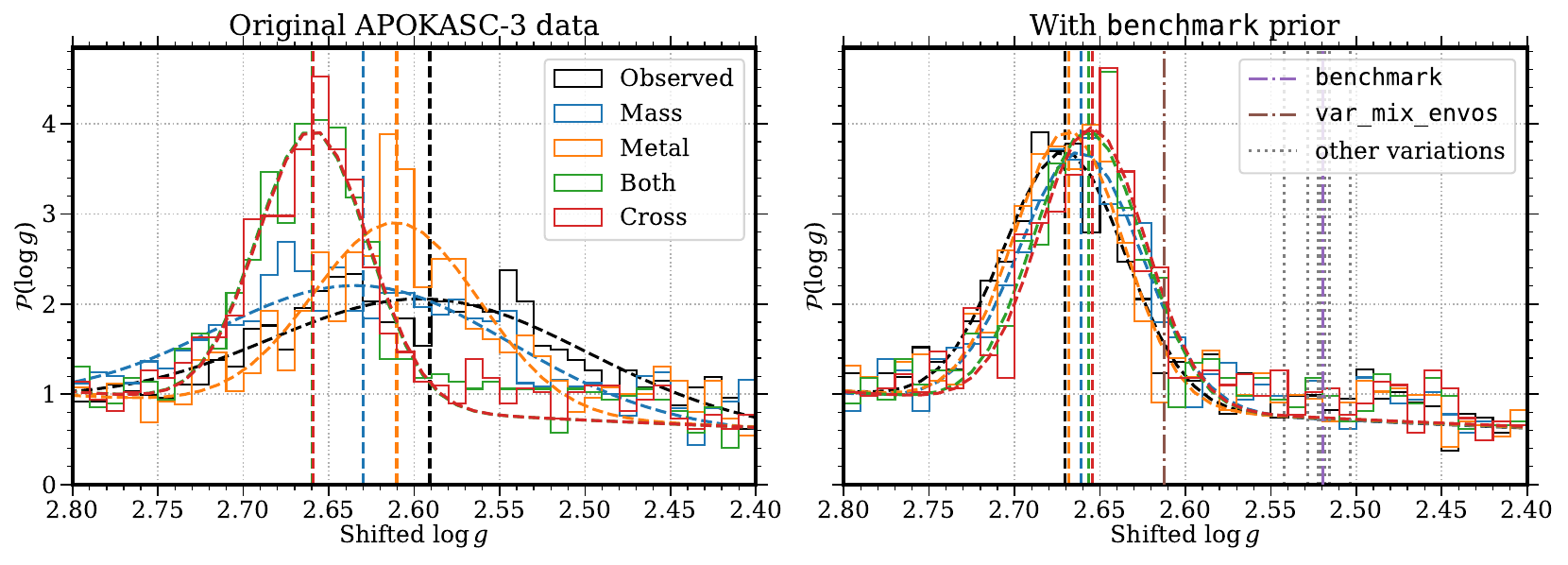}
    \caption{\label{fig:rgbb_obs}Observed RGBB location in the APOKASC-3 sample in terms of probability density functions, which are all normalized to $\mathcal{P} (3.0 > \log g > 2.0) = 1$. Each color represents a set of results: Solid histogram is the data, dashed curve is the exponential background plus Gaussian bump fit Equation~(\ref{eq:bump_fit}), and vertical dashed line is the bump center $c_{\rm b}$. On each panel, black does not involve any (further) $\log g$ correction, blue and orange assume a linear dependence on mass and metallicity, respectively, while green assumes a linear dependence on both, and red assumes a cross term in addition to green. The left panel shows results based on raw APOKASC-3 $\log g$ values, and while right panel ``correct'' them using a theoretical prior for mass and metallicity trends based on {\tt benchmark} predictions. In addition, on the right panel, theoretical predictions for RGBB location at $M = 1.0 \,{\rm M}_\odot$, $[{\rm Fe}/{\rm H}] = 0.0$ are shown as vertical lines: The {\tt benchmark} ({\tt var\_mix\_envos}) prediction is shown as a purple (brown) dash-dotted line, while other variations are shown as gray dotted lines.}
\end{figure*}

Specifically, we assign a shift in surface gravity space as a function of mass and metallicity, i.e., $\Delta \log g (M, [{\rm Fe}/{\rm H}])$, and try to maximize the ratio of the second term in Equation~(\ref{eq:bump_fit}) to the first term, both evaluated at the bump center $c_{\rm b}$, i.e., $(h_{\rm b} / w_{\rm b} \sqrt{2 \pi}) / \exp (k \log c_{\rm b})$. Note that a different shift function $\Delta \log g (M, [{\rm Fe}/{\rm H}])$ results in a slightly different sample selection, as $\log g$ cuts at $3.0$ and $2.0$ are applied to shifted $\log g$. Figure~\ref{fig:rgbb_obs} presents the RGBB location derived from this stacking technique. After trying various functional forms of this shift, including a linear dependence on mass (blue) or metallicity (orange), a linear dependence on both (green), and a cross term in addition (red), we find that the RGBB location depends on both mass and metallicity, but the marginal benefit of introducing a cross term is limited. Therefore, we adopt the green results as our benchmark observational results: $\log g = 2.660418 - 0.183181 (M/{\rm M}_\odot-1) + 0.282392 \,[{\rm Fe}/{\rm H}]$. Since the relationship between $\log g$ and $T_{\rm eff}$ can be well approximated by a linear function in the vicinity of the RGBB (see the lower panel of Figure~\ref{fig:selection}), we do not separately address the latter in this context.

Instead of introducing quadratic or higher-order terms in the above method (which could lead to overfitting), on the right panel of Figure~\ref{fig:rgbb_obs}, we use {\tt benchmark} predictions as theoretical prior (see Section~\ref{ss:rgbb_base}) for $\Delta \log g (M, [{\rm Fe}/{\rm H}])$. After accounting for mass and metallicity dependence in this way, we see that an additional $\Delta \log g$ term (different colors on the right panel) only makes minor differences. We conclude that our theoretical predictions successfully replicate the observed trends in RGBB location, although the zero-point is significantly off.

\subsection{RGBB: benchmark results} \label{ss:rgbb_base}

\begin{figure*}
    \centering
    \includegraphics[width=0.95\textwidth]{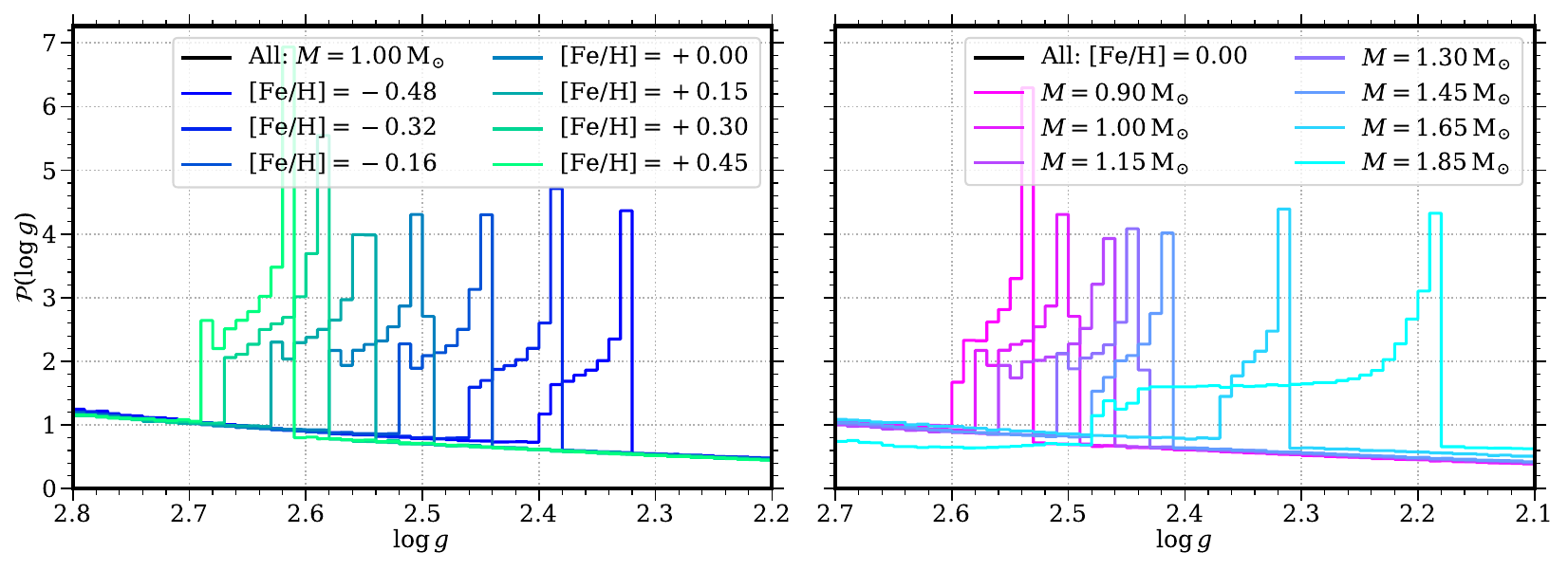}
    \caption{\label{fig:rgbb_pred}Predicted RGBB shapes according to the {\tt benchmark} grid, again in terms of probability density functions normalized to $\mathcal{P} (3.0 > \log g > 2.0) = 1$. Left panel: RGBB shapes at $M = 1.00 \,{\rm M}_\odot$, color-coded by metallicity; right panel: RGBB shapes at $[{\rm Fe}/{\rm H}] = 0.00$, color-coded by mass. Note that all mass and metallicity values quoted here are {\bf birth values}. Besides, these step functions are based on solar mixing length parameter $\alpha_{{\rm MLT}, \odot}$ (see Table~\ref{tab:SunCalibr}); with RGB calibration (see Table~\ref{tab:RgbCalibr}), the shapes would be shifted.}
\end{figure*}

Figure~\ref{fig:rgbb_pred} illustrates the predicted RGBB shapes in our {\tt benchmark} evolutionary tracks. Qualitatively, the metallicity and mass trends of the RGBB location agree with those displayed by the observed sample (see Figure~\ref{fig:rgbb_obs}). However, the bump shape is clearly non-Gaussian, with a sharp edge at the beginning of the ``retrograde motion.'' This is caused by the fact that the evolutionary speed on the Kiel diagram is significantly slower when the hydrogen-burning shell and the outer convective zone are about to converge \citep[see Figure~3 of][]{Christensen-Dalsgaard2015MNRAS}.

\begin{figure*}
    \centering
    \includegraphics[width=0.95\textwidth]{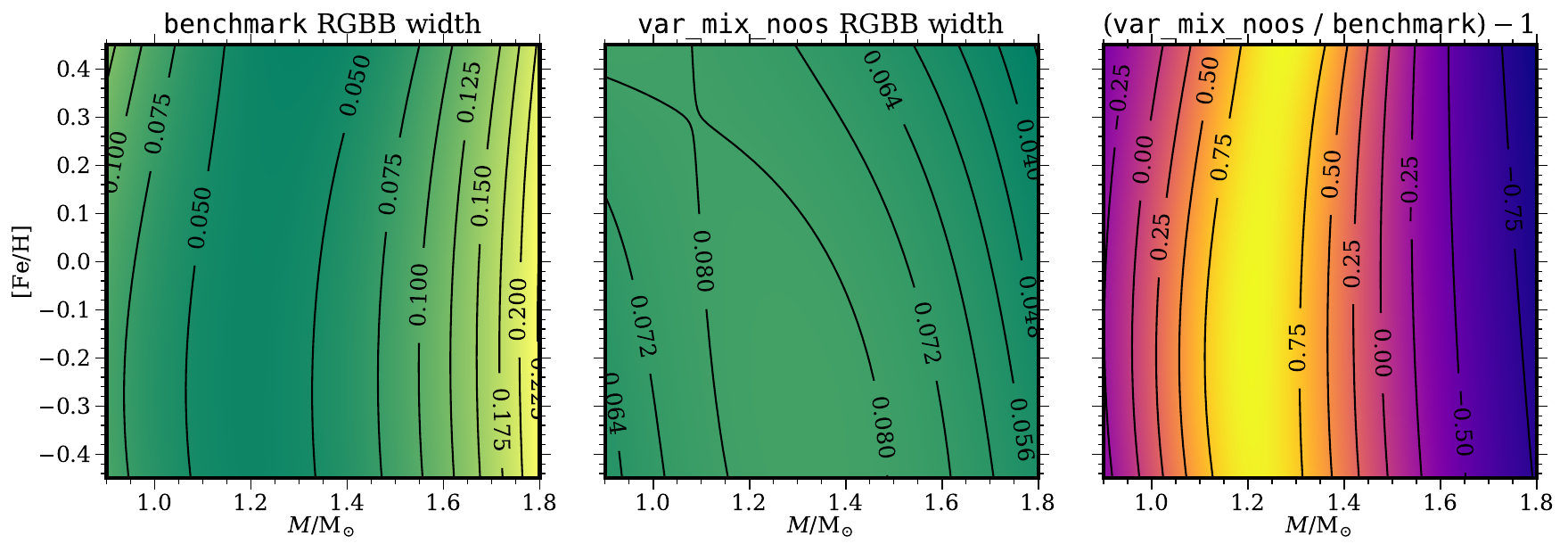}
    \caption{\label{fig:RGBB_width}Predicted RGBB width in terms of $\log g$ discrepancy between the two turning points (see Section~\ref{ss:rgbb_base} for definition). Each panel visualizes a bivariate function of mass and metallicity as a map with contours. Our {\tt benchmark} ({\tt var\_mix\_noos}) predictions are shown on the left (middle) panel, while the fractional discrepancy is shown on the right panel. Note that the coloring scheme is unified for the first two panels.}
\end{figure*}

Interestingly, the RGBB is much wider at $1.85 \,{\rm M}_\odot$ than at $1.65 \,{\rm M}_\odot$. Figure~\ref{fig:RGBB_width} compares the RGBB width predicted by our {\tt benchmark} and {\tt var\_mix\_noos} grids in terms of the $\log g$ discrepancy between the two turning points, i.e., where $\log g$ stops decreasing and starts to increase (first turning point) and where $\log g$ stops increasing and restarts to decrease (second turning point). Fitting coefficients for all grids are tabulated in Table~\ref{tab:RGBB_width}. It is clear that without core overshooting, the RGBB width is much less sensitive to the mass, and the quadratic dependence is inverted: At around $1.20 \,{\rm M}_\odot$, the RGBB width reaches its minimum (maximum) with (without) core overshooting. The ``widening'' effect is only significant at relatively high mass, where the APOKASC-3 sample is small, so we cannot directly test it with this data set.

The RGBB location depends on the mixing length parameter $\alpha_{\rm MLT}$. This does not impact our results because the choice of mixing length is fixed by the requirement that our models reproduce the observed APOKASC-3 locus in the Kiel diagram. As a validation of our predictions, we note that for a $M = 1.0 \,{\rm M}_\odot$, $[{\rm Fe}/{\rm H}] = 0.0$ track, the RGBB center in luminosity space is at $L = 31.03 \,{\rm L}_\odot$  according to our {\tt benchmark} grid, in good agreement with Figure~5 in \citet{SilvaAguirre2020A&A}. For a $1.5 \,{\rm M}_\odot$, $[{\rm Fe}/{\rm H}] = 0.0$ track, our prediction is $L = 62.31 \,{\rm L}_\odot$, relatively high but still within the cohort of their Figure~9.

\begin{figure*}
    \centering
    \includegraphics[width=0.95\textwidth]{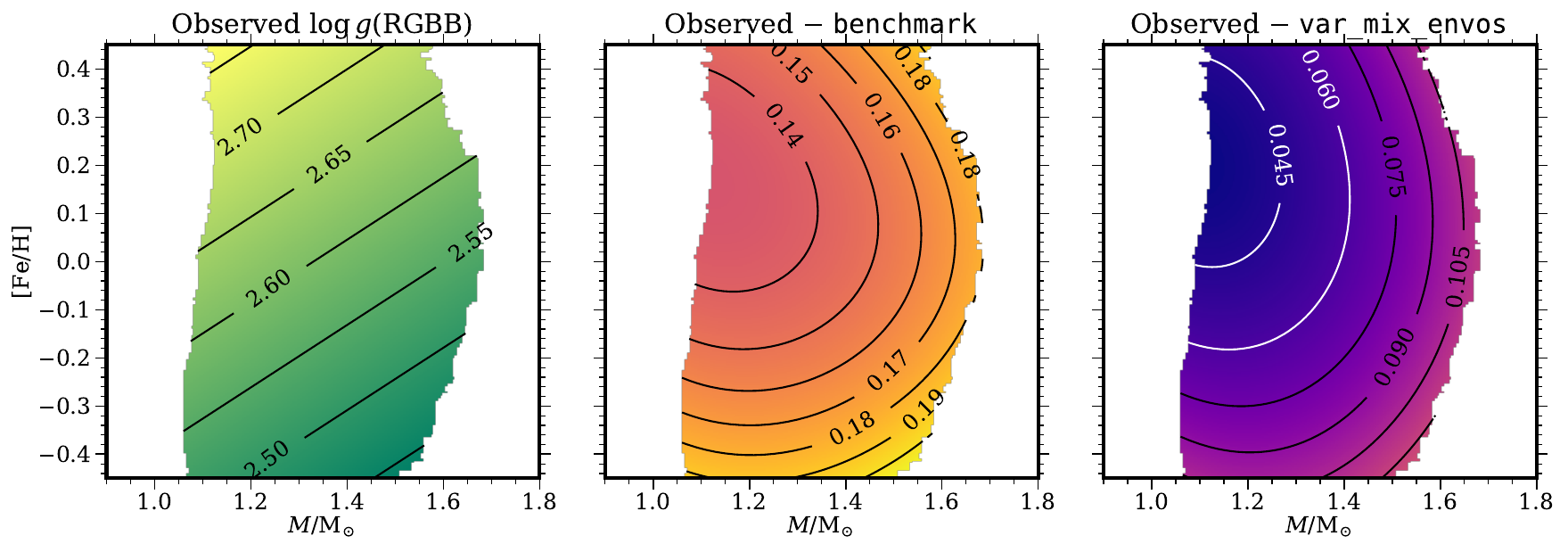}
    \caption{\label{fig:rgbb_base_envos}Comparisons between observed and predicted at RGBB locations. Each panel visualizes a bivariate function of mass and metallicity as a map with contours. Bivariate linear fits to APOKASC-3 data are shown on the left panel, while discrepancies (observed minus predicted) between observational results and {\tt benchmark} ({\tt var\_mix\_envos}) predictions are shown on the middle (right) panel. All three functions are only shown within the contour derived in the middle panel of Figure~\ref{fig:selection}. Note that the coloring scheme is unified for the last two panels.}
\end{figure*}

To facilitate systematic comparisons with the observational data, we take the arithmetic mean of the two $\log g$ values at the two RGBB turning points as the ``predicted bump location.'' Figure~\ref{fig:rgbb_base_envos} compares our {\tt benchmark} and {\tt var\_mix\_envos} predictions to observational results. Since the observed RGBB location assumes a linear dependence on mass and metallicity, contours on the left panel (observations) are simply uniformly spaced straight lines, while those on the middle and right panels (discrepancies) are non-uniformly spaced curves. This difference produces a non-linear pattern on the discrepancy panels. However, as shown on the right panel of Figure~\ref{fig:rgbb_obs}, calibrated {\tt benchmark} models can largely replicate the trends in both mass and metallicity, and the overall offset is significantly larger than the modest trends seen in the central panel. This supports the idea that the difference can be explained by a zero-point shift in the locus of the RGBB. According to the right panel, MIST envelope undershooting is not sufficient to account for this zero-point shift; a quantitative quote is included in the discussion below.

\subsection{RGBB: variations} \label{ss:rgbb_var}

\begin{figure*}
    \centering
    \includegraphics[width=0.95\textwidth]{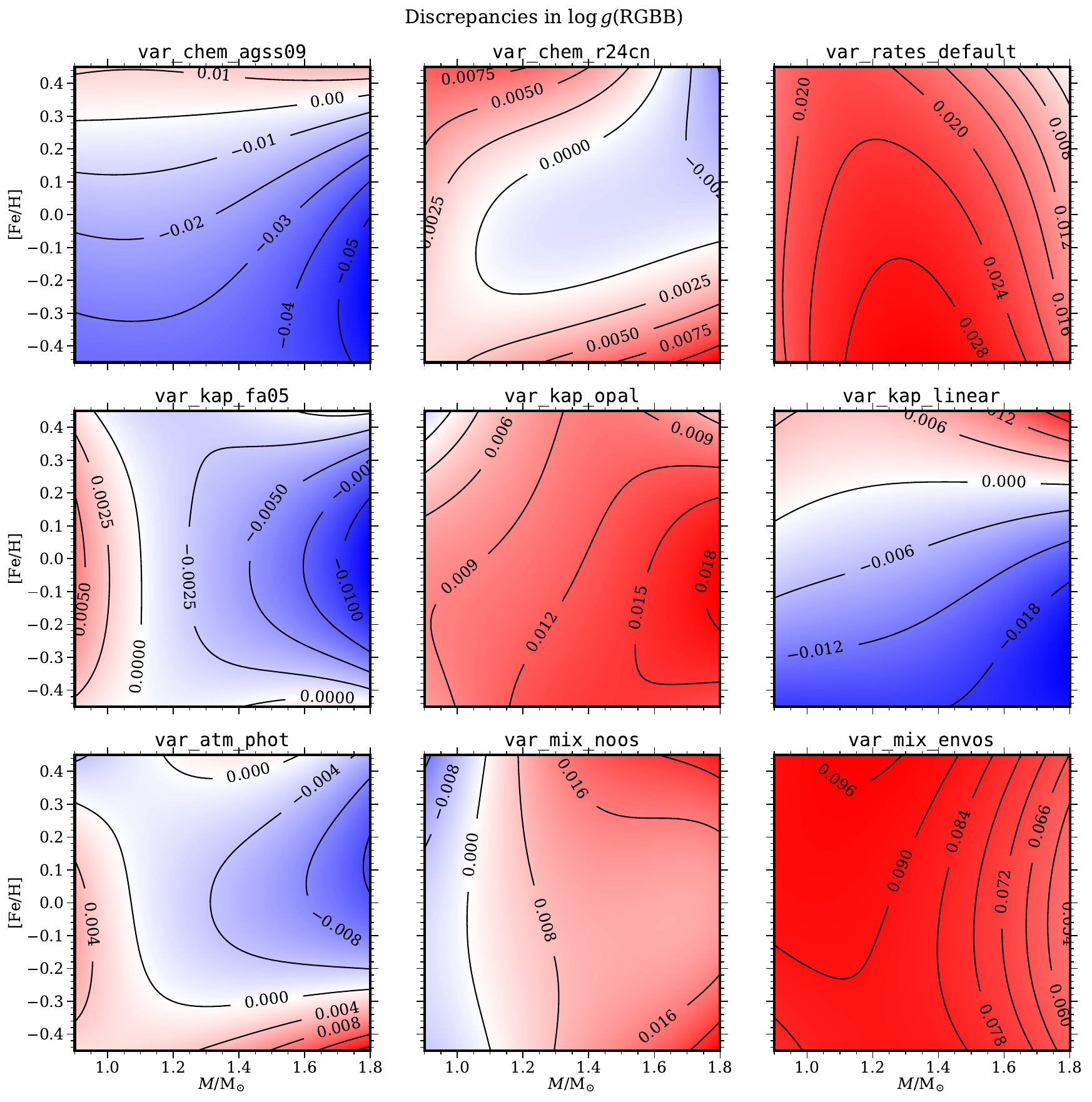}
    \caption{\label{fig:rgbb_var}Similar to Figure~\ref{fig:FDU_DCN_var}, but for $\log g$ at RGBB location.}
\end{figure*}

Figure~\ref{fig:rgbb_var} shows discrepancies in the predicted $\log g$ at the RGBB location. Fitting coefficients for observed and predicted $\log g$ at the RGBB location are tabulated in Table~\ref{tab:RGBB_coeffs}. Like in Section~\ref{ss:fdu_var}, below we comment on some of the variation grids.

\paragraph{{\tt var\_chem\_agss09}} Using the \citet{Asplund2009ARA&A} solar composition leads to modest differences, of order $0.05 \,{\rm dex}$ compared with using the \citet{Grevesse1998SSRv} mixture. However, in most regions of the map ($[{\rm Fe}/{\rm H}] \lesssim 0.3$), the discrepancy only makes the disagreement with observational results worse.

\paragraph{{\tt var\_rates\_default}} This grid leads to slightly better agreement with observational results than {\tt benchmark}. We still trust NACRE II \citep{Xu2013NuPhA} nuclear reaction rates (adopted by most grids) more than the MESA default, but it is worth keeping in mind that the choice of rates can have a significant impact on RGBB location. We leave in-depth investigation to future work.

\paragraph{{\tt var\_kap\_opal} and {\tt var\_kap\_linear}} Interestingly, both high-$T$ opacity and opacity interpolation scheme matter at the $0.01 \,{\rm dex}$ level.

\paragraph{{\tt var\_mix\_noos}} This grid also differs from {\tt benchmark} at the $0.01 \,{\rm dex}$ level. Lack of core overshooting changes the abundance profiles, causing the ``retrograde motion'' to happen at higher gravity. Although this pulls theoretical predictions towards the ``right'' direction at relatively high mass, core overshooting is favored by MS observations. Hence we do not think excluding core overshooting is a promising solution to the RGBB tension between theory and observation.

\paragraph{{\tt var\_mix\_envos}} This is the only grid which differs from {\tt benchmark} at the $0.1 \,{\rm dex}$ level --- and yes, in the ``right'' direction. Figure~\ref{fig:rgbb_base_envos} displays the better agreement between this grid and observational results. Although the agreement is still not perfect, it is possible to make it still better by fine-tuning the overshooting parameter. However, in the next section, we argue that envelope undershooting creates tension with ${\rm Li}$ data in RGB stars. This is in addition with complications for pre-MS ${\rm Li}$ depletion that have been previously discussed in the literature. Here we emphasize that, according to the $a$ column of Table~\ref{tab:rgbb_priors}, the envelope undershooting parameter from MIST (adopted for this grid) probably needs to be multiplied a factor of $1.7076$ to replicate observed RGBB location, which would destroy even more ${\rm Li}$.

\begin{table*}
    \caption{\label{tab:rgbb_priors}Fitting coefficients for the observed RGBB with different grids as theoretical priors. $\mathcal{N}$, $k$, $\mathcal{N} h_{\rm b}$, $c_{\rm b}$, and $w_{\rm b}$ are parameters in Equation~(\ref{eq:bump_fit}); $\Delta c_{\rm b}$ is the discrepancy between $c_{\rm b}$ and predicted RGBB center at $M = 1.0 \,{\rm M}_\odot$, $[{\rm Fe}/{\rm H}] = 0.0$ (observed minus predicted).}
    \centering
    \begin{tabular}{lcccccc}
    \hline
        Grid name & $\mathcal{N}$ & $k$ & $\mathcal{N} h_{\rm b}$ & $c_{\rm b}$ & $w_{\rm b}$ & $\Delta c_{\rm b}$ \\
    \hline
        {\tt benchmark} & $0.0310 \pm 0.0096$ & $1.2509 \pm 0.1146$ & $0.2511 \pm 0.0113$ & $2.6706 \pm 0.0016$ & $0.0356 \pm 0.0017$ & $0.1509$ \\
        {\tt var\_chem\_agss09} & $0.0300 \pm 0.0098$ & $1.2551 \pm 0.1199$ & $0.2671 \pm 0.0134$ & $2.6745 \pm 0.0022$ & $0.0443 \pm 0.0023$ & $0.1712$ \\
        {\tt var\_chem\_r24cn} & $0.0307 \pm 0.0102$ & $1.2557 \pm 0.1229$ & $0.2502 \pm 0.0121$ & $2.6719 \pm 0.0017$ & $0.0352 \pm 0.0018$ & $0.1509$ \\
        {\tt var\_rates\_default} & $0.0319 \pm 0.0102$ & $1.2416 \pm 0.1178$ & $0.2494 \pm 0.0117$ & $2.6674 \pm 0.0017$ & $0.0356 \pm 0.0018$ & $0.1254$ \\
        {\tt var\_kap\_fa05} & $0.0312 \pm 0.0098$ & $1.2485 \pm 0.1158$ & $0.2528 \pm 0.0116$ & $2.6760 \pm 0.0017$ & $0.0361 \pm 0.0018$ & $0.1537$ \\
        {\tt var\_kap\_opal} & $0.0320 \pm 0.0103$ & $1.2385 \pm 0.1188$ & $0.2533 \pm 0.0117$ & $2.6677 \pm 0.0017$ & $0.0357 \pm 0.0018$ & $0.1395$ \\
        {\tt var\_kap\_linear} & $0.0314 \pm 0.0105$ & $1.2413 \pm 0.1239$ & $0.2611 \pm 0.0130$ & $2.6742 \pm 0.0020$ & $0.0397 \pm 0.0021$ & $0.1583$ \\
        {\tt var\_atm\_phot} & $0.0312 \pm 0.0096$ & $1.2489 \pm 0.1137$ & $0.2517 \pm 0.0112$ & $2.6751 \pm 0.0016$ & $0.0355 \pm 0.0017$ & $0.1535$ \\
        {\tt var\_mix\_noos} & $0.0350 \pm 0.0115$ & $1.2028 \pm 0.1216$ & $0.2548 \pm 0.0122$ & $2.6615 \pm 0.0018$ & $0.0366 \pm 0.0019$ & $0.1431$ \\
        {\tt var\_mix\_envos} & $0.0297 \pm 0.0097$ & $1.2675 \pm 0.1211$ & $0.2518 \pm 0.0124$ & $2.6749 \pm 0.0019$ & $0.0377 \pm 0.0020$ & $0.0625$ \\
        \hline
    \end{tabular}
\end{table*}

Other variation grids only differ from {\tt benchmark} at the $0.001 \,{\rm dex}$ level. Table~\ref{tab:rgbb_priors} summarizes RGBB fitting results with different grids as theoretical priors. Like at the end of Section~\ref{sec:fdu}, we estimate theoretical systematics by summing up squares of half the discrepancies between variations and {\tt benchmark}, but we exclude {\tt var\_mix\_envos} here. The zero-point shift of RGBB location in $\log g$ space is:
\begin{itemize}
    \item RGBB center: $0.1509 \pm 0.0017 \,({\rm obs}) \pm 0.0182 \,({\rm sys})$.
\end{itemize}

\subsection{Envelope undershooting and post-MS ${\rm Li}$ depletion} \label{ss:envos_li}

In this section, we investigate envelope undershooting as a proposed solution to the RGBB tension by considering post-main sequence lithium data.

\begin{figure*}
    \centering
    \includegraphics[width=0.95\textwidth]{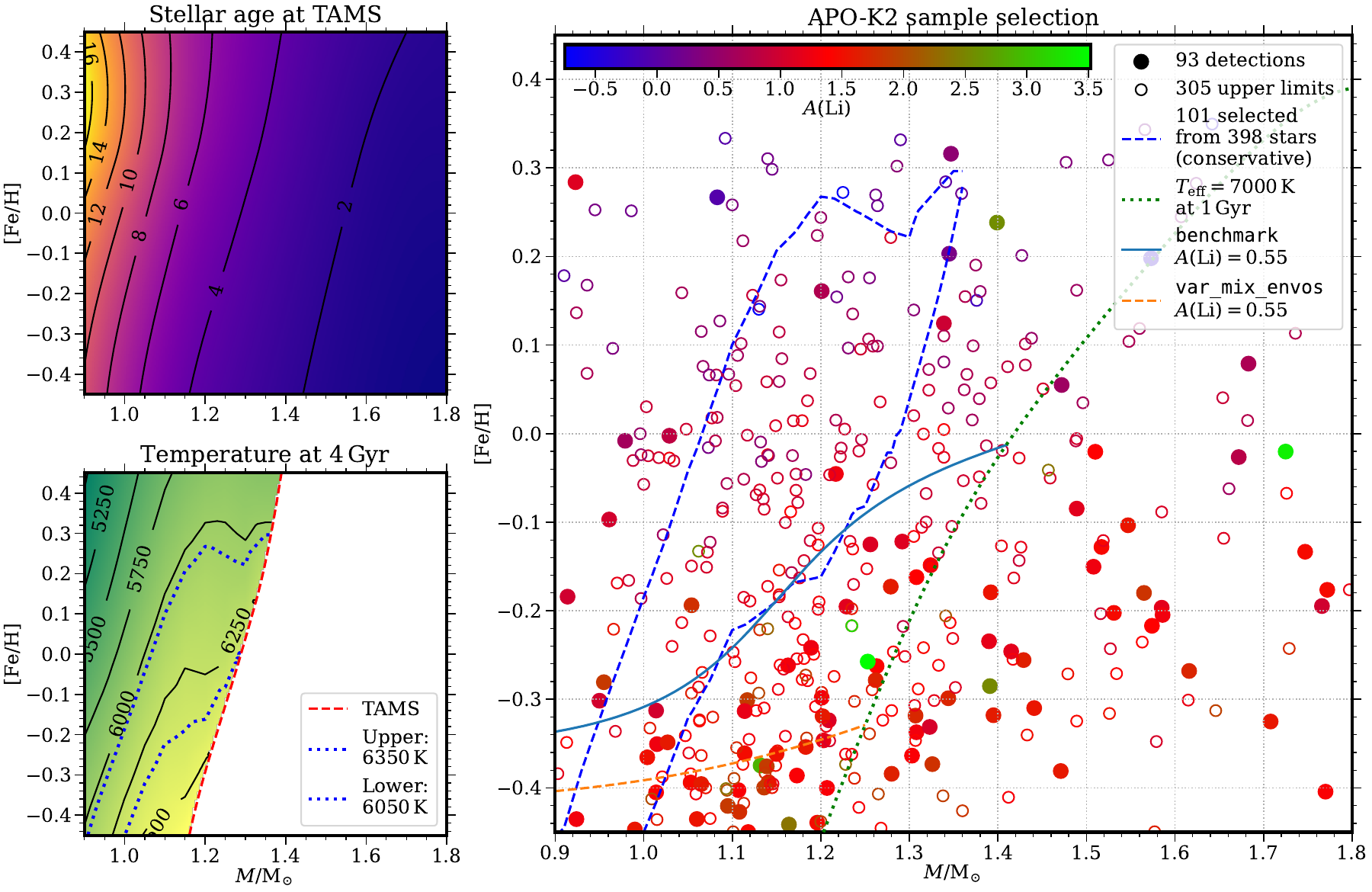}
    \caption{\label{fig:RGBB_Li_select}Sample selection for post-MS ${\rm Li}$ depletion comparisons. Each panel visualizes a bivariate function of mass and metallicity as a map with contours based on our {\tt benchmark} predictions. Upper left panel: stellar age at TAMS (in ${\rm Gyr}$). Lower left panel: effective temperature at the age of $4 \,{\rm Gyr}$, MS stars only; see the text for explanation for the $6050 \,{\rm K}$ and $6350 \,{\rm K}$ contours. Right panel: among $398$ $\alpha$-poor red giants with lithium data in the APO-K2 catalog, $101$ are within the mass-metallicity band found in the first two panels; the data points are color-coded by observed lithium abundance, with filled and empty dots representing detections and upper limits, respectively. In addition, on the right panel, the green dotted curve represents where $T_{\rm eff} = 7000 \,{\rm K}$ at the age of $1 \,{\rm Gyr}$, and the blue solid (orange dashed) curve corresponds to post-MS $A({\rm Li}) = 0.55$ (assuming TAMS $A({\rm Li}) = 2.55$) according to our {\tt benchmark} ({\tt var\_mix\_envos}) grid; see the text for discussion.}
\end{figure*}

There are strong trends in MS ${\rm Li}$ depletion as a function of mass and age. Cool stars experience severe depletion; as an example, the current solar ${\rm Li}$ abundance is more than $100$ times lower than the meteoritic one. \citet{Sestito2005A&A} reported that in Messier 67, MS stars within the temperature range of $6050 \,{\rm K} < T_{\rm eff} < 6350 \,{\rm K}$ at the age of $4 \,{\rm Gyr}$ are observed to preserve an $A({\rm Li})$ value of $2.55 \pm 0.18$. For hotter stars, there is a window in $T_{\rm eff}$ where ${\rm Li}$ is severely depleted, which is referred to as the ${\rm Li}$ dip. At least some hotter stars preserve much more of their ${\rm Li}$, through a mechanism that is still being debated. To establish a reliable comparison between theory and data, we locate the mass-metallicity domain of stars within the range of main sequence $T_{\rm eff}$ where ${\rm Li}$ is preserved in M67. The sample selection process is performed using our {\tt benchmark} results and is illustrated in Figure~\ref{fig:RGBB_Li_select}. On the upper left panel, we set the right boundary of the band by identifying stars which reach their maximum $T_{\rm eff}$ before $4 \,{\rm Gyr}$. The maximum $T_{\rm eff}$ is usually reached right after the terminal age main sequence (TAMS), i.e., core ${\rm H}$ exhaustion, and the change in $A({\rm Li})$ in between is negligible. Then on the lower left panel, we set the upper left and lower right boundaries by considering the surface temperature at $4 \,{\rm Gyr}$. Finally, we apply these boundaries to our APO-K2 sample (see Section~\ref{ss:apok2}) on the right panel, obtaining a subsample of $101$ stars selected from $398$. 

We note that the precise location of the ${\rm Li}$ dip is only measured close to solar metallicity, so hotter stars can still preserve ${\rm Li}$ at a level comparable to the M67 plateau. There is evidence for this in GALAH subgiant data (Smedile et al. 2025, in preparation). As a secondary test, we therefore include stars with an MS effective temperature hotter than the M67 plateau. We truncate this range at an MS $T_{\rm eff}$ of $7000 \,{\rm K}$. Hotter than this, stars do not spin down from magnetized winds and they can leave the MS with much higher ${\rm Li}$ than cool stars \citep{Gao2020MNRAS}; however, there are also many such stars that leave the MS with severe depletion. This requires a different treatment of the turnoff ${\rm Li}$ for the FDU.

In addition, we use three curves to divide the mass-metallicity space into four regions. On the right, stars are predicted to have $T_{\rm eff} > 7000 \,{\rm K}$ at the MS age of $1\,{\rm Gyr}$; $35/92 = 38.04\%$ of these stars have preserved ${\rm Li}$. As discussed above, this may be traced in part to a different MS depletion pattern, so we do not use them to test the FDU. No ${\rm Li}$ detections are expected beyond the two solid curves according to our {\tt benchmark} and {\tt var\_mix\_envos} models, respectively. In the upper left region, only $16/176 = 9.09\%$ of the stars have ${\rm Li}$ detections, which require post-main sequence ${\rm Li}$ production. There is an extensive literature discussing this interesting background population \citep[see][for a discussion]{Martell2021MNRAS}; however, the vast majority of stars are in good agreement with standard theory. In the lower left region, $21/43 = 48.84\%$ of the stars have ${\rm Li}$ detections, in agreement with both our grids. The middle left region corresponds to where ${\rm Li}$ detections are expected by {\tt benchmark} models but not {\tt var\_mix\_envos} models; actually, $21/87 = 24.14\%$ of stars therein have ${\rm Li}$ detections, showing how envelope undershooting deteriorates the agreement with observations.

\begin{figure}
    \centering
    \includegraphics[width=\columnwidth]{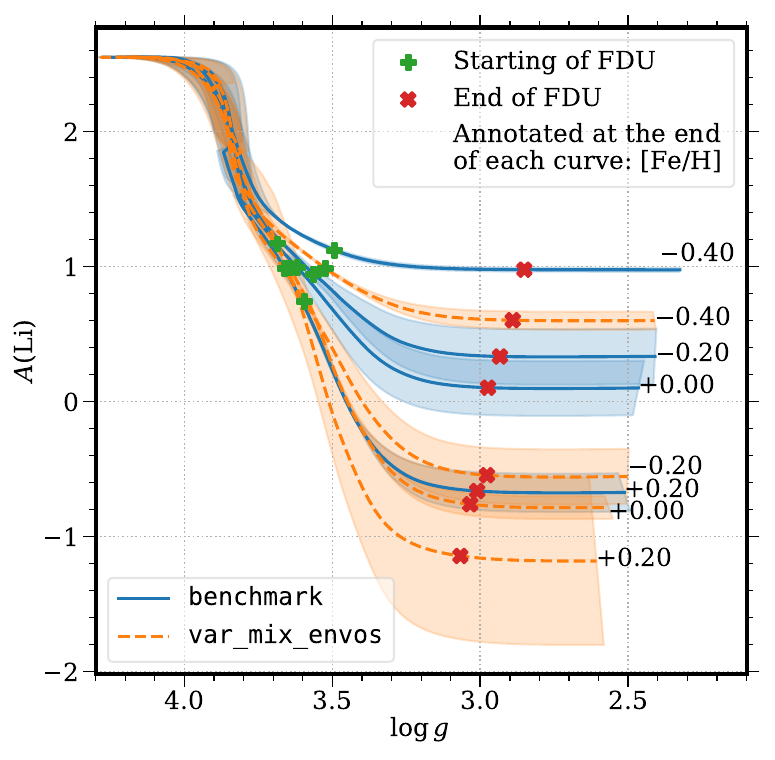}
    \caption{\label{fig:RGBB_ALi_logg}Predicted post-MS lithium abundances as functions of $\log g$. Following \citet{Sestito2005A&A}, the TAMS value is set to $A({\rm Li}) = 2.55$. Solid and dashed curves correspond to our {\tt benchmark} and {\tt var\_mix\_envos} grids, respectively. These curves are arithmetic means of maximum and minimum $A({\rm Li})$ values within the mass range at each metallicity (see the middle panel of Figure~\ref{fig:RGBB_Li_select}), which are shown as shaded regions. Metallcities of the curves are annotated at the end of each curve. Starting (i.e., bottom of RGB) and end points of FDU are marked with green and red crosses, respectively.}
\end{figure}

Figure~\ref{fig:RGBB_ALi_logg} visualizes predicted post-MS $A({\rm Li}) = \log_{10} (n_{\rm Li}/n_{\rm H})$ values as functions of $\log g$. We presume that $A ({\rm Li}) = 2.55$ for stars within the mass-metallicity band found in Figure~\ref{fig:RGBB_Li_select}. At each step on the osculating tracks given by each grid, we derive the maximum and minimum $A({\rm Li})$ values at each given metallicity, and plot the arithmetic mean of these two extrema to avoid cluttering. With starting and end points of the first dredge-up marked on each curve, we see that differences in post-MS ${\rm Li}$ depletion mostly occur during FDU: Before FDU (i.e., on the subgiant branch), the ${\rm Li}$ depletion is largely insensitive to both metallicity and whether we include envelope undershooting; after FDU, the ${\rm Li}$ abundance basically remains constant on each curve.

\begin{figure*}
    \centering
    \includegraphics[width=0.95\textwidth]{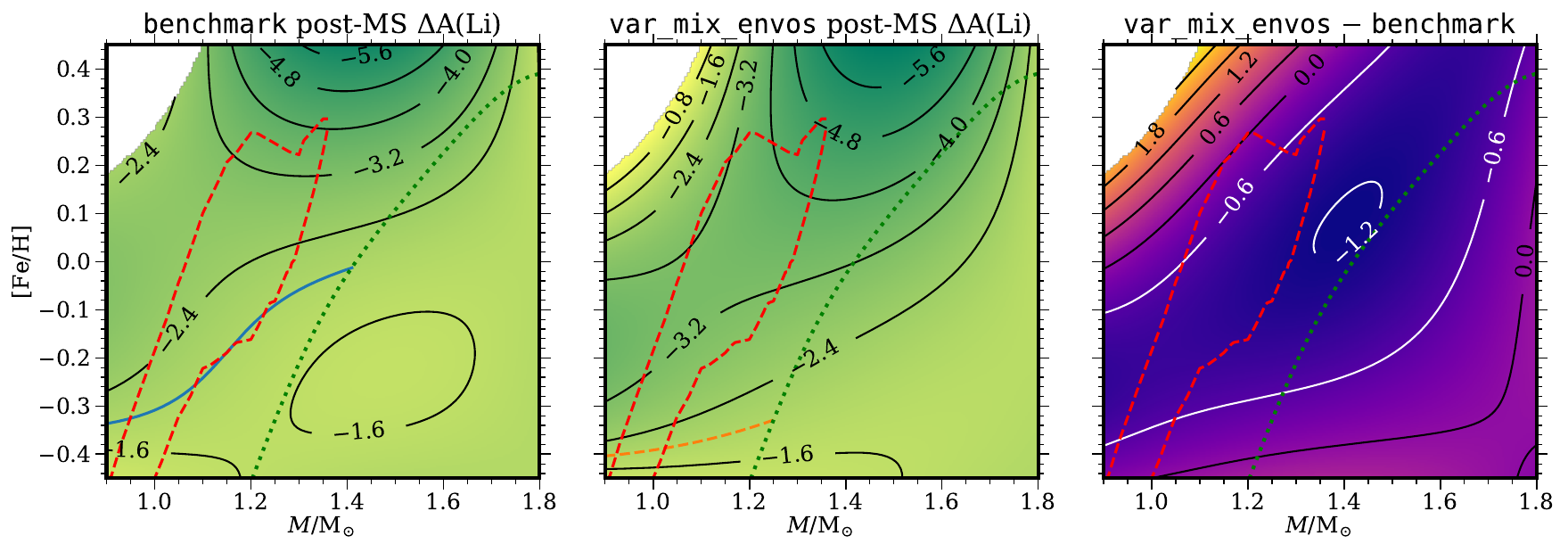}
    \caption{\label{fig:RGBB_DALi}Predicted post-MS surface lithium abundance change $\Delta A({\rm Li})$. Each panel visualizes a bivariate function of mass and metallicity as a map with contours; the upper left corners correspond to artifacts and are omitted. Our {\tt benchmark} ({\tt var\_mix\_envos}) predictions are shown on the left (middle) panel, while the discrepancy is shown on the right panel. Note that the coloring scheme is unified for the first two panels. Boundaries of the mass-metallicity band found in Figure~\ref{fig:RGBB_Li_select} are shown as a red dashed contour; the green dotted, blue solid, and orange dashed curves are the same as those in the right panel of Figure~\ref{fig:RGBB_Li_select}.}
\end{figure*}

Figure~\ref{fig:RGBB_DALi} compares our {\tt benchmark} and {\tt var\_mix\_envos} grids in terms of post-MS ${\rm Li}$ depletion, defined as change in surface ${\rm Li}$ abundance $A({\rm Li})$ values from TAMS\footnote{MESA defines this as when the central hydrogen mass fraction reaches $X = 0.1$, which is not quite ``terminal'' yet. However, this does not affect our results, as $A({\rm Li})$ basically does not change on the final segment of MS in our models.} to RGBB; fitting coefficients for all grids are tabulated in Table~\ref{tab:DALi_coeffs}. Positive $\Delta A({\rm Li})$ values at the low-mass, high-metallicity corner are artifacts and are omitted from the maps. In most regions of the mass-metallicity space, envelope undershooting destroys more ${\rm Li}$ between TAMS and RGBB by about an order of magnitude, which would substantially reduce the chance of any ${\rm Li}$ detection on the RGB.

\begin{figure*}
    \centering
    \includegraphics[width=0.95\textwidth]{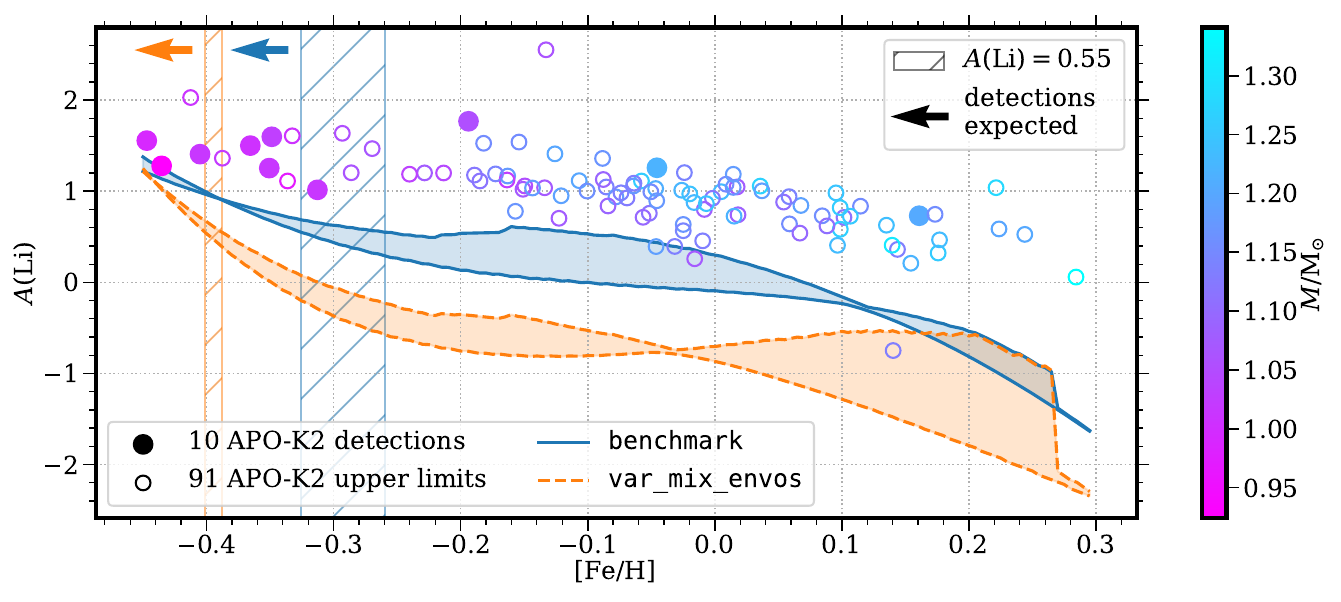}
    \caption{\label{fig:RGBB_ALi_FeH} Lithium abundance $A({\rm Li})$ at lower RGB as a function of metallicity $[{\rm Fe}/{\rm H}]$. The data points are stars in the APO-K2 catalog with ${\rm Li}$ detections (filled) or upper limits (empty) from GALAH (see Section~\ref{ss:apok2}), color-coded by mass. Blue solid (orange dashed) curves shows maximum and minimum $A({\rm Li})$ values as functions of metallicity given by the {\tt benchmark} ({\tt var\_mix\_envos}) grid. Metallicity ranges corresponding to $A({\rm Li}) = 0.55$, beyond which detections are expected, are shown as hatched regions.}
\end{figure*}

\begin{figure*}
    \centering
    \includegraphics[width=0.95\textwidth]{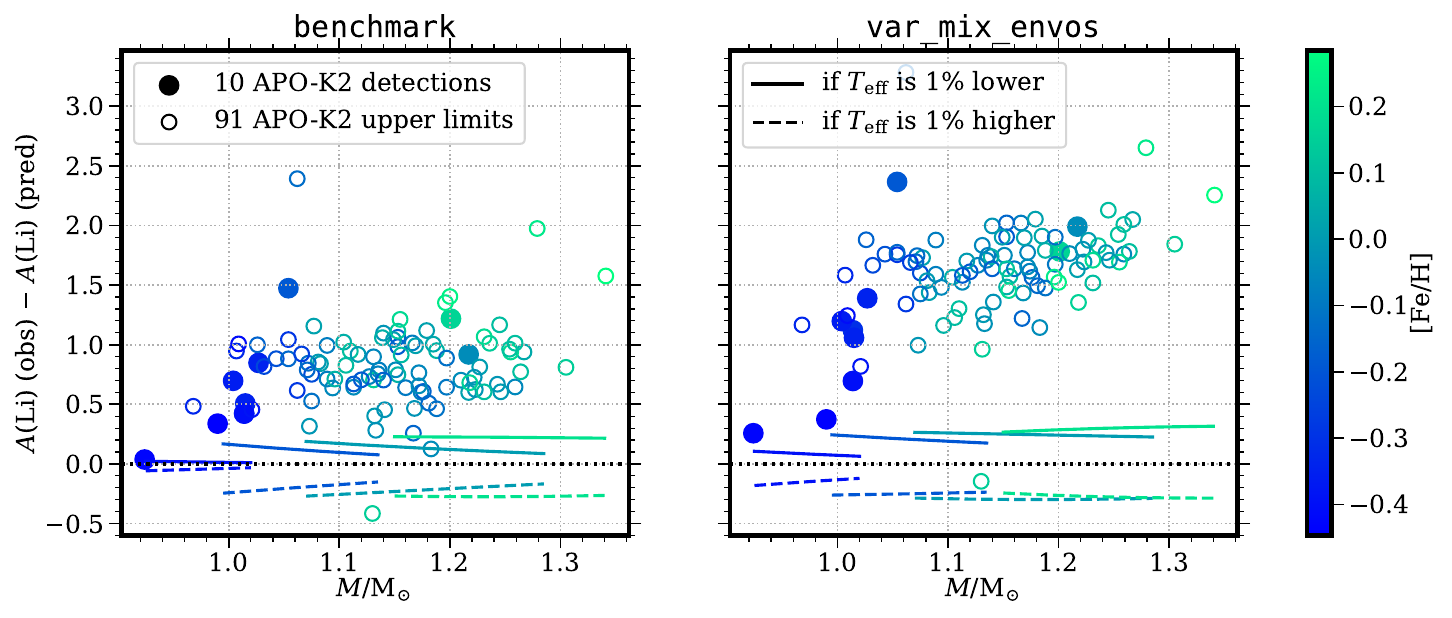}
    \caption{\label{fig:RGBB_ALi_recast}Discrepancies between observed and predicted lithium abundances. Results based on the {\tt benchmark} and {\tt var\_mix\_envos} grid are shown in the left and right panels, respectively. In each panel, ${\rm Li}$ detections and upper limits in the APO-K2 catalog are shown in filled and empty dots, respectively. Systematics due to temperature uncertainties propagated though mixing length calibration (see Section~\ref{ss:cal_synth}) are represented as solid (if $T_{\rm eff}$ is lower) and dashed (if $T_{\rm eff}$ is higher). Both dots and curves are color-coded by metallicity.}
\end{figure*}

In Figure~\ref{fig:RGBB_ALi_FeH}, we compare observed and predicted ${\rm Li}$ abundances by plotting $A({\rm Li})$ versus metallicity. The APO-K2 sample and both our grids agree on the statement that $A({\rm Li})$ decreases with metallicity. Both grids predict severe ${\rm Li}$ depletion at metallicities above $\sim -0.3$, in agreement with the data. However, ${\rm Li}$ is detected between $-0.3$ and $-0.4$, in agreement with standard models but in conflict with the envelope undershooting ones. In Figure~\ref{fig:RGBB_ALi_recast}, we recast these comparisons by plotting the discrepancy between observed and predicted $A({\rm Li})$ for each star versus mass, and consider the impact of the percent-level temperature uncertainties in the APOKASC-3 catalog. During the RGB mixing length calibration process (see Section~\ref{ss:cal_synth}), we found that $A({\rm Li})$ values predicted for low-mass stars are highly sensitive to $\alpha_{\rm MLT}$, with about an order of magnitude less (more) destruction if $\alpha_{\rm MLT}$ is smaller (larger) by $0.3$. Therefore, ${\rm Li}$ predictions are related to temperature measurements. Although $A({\rm Li})$ uncertainties at the $0.1 \,{\rm dex}$ level are not large enough to explain the discrepancies between APO-K2 and {\tt benchmark}, we note that {\tt var\_mix\_envos} has much larger discrepancies as well as a mass trend, and thus is harder to justify.

If envelope undershooting is not a promising solution, how to address the mismatch between observed and predicted RGBB locations? We think physicists may need to resort to other components of stellar interiors. For instance, different surface boundary conditions or alternative models for degenerate material \citep[e.g.,][]{Caplan2022MNRAS} could make the ${\rm He}$ core grow faster, so that it would meet the bottom of the convective envelope earlier, even the latter is not deeper than currently predicted. We leave such investigation to future work.

\section{Summary and discussion} \label{sec:discuss}

The first dredge-up is a deep prediction of stellar theory, and we are now in a position to test this theory rigorously. In this work, we have modeled APOKASC-3 \citep{Pinsonneault2025ApJS} $\alpha$-poor red giant branch (RGB) stars with MESA \citep{Paxton2011, Paxton2013, Paxton2015, Paxton2018, Paxton2019, Jermyn2023ApJS} r22.11.1. We have systematically explored different input physics options (see Sections~\ref{ss:comp_nuc} and \ref{ss:variation}), and calibrated Solar composition and RGB mixing length for all our models (see Section~\ref{ss:cal_synth}).

The overall change in the surface ${\rm C}$ to ${\rm N}$ abundance ratio as seen in the data is systematically smaller than that predicted in models, but has relative mass and birth composition trends similar to theoretical predictions (see Section~\ref{sec:fdu}). This is surprisingly resilient against changes in the input physics, but does depend on the assumed birth ${\rm C}$ to ${\rm N}$. Unfortunately, adopting the observed pre-dredge-up ${\rm C}$ to ${\rm N}$ trends with ${\rm Fe}$ to ${\rm H}$ partially worsens agreement with the data. However, the magnitude of this effect is comparable to systematic errors in the absolute abundance scale, so we conclude that it is too early to use this as evidence for a serious defect in our stellar interiors models. Instead, we encourage further work constraining the absolute abundances, which could yield powerful insights and challenges to theory if they can be improved to better than the $0.1 \,{\rm dex}$ level.

This pattern --- a zero-point offset between theory and data, but good agreement on mass and birth composition trends --- is replicated in the location of the RGBB, but in the opposite sense: The true RGBB is fainter than the predicted one (see Section~\ref{sec:rgbb}). Our result here is similar to that obtained by prior studies in the literature. There is an interesting tension between these two results. Relative to theory, the FDU results suggest less nuclear processed material has been mixed into the envelope, while the RGBB results imply that the surface convection zone reached deeper into the star. The standard explanation for the RGBB in the literature is to add envelope undershooting, which moves the RGBB while having only a small effect on the predicted surface ${\rm C}$ to ${\rm N}$. This appears at first glance to solve the problem. But, as established in the prior literature, envelope undershooting causes significant problems in other, related astrophysical contexts. Of these, the most severe is that it predicts greatly increased pre-MS destruction of ${\rm Li}$ in stars with a mechanical structure similar to evolved red giants. Our work adds another, and more direct, challenge to envelope undershooting as an explanation. ${\rm Li}$ is not merely dredged-up in low-mass and high-metallicity giants: It is also directly destroyed in the envelope in addition to passive dilution. This creates a ``forbidden zone'' where any ${\rm Li}$ detections require ${\rm Li}$ production during the post-MS, most likely through interactions with companion stars or giant planets. We see clear evidence for such a boundary in GALAH ${\rm Li}$ data, and standard models without envelope undershooting are consistent with the data. Envelope undershooting models, however, predict that this zone should reach much further down in metallicity than is supported by the data.

We caution that the observed ${\rm Li}$ depletion pattern is complex, and that ${\rm Li}$ predictions are more sensitive to model input physics than ${\rm CN}$ ones are. A deeper direct investigation of ${\rm Li}$ depletion, destruction, and production in red giants would therefore be of great value. However, the physical problems with envelope undershooting lead us to question whether another physical effect, not included in current models, might be responsible instead for the clear challenges to theory that we find. Such a mechanism could include, for example, a significant change in the mapping between core mass and surface gravity, or a change in the rate of core growth relative to the deepening of the envelope. At minimum, our work indicates that envelope undershooting should be carefully re-evaluated and tested against the full set of data before being adopted as a convincing physical explanation for the clear mismatches that we see with theoretical expectations.

In general, we think self-consistent stellar models are preferred for making reliable predictions for RGB stars in the era of precision asteroseismology. Thenceforth, we are working on a public release of the Yale Rotating Stellar Evolution Code \citep[YREC; carefully described in][Pinsonneault et al. 2025, in preparation]{vanSaders2012ApJ}. Meanwhile, a follow-up paper with MESA (Cao \& Pinsonneault 2025, in preparation) will measure mass loss using APOKASC-3 data and make age predictions based on the resulting synthetic models.

\section*{Acknowledgments}

We are grateful for judicious feedback from Jamie Tayar and the anonymous referee. We thank Joel C. Zinn, Jessica Schonhut-Stasik and Jack Warfield for their help regarding the APOKASC-3 and APO-K2 data catalogs. We acknowledge insightful comments from our colleagues at OSU, especially Jennifer A. Johnson, Amanda Ash, and Jack Roberts. We appreciate helpful communications with the MESA developers and other users, especially Meridith Joyce, Warrick Ball, and Ebraheem Farag. We thank Paola Marigo and Franck Delahaye for their help regarding the opacity tables they developed.

KC acknowledges support from NASA under subaward AWP-10019534 from JPL. MHP is supported by NASA grant 80NSSC24K0637. Last but not least, KC thanks Annika H. G. Peter for her support as his co-advisor during the first year of this project.

\vspace{5mm}
\facilities{CCAPP Condo on Pitzer Cluster of the Ohio Supercomputer Center (\url{https://www.osc.edu/ccapp_condo_on_pitzer_cluster})}

\software{MESA \citep{Paxton2011, Paxton2013, Paxton2015, Paxton2018, Paxton2019, Jermyn2023ApJS} r22.11.1., MESA software development kit \citep[MESA SDK;][]{Townsend2021zndo} for Linux, {\sc MESA Reader} \citep{Wolf2017zndo};
          {\sc NumPy} \citep{Harris2020Natur}, {\sc SciPy} \citep{Virtanen2020NatMe}, {\sc Matplotlib} \citep{Hunter2007CSE}, {\sc pandas} \citep{Reback2022zndo}}

\section*{Data Availability}

The APOKASC-3 data catalog is available in the published paper. The MESA software is open source; our MESA inlists and visualization notebooks are available on Zenodo under an open-source Creative Commons Attribution license: \dataset[doi: 10.5281/zenodo.15863879]{https://doi.org/10.5281/zenodo.15863879}; the simulation results are available on request. The Python programs for this project are available in the GitHub repository:
\begin{itemize}
\item \url{https://github.com/kailicao/mesa_apokasc.git}
\end{itemize}

\appendix

\section{Rescaling $[{\rm C}/{\rm Fe}]$ and $[{\rm N}/{\rm Fe}]$ errors in APOKASC-3} \label{app:CN_errors}

As mentioned in the APOKASC-3 catalog paper \citep{Pinsonneault2025ApJS}, spectroscopic errors in APOGEE DR17 \citep{Abdurrouf2022ApJS} are underestimated. While we have adopted a set of more sensible uncertainties for $T_{\rm eff}$, $\log g_{\rm spec}$, $[{\rm Fe}/{\rm H}]$, and $[\alpha/{\rm Fe}]$ \citep[see Table~3 of][]{Pinsonneault2025ApJS}, $[{\rm C}/{\rm Fe}]$ and $[{\rm N}/{\rm Fe}]$ errors are simply DR17 values, with medians of $0.0147$ and $0.0179$, respectively. In this appendix, we rescale these two errors to reproduce the observed random scatter in abundance and investigate the correlation between them.

For $[{\rm X}/{\rm Fe}]$, where ${\rm X} = {\rm C}, {\rm N}$, we perform a bivariate fit to our APOKASC-3 lower RGB sample (see Section~\ref{ss:select})
\begin{align}
    \nonumber [{\rm X}/{\rm Fe}]_{\rm fit} (M, [{\rm Fe}/{\rm H}]) = a + b_1 (M/{\rm M}_\odot-1) + b_2 (M/{\rm M}_\odot-1)^2 \\
    + c_1 [{\rm Fe}/{\rm H}] + c_2 [{\rm Fe}/{\rm H}]^2 + d (M/{\rm M}_\odot-1) [{\rm Fe}/{\rm H}]
\end{align}
and compute the reduced $\chi^2$ as
\begin{align}
    \chi^2 ({[{\rm X}/{\rm Fe}]}) = \frac{1}{n-6} \sum_i \left[ \frac{[{\rm X}/{\rm Fe}] - [{\rm X}/{\rm Fe}]_{\rm fit} (M_i, [{\rm Fe}/{\rm H}]_i)}{f_{[{\rm X}/{\rm Fe}]} \sigma_{[{\rm X}/{\rm Fe}], i}} \right]^2,
\end{align}
where $n$ is the number of stars, $6$ is the number of parameters in our fitting function, $i$ is the star index, $f_{[{\rm X}/{\rm Fe}]}$ is a global scaling factor, and $\sigma_{[{\rm X}/{\rm Fe}]}$ is the DR17 error. By tuning $f_{[{\rm X}/{\rm Fe}]}$ so that $\chi^2_{[{\rm X}/{\rm Fe}]} = 1$, we obtain $f_{[{\rm C}/{\rm Fe}]} = 3.0278$ and $f_{[{\rm N}/{\rm Fe}]} = 2.7109$, resulting in median errors of $0.0445$ and $0.0485$, respectively. These scaling factors are recommended for users interested in $[{\rm C}/{\rm Fe}]$ and $[{\rm N}/{\rm Fe}]$ uncertainties in APOKASC-3.

Since $[{\rm C}/{\rm N}] = [{\rm C}/{\rm Fe}] - [{\rm N}/{\rm Fe}]$, the associated uncertainty can be written as
\begin{align}
    \sigma_{[{\rm C}/{\rm N}], i}^2 = \sigma_{[{\rm C}/{\rm Fe}], i}^2 + \sigma_{[{\rm N}/{\rm Fe}], i}^2 - 2\rho_{\rm C,N} \sigma_{[{\rm C}/{\rm Fe}], i} \sigma_{[{\rm N}/{\rm Fe}], i}.
\end{align}
Assuming no correlation between $[{\rm C}/{\rm Fe}]$ and $[{\rm N}/{\rm Fe}]$ (i.e., $\rho_{\rm C,N} = 0$), we would have $\chi^2 ({[{\rm C}/{\rm N}])} = 1.0941$. This indicates that $[{\rm C}/{\rm Fe}]$ and $[{\rm N}/{\rm Fe}]$ errors are actually negatively correlated, which agrees with the understanding that both carbon and nitrogen abundances are largely based molecular spectral features. By tuning $\rho_{\rm C,N}$ so that $\chi^2 ({[{\rm C}/{\rm N}]}) = 1$, we obtain $\rho_{\rm C,N} = -0.0946$, giving a median $[{\rm C}/{\rm N}]$ error of $0.0658$.

We have applied results in this appendix to both APOKASC-3 and APO-K2 samples for deriving error bars in Section~\ref{ss:fdu_base}.

\section{Details of modeling} \label{app:models}

In this appendix, we provide additional details of modeling to supplement Section~\ref{sec:models}. In Sections~\ref{ss:comp_nuc} and \ref{ss:variation}, we discuss the key ingredients for models: reference solar mixtures, nuclear reaction rates, radiative opacities, surface boundary conditions, and mixing settings. Then in Sections~\ref{ss:numerics} and \ref{ss:cal_synth}, we present our numerical choices, calibrate our models, and synthesize tracks used in comparisons.

\subsection{Chemical composition and nuclear reactions} \label{ss:comp_nuc}

The relative abundances of heavy elements in the Sun are well-constrained in meteorites, but the light species (${\rm CNO}$ and ${\rm Ne}$) and the $Z/X$ ratio require stellar atmospheres theory. As a result, there are ongoing controversies related to the solar mixture: \citet{Anders1989GeCoA, Grevesse1993oee, Grevesse1998SSRv, Holweger2001AIPC, Lodders2003ApJ, Grevesse2007SSRv, Lodders2009LanB, Asplund2009ARA&A, Caffau2011SoPh, Asplund2021A&A, Magg2022A&A}. Recent papers have converged on two distinct families of solutions: a ``high scale'' (e.g., GS98 and MBS22) and a ``low scale'' (e.g., AGSS09 and AAG21). The high scale is in better agreement with helioseismic data \citep[see][]{Magg2022A&A}, so we adopt it as our base case.

For theoretical purposes, the reference solar composition is used to: i) set the initial composition of the models, ii) make the high-$T$ opacity tables, and iii) make the low-$T$ opacity tables. In principle, these should be consistent with each other, yet exact consistency is hard to achieve for technical reasons. Therefore, we simply use reference mixtures adopted by MESA developers and opacity providers (see Section~\ref{ss:variation}), which can be slightly different. An exception to this approximate consistency is that Opacity Project tables for standard GS98 composition are used for the variation grid following \citet{Roberts2024MNRAS}, as the ${\rm CNO}$ abundances have a small impact on high-$T$ opacities \citep{Bahcall1988RvMP}.

Our focus is on ${\rm H}$-shell burning stars. As for nuclear reaction network, the default {\tt \textquotesingle basic.net\textquotesingle} only provides a reasonable approximation in the scenario where the temperature is so low that advanced burning and hot ${\rm CNO}$ issues can be safely ignored. Therefore, we adopt {\tt \textquotesingle pp\_and\_cno\_extras.net\textquotesingle} for all our simulations. MESA's default nuclear reaction rates are from JINA REACLIB \citep{Cyburt2010ApJS}, NACRE \citep{Angulo1999NuPhA} and additional tabulated weak reaction rates \citep{Fuller1985ApJ, Oda1994ADNDT, Langanke2000NuPhA}. Screening is included via the prescription of \citet{Chugunov2007}. Thermal neutrino loss rates are from \citet{Itoh1996}.

\subsection{Alternative physics routines} \label{ss:variation}

In addition to chemical composition and nuclear reactions, this work systematically tests the input physics in terms of radiative opacities, surface boundary conditions, and treatment of convection and convective zones.

For high-$T$ opacities, we note that OPAL \citep{Iglesias1993ApJ, Iglesias1996ApJ} fully considers the collective effects in atomic physics but adopts simplified models for individual species, while OP \citep{Badnell2005MNRAS} models the ionization states in a more sophisticated way but has a more simplified treatment of collective plasma effects.

In the low-$T$ regime, variations in ${\rm CNO}$ abundances are important for opacity calculations, as these elements are abundant and can form a wide variety of molecules. Therefore, this work adopts {\AE}SOPUS 1.0 opacities \citep{Marigo2009A&A} for most grids. According to \citet{DiazReeve2023poster, DiazReeve2023RNAAS}, {\AE}SOPUS 1.0 and the default \citet{Ferguson2005ApJ} are intrinsically different at the $10\%$ level for solar models.

Different surface boundary conditions, may lead to large differences in the internal structure of stars. Most of our grids adopt the MESA default {\tt \textquotesingle Eddington\textquotesingle} grey atmosphere. To accompany this choice, we set {\tt atm\_T\_tau\_opacity} to {\tt \textquotesingle varying\textquotesingle}, which is {\tt \textquotesingle fixed\textquotesingle} by default. Full model atmosphere calculations can also be used to infer the relationship between pressure and temperature at the base of the atmosphere. In MESA, this is enabled with the {\tt \textquotesingle photosphere\textquotesingle} boundary condition, which is based on a combination of PHOENIX \citep{Hauschildt1999ApJa, Hauschildt1999ApJb} and ATLAS9 \citep{Castelli2003IAUS} models. This is used in our {\tt var\_atm\_phot grid.}

In MLT \citep{Henyey1965ApJ, Cox1968pss, Kuhfuss1986A&A}, the dimensionless mixing length parameter, $\alpha_{\rm MLT}$, defined as the mixing length divided by the pressure scale height, is a free parameter in the theory that we can calibrate to obtain a first-order agreement with observations (see Section~\ref{ss:cal_synth}).

\subsection{Numerical choices} \label{ss:numerics}

We adopt stringent (in MESA terms, ``gold'') tolerances for numerical convergence. The numerical resolutions are set to establish a reasonable balance between precision and time consumption. We found that the combination of {\tt time\_delta\_coeff = 0.3d0} and {\tt mesh\_delta\_coeff = 1.0d0} yields a absolute relative error in post-FDU surface $[{\rm C}/{\rm N}]$ of $\sim 0.002$ compared to a counterpart with {\tt time\_delta\_coeff = 0.1d0} and {\tt mesh\_delta\_coeff = 0.2d0}, but takes $\sim 15$ times less time. Therefore, we adopt these resolution parameters for all the models.

We initialize our models at central deuterium exhaustion in the pre-main sequence phase of evolution. We infer the solar helium and mixing length by requiring that solar models match the solar luminosity and radius at the age of the Sun, which we take as $4.568 \,{\rm Gyr}$ \citep{Soderblom2010ARA&A, Desch2023Icar}. We end our RGB models when the surface $\log g$ reaches $1.0$.

To match our observational sample selection (see Section~\ref{ss:select}), for each set of MESA settings, we run a 3D simulation grid of RGB models in mass, metallicity, and mixing length parameter $\alpha_{\rm MLT}$. Through testing, we found that the configuration of nodes (i.e., parameter combinations) with initial $M/{\rm M}_\odot$ values $\{ 0.90$, $1.00$, $1.15$, $1.30$, $1.45$, $1.65$, $1.85 \}$ and initial $[{\rm Fe}/{\rm H}]$ values $\{ -0.48$, $-0.32$, $-0.16$, $0.00$, $0.15$, $0.30$, $0.45 \}$ is a reasonable sampling of the mass-metallicity space.\footnote{In the rare cases where a model fails due to numerical glitches, we manually ``nudge'' the corresponding node to produce a substitute very close to the desired model.} For each grid, the mixing length dimension has three nodes, $\{ \alpha_{{\rm MLT}, \odot} - 0.3$, $\alpha_{{\rm MLT}, \odot}$, $\alpha_{{\rm MLT}, \odot} + 0.3 \}$; see below for the solar calibrated $\alpha_{{\rm MLT}, \odot}$.

\subsection{Calibrations and synthesis} \label{ss:cal_synth}

\begin{table}
    \caption{\label{tab:SunCalibr}Solar calibration results. These include birth metal ($Z_{\odot, {\rm birth}}$) and ${\rm He}$ ($Y_{\odot, {\rm birth}}$) mass fractions, and optimal mixing length parameter ($\alpha_{{\rm MLT}, \odot}$) of the solar model. The {\tt var\_chem\_r24cn} grid does not include a solar model, hence {\tt benchmark} results are used.}
    \centering
    \begin{tabular}{lccc}
    \hline
        Grid name & $Z_{\odot, {\rm birth}}$ & $Y_{\odot, {\rm birth}}$ & $\alpha_{{\rm MLT}, \odot}$ \\
    \hline
        {\tt benchmark} & $0.01857$ & $0.2696$ & $1.7547$ \\
        {\tt var\_chem\_agss09} & $0.01500$ & $0.2642$ & $1.7533$ \\
        {\tt var\_chem\_r24cn} & --- & --- & --- \\
        {\tt var\_rates\_default} & $0.01857$ & $0.2697$ & $1.7547$ \\
        {\tt var\_kap\_fa05} & $0.01859$ & $0.2698$ & $1.8202$ \\
        {\tt var\_kap\_opal} & $0.01855$ & $0.2711$ & $1.7453$ \\
        {\tt var\_kap\_linear} & $0.01864$ & $0.2684$ & $1.7518$ \\
        {\tt var\_atm\_phot} & $0.01858$ & $0.2697$ & $1.8399$ \\
        {\tt var\_mix\_noos} & $0.01857$ & $0.2696$ & $1.7546$ \\
        {\tt var\_mix\_envos} & $0.01835$ & $0.2686$ & $1.7438$ \\
    \hline
    \end{tabular}
\end{table}

We calibrate the initial solar composition (in terms of helium abundance $Y_{\odot, {\rm birth}}$ and metallicity $Z_{\odot, {\rm birth}}$) and the mixing length parameter $\alpha_{{\rm MLT}, \odot}$, so that the solar model yields the solar effective temperature $T_{\rm eff, \odot} = 5772 \,{\rm K}$, the solar luminosity $L = 1.0 \,{\rm L}_\odot = 3.828 \times 10^{33} \,{\rm erg}/{\rm s}$ \citep{Mamajek2015arXiv}, and the solar $Z/X$ ratio ($0.02292$ for GS98 and $0.01812$ for AGSS09) at solar age. Note that the birth mixture of solar models described above do not correspond to $[{\rm Fe}/{\rm H}] = 0.0$ because of gravitational settling \citep[e.g.,][]{Souto2019ApJ}. The calibrated $Y_{\odot, {\rm birth}}$, $Z_{\odot, {\rm birth}}$, and $\alpha_{{\rm MLT}, \odot}$ are tabulated in Table~\ref{tab:SunCalibr}. We then use Equation~(\ref{eq:helium}) to calculate the initial $Y$ for any given $Z$.

It is standard practice to use a solar-calibrated mixing length for isochrones. However, with asteroseismic data, we have higher precision and require a calibration consistent with the observed HR diagram location of stars as a function of mass and metallicity. Therefore, we perform a separate mixing length calibration for the advanced stage of our models. For comparison purposes, we define two key equal evolutionary points (EEPs): the starting point of the first dredge-up, defined as when the surface $[{\rm C}/{\rm N}]$ starts to change, and the first turning point of the RGBB, which is a local minimum in surface $\log g$. Between these two EEPs (including both ends), we use the helium core mass as a coordinate, and extract other quantities via linear interpolation. Then for each star in our RGB calibration sample (see Section~\ref{ss:select}), we use its mass, surface metallicity, and surface gravity to find an ``optimal'' mixing length parameter which gives the observed effective temperature.

\begin{figure*}
    \centering
    \includegraphics[width=0.95\textwidth]{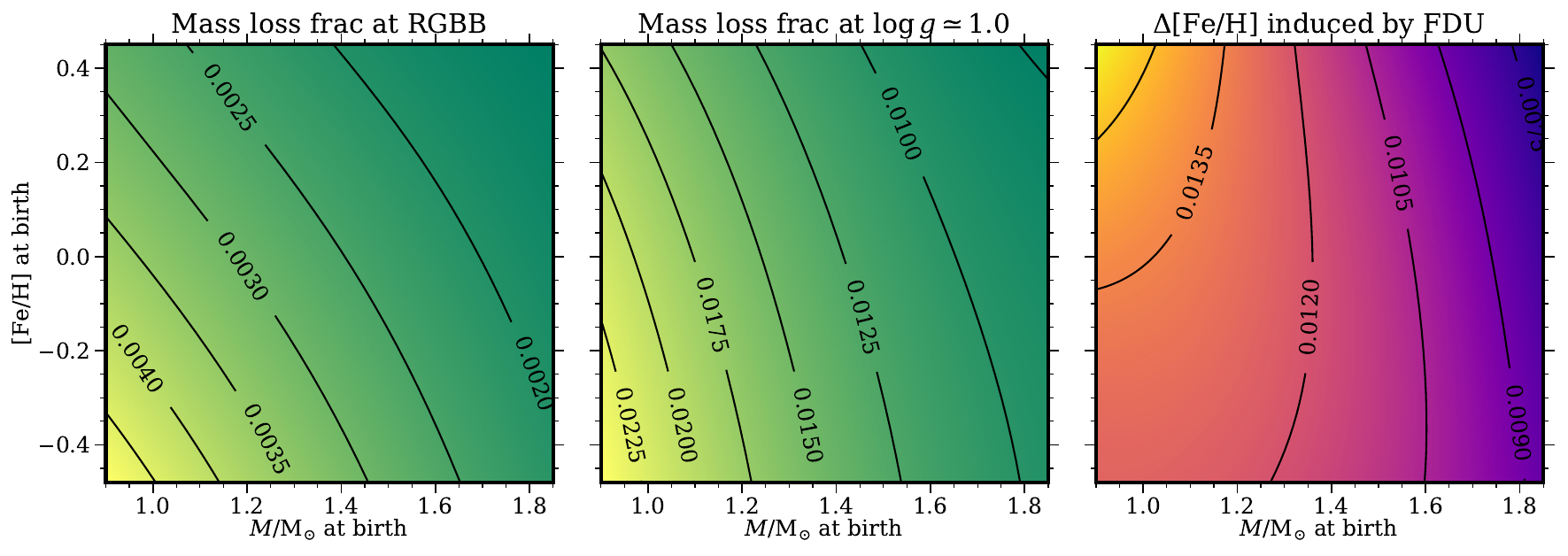}
    \caption{\label{fig:oscu_motivation}Motivation for osculating tracks. Each panel visualizes a bivariate function of mass and metallicity {\bf at birth} as a color map with associated contours. Left and middle panels: mass loss fractions (fractional differences between ``current'' masses and birth masses) at red giant branch bump and $\log g \simeq 1.0$, respectively; right panel: change in surface metallicity induced by first dredge-up. See the text for further explanations.}
\end{figure*}

There is a subtle issue involved here. In light of the deep surface convection zones, the gravitational settling \citep[see][for implementation]{Paxton2018} during the MS stage is largely undone, and the surface mixture of red giants is close to their birth mixture, except for elements affected by the first dredge-up. However, the birth and post-FDU metallicities (in terms of $[{\rm Fe}/{\rm H}]$) of a model are slightly different, mainly because the dredge-up of helium dilutes hydrogen, while the iron abundance is largely recovered through mixing. To compare theoretical predictions to observational data, we need to use the ``current'' $[{\rm Fe}/{\rm H}]$ of each model instead of the birth value; similarly, the ``current'' mass is also different from the birth value due to mass loss, although it is not significant until upper RGB. To avoid biases that would be introduced by comparing birth properties of theoretical models with observed properties of evolved stars, here we introduce the concept of ``osculating tracks'': We combine multiple models to synthesize tracks with constant {\bf observed} mass and metallicity, so that they are directly comparable with observed stars. Figure~\ref{fig:oscu_motivation} visualizes the motivation for osculating tracks.

\begin{table}
    \caption{\label{tab:RgbCalibr}RGB mixing length calibration results. For an RGB model, the optimal mixing length parameter $\alpha_{\rm MLT}$ is written as a polynomial of mass $M$ and metallicity $[{\rm Fe}/{\rm H}]$: $\alpha_{\rm MLT} (M, [{\rm Fe}/{\rm H}]) = a + b_1 (M/{\rm M}_\odot-1) + b_2 (M/{\rm M}_\odot-1)^2 + c [{\rm Fe}/{\rm H}]$.}
    \centering
    \begin{tabular}{lcccc}
    \hline
        Grid name & $a$ & $b_1$ & $b_2$ & $c$ \\
    \hline
        {\tt benchmark} & $1.7316$ & $0.2303$ & $-0.3616$ & $0.1101$ \\
        {\tt var\_chem\_agss09} & $1.7566$ & $0.2046$ & $-0.3496$ & $0.1633$ \\
        {\tt var\_chem\_r24cn} & $1.7383$ & $0.2003$ & $-0.3383$ & $0.1031$ \\
        {\tt var\_rates\_default} & $1.7554$ & $0.2148$ & $-0.4064$ & $0.1232$ \\
        {\tt var\_kap\_fa05} & $1.8356$ & $0.2409$ & $-0.4226$ & $0.1579$ \\
        {\tt var\_kap\_opal} & $1.7317$ & $0.2326$ & $-0.3824$ & $0.0830$ \\
        {\tt var\_kap\_linear} & $1.7416$ & $0.2061$ & $-0.3476$ & $0.1278$ \\
        {\tt var\_atm\_phot} & $1.8227$ & $0.2290$ & $-0.3878$ & $0.0887$ \\
        {\tt var\_mix\_noos} & $1.7139$ & $0.2572$ & $-0.3224$ & $0.1168$ \\
        {\tt var\_mix\_envos} & $1.7421$ & $0.1740$ & $-0.3013$ & $0.1158$ \\
    \hline
    \end{tabular}
\end{table}

We perform bivariate fits to the optimal $\alpha_{\rm MLT}$ values of all stars in our calibration sample (see Section~\ref{ss:select}). Based on chi-squared tests, the combination of a linear dependence on metallicity and a quadratic dependence on mass provides a reasonable balance between $\chi^2$ and the degrees of freedom. The resulting coefficients are tabulated in Table~\ref{tab:RgbCalibr}. Note that for each grid, the zeroth-order term $a$ is close to $\alpha_{{\rm MLT}, \odot}$ in Table~\ref{tab:SunCalibr}, but slightly different.

\begin{figure*}
    \centering
    \includegraphics[width=0.95\textwidth]{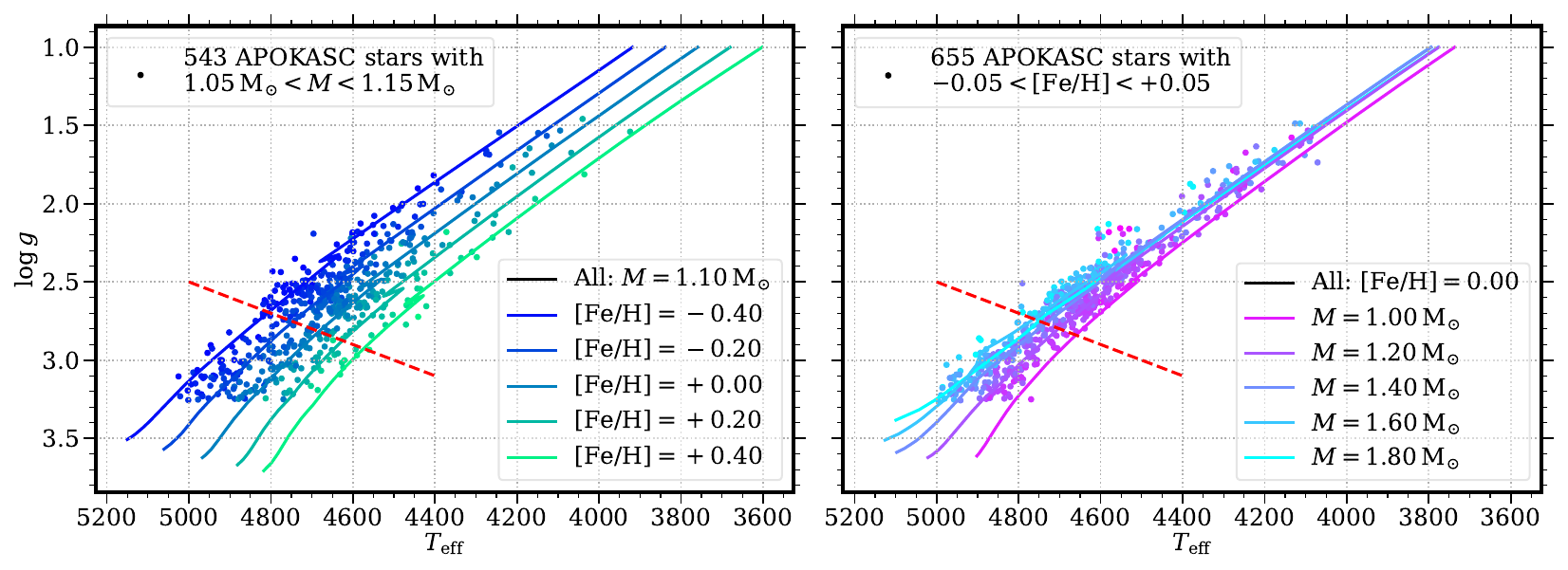}
    \caption{\label{fig:synthesis}Synthetic RGB tracks of the {\tt benchmark} grid (curves) on Kiel diagrams, compared to corresponding APOKASC-3 samples (data points). The format of the curves is the same as in Figure~\ref{fig:FDU_gravity_base}. The lower RGB cut (same as that in the lower panel of Figure~\ref{fig:selection}) is shown as red dashed lines.}
\end{figure*}

Finally, we use the $\alpha_{\rm MLT} (M, [{\rm Fe}/{\rm H}])$ polynomials and further interpolations to synthesize a grid spanning the mass range $0.9 \,{\rm M}_\odot \leq M \leq 1.8 \,{\rm M}_\odot$ and the metallicity range $-0.45 \leq [{\rm Fe}/{\rm H}] \leq 0.45$, both in steps of $0.05$. To testify to the effectiveness of our RGB calibration and synthesis pipeline, in Figure~\ref{fig:synthesis} we plot some of our {\tt benchmark} synthetic RGB tracks on the Kiel diagram, together with corresponding stars in the APOKASC-3 sample. Note that our tracks are only calibrated using lower RGB stars (below red dashed lines), while most outliers on the upper RGB are likely AGB stars.

\section{Fitting coefficients} \label{app:tables}

This appendix tabulates fitting coefficients for theoretical predictions --- post-FDU $^{12}{\rm C}/^{13}{\rm C}$, $\Delta [{\rm C}/{\rm N}]$, $\log g({\rm RGBB})$, RGBB width, and post-MS $\Delta A({\rm Li})$ --- for reference.

\begin{table*}
    \caption{\label{tab:Ciso_coeffs}Fitting coefficients for post-FDU $^{12}{\rm C}/^{13}{\rm C}$. For an RGB star, the post-FDU surface carbon isotopic ratio $^{12}{\rm C}/^{13}{\rm C}$ is predicted to be a polynomial of mass $M$ and metallicity $[{\rm Fe}/{\rm H}]$: $^{12}{\rm C}/^{13}{\rm C} (M, [{\rm Fe}/{\rm H}]) = a + b_1 (M/{\rm M}_\odot-1) + b_2 (M/{\rm M}_\odot-1)^2 + c_1 [{\rm Fe}/{\rm H}] + c_2 [{\rm Fe}/{\rm H}]^2 + d (M/{\rm M}_\odot-1) [{\rm Fe}/{\rm H}]$.}
    \centering
    \begin{tabular}{lcccccc}
    \hline
        Grid name & $a$ & $b_1$ & $b_2$ & $c_1$ & $c_2$ & $d$ \\
    \hline
        {\tt benchmark} & $33.753896$ & $-18.872825$ & $12.046570$ & $1.065588$ & $-0.979091$ & $-0.321176$ \\
        {\tt var\_chem\_agss09} & $33.902498$ & $-19.048381$ & $11.955210$ & $1.721946$ & $-0.550412$ & $-1.156475$ \\
        {\tt var\_chem\_r24cn} & $33.735232$ & $-18.806002$ & $12.012956$ & $0.989056$ & $-0.898037$ & $-0.084608$ \\
        {\tt var\_rates\_default} & $30.603291$ & $-18.154721$ & $11.950051$ & $1.166664$ & $-0.890989$ & $-0.156170$ \\
        {\tt var\_kap\_fa05} & $33.753227$ & $-18.815450$ & $11.972712$ & $1.045977$ & $-0.982529$ & $-0.287422$ \\
        {\tt var\_kap\_opal} & $33.657018$ & $-18.954118$ & $12.180246$ & $0.872715$ & $-1.033285$ & $-0.193558$ \\
        {\tt var\_kap\_linear} & $33.726958$ & $-18.817124$ & $12.011608$ & $1.256854$ & $-0.913317$ & $-0.543882$ \\
        {\tt var\_atm\_phot} & $33.761083$ & $-18.837626$ & $11.989942$ & $1.054647$ & $-0.990961$ & $-0.314932$ \\
        {\tt var\_mix\_noos} & $33.792362$ & $-18.012779$ & $11.004356$ & $1.080486$ & $-0.968091$ & $-0.261694$ \\
        {\tt var\_mix\_envos} & $33.779515$ & $-18.900845$ & $12.042190$ & $1.118480$ & $-0.949710$ & $-0.373363$ \\
    \hline
    \end{tabular}
\end{table*}

\begin{table*}
    \caption{\label{tab:CN_coeffs}Similar to Table~\ref{tab:Ciso_coeffs}, but for $\Delta [{\rm C}/{\rm N}]$, the change in surface ${\rm C}$ to ${\rm N}$ abundance ratio from birth to post-first dredge-up.}
    \centering
    \begin{tabular}{lcccccc}
    \hline
        Grid name & $a$ & $b_1$ & $b_2$ & $c_1$ & $c_2$ & $d$ \\
    \hline
        APOKASC-3 & $-0.074176$ & $-0.665410$ & $0.405058$ & $0.325135$ & $0.275523$ & $-0.261910$ \\
    \hline
        {\tt benchmark} & $-0.229264$ & $-0.658554$ & $0.327720$ & $0.026453$ & $-0.036522$ & $0.097472$ \\
        {\tt var\_chem\_agss09} & $-0.227594$ & $-0.670046$ & $0.325442$ & $0.049640$ & $-0.021314$ & $0.072916$ \\
        {\tt var\_chem\_r24cn} & $-0.217437$ & $-0.637150$ & $0.315126$ & $0.140420$ & $-0.006651$ & $0.213751$ \\
        {\tt var\_rates\_default} & $-0.218515$ & $-0.629249$ & $0.302218$ & $0.034416$ & $-0.028879$ & $0.100077$ \\
        {\tt var\_kap\_fa05} & $-0.229404$ & $-0.662996$ & $0.335763$ & $0.029651$ & $-0.034985$ & $0.090795$ \\
        {\tt var\_kap\_opal} & $-0.231177$ & $-0.653720$ & $0.320787$ & $0.023739$ & $-0.041677$ & $0.100580$ \\
        {\tt var\_kap\_linear} & $-0.229098$ & $-0.657151$ & $0.325426$ & $0.030956$ & $-0.036130$ & $0.091885$ \\
        {\tt var\_atm\_phot} & $-0.229384$ & $-0.661233$ & $0.334077$ & $0.027692$ & $-0.037739$ & $0.093137$ \\
        {\tt var\_mix\_noos} & $-0.231272$ & $-0.687270$ & $0.383615$ & $0.034423$ & $-0.042901$ & $0.080585$ \\
        {\tt var\_mix\_envos} & $-0.238928$ & $-0.638929$ & $0.311983$ & $0.028968$ & $-0.034915$ & $0.094706$ \\
    \hline
    \end{tabular}
\end{table*}

\begin{table*}
    \caption{\label{tab:RGBB_coeffs}Similar to Table~\ref{tab:Ciso_coeffs}, but for $\log g({\rm RGBB})$, the logarithmic surface gravity at RGBB center. Second-order coefficients are not included in our analysis of observational data.}
    \centering
    \begin{tabular}{lcccccc}
    \hline
        Grid name & $a$ & $b_1$ & $b_2$ & $c_1$ & $c_2$ & $d$ \\
    \hline
        APOKASC-3 & $2.660418$ & $-0.183181$ & --- & $0.282392$ & --- & --- \\
    \hline
        {\tt benchmark} & $2.522144$ & $-0.122316$ & $-0.189352$ & $0.341392$ & $-0.188581$ & $-0.056193$ \\
        {\tt var\_chem\_agss09} & $2.502833$ & $-0.113736$ & $-0.234694$ & $0.388369$ & $-0.125810$ & $-0.026268$ \\
        {\tt var\_chem\_r24cn} & $2.522779$ & $-0.128605$ & $-0.184822$ & $0.349434$ & $-0.159290$ & $-0.084085$ \\
        {\tt var\_rates\_default} & $2.544669$ & $-0.091059$ & $-0.247818$ & $0.335761$ & $-0.196380$ & $-0.074462$ \\
        {\tt var\_kap\_fa05} & $2.522630$ & $-0.137827$ & $-0.182909$ & $0.339822$ & $-0.171332$ & $-0.053777$ \\
        {\tt var\_kap\_opal} & $2.530425$ & $-0.107057$ & $-0.196605$ & $0.333268$ & $-0.203381$ & $-0.054096$ \\
        {\tt var\_kap\_linear} & $2.516380$ & $-0.127384$ & $-0.190307$ & $0.364442$ & $-0.171884$ & $-0.032362$ \\
        {\tt var\_atm\_phot} & $2.520833$ & $-0.131511$ & $-0.186713$ & $0.339148$ & $-0.154205$ & $-0.071708$ \\
        {\tt var\_mix\_noos} & $2.518021$ & $-0.073159$ & $-0.220889$ & $0.340494$ & $-0.167384$ & $-0.044817$ \\
        {\tt var\_mix\_envos} & $2.612386$ & $-0.105984$ & $-0.263292$ & $0.352256$ & $-0.170093$ & $-0.065354$ \\
    \hline
    \end{tabular}
\end{table*}

\begin{table*}
    \caption{\label{tab:RGBB_width}Similar to Table~\ref{tab:Ciso_coeffs}, but for RGBB width in terms of $\log g$ discrepancy between the two turning points (see Section~\ref{ss:rgbb_base} for definition).}
    \centering
    \begin{tabular}{lcccccc}
    \hline
        Grid name & $a$ & $b_1$ & $b_2$ & $c_1$ & $c_2$ & $d$ \\
    \hline
        {\tt benchmark} & $0.070700$ & $-0.256496$ & $0.552080$ & $0.023836$ & $-0.013497$ & $-0.079390$ \\
        {\tt var\_chem\_agss09} & $0.070700$ & $-0.243736$ & $0.571289$ & $0.019035$ & $-0.026214$ & $-0.063866$ \\
        {\tt var\_chem\_r24cn} & $0.071125$ & $-0.255896$ & $0.551287$ & $0.019718$ & $-0.025215$ & $-0.070895$ \\
        {\tt var\_rates\_default} & $0.065505$ & $-0.362975$ & $0.764073$ & $0.022435$ & $-0.011432$ & $-0.091324$ \\
        {\tt var\_kap\_fa05} & $0.070507$ & $-0.226609$ & $0.506472$ & $0.008615$ & $-0.036904$ & $-0.044788$ \\
        {\tt var\_kap\_opal} & $0.068731$ & $-0.270899$ & $0.581471$ & $0.023123$ & $-0.005357$ & $-0.085270$ \\
        {\tt var\_kap\_linear} & $0.070920$ & $-0.260102$ & $0.564430$ & $0.024767$ & $-0.017406$ & $-0.080592$ \\
        {\tt var\_atm\_phot} & $0.071526$ & $-0.241531$ & $0.523029$ & $0.020094$ & $-0.021796$ & $-0.062363$ \\
        {\tt var\_mix\_noos} & $0.076399$ & $0.049361$ & $-0.112393$ & $0.009129$ & $-0.011908$ & $-0.054942$ \\
        {\tt var\_mix\_envos} & $0.066960$ & $-0.284638$ & $0.608137$ & $0.022118$ & $-0.015340$ & $-0.072940$ \\
    \hline
    \end{tabular}
\end{table*}

\begin{table*}
    \caption{\label{tab:DALi_coeffs}Similar to Table~\ref{tab:Ciso_coeffs}, but for post-MS (from TAMS to RGBB) surface lithium abundance change $\Delta A({\rm Li})$.}
    \centering
    \begin{tabular}{lcccccc}
    \hline
        Grid name & $a$ & $b_1$ & $b_2$ & $c_1$ & $c_2$ & $d$ \\
    \hline
        {\tt benchmark} & $-2.059759$ & $-3.113070$ & $4.664477$ & $-1.468260$ & $-2.173902$ & $-2.179916$ \\
        {\tt var\_chem\_agss09} & $-1.961407$ & $-2.576188$ & $3.847494$ & $-1.130959$ & $-2.396550$ & $-2.486363$ \\
        {\tt var\_chem\_r24cn} & $-1.964800$ & $-3.950974$ & $5.479746$ & $-0.669844$ & $-1.372084$ & $-3.824105$ \\
        {\tt var\_rates\_default} & $-2.280181$ & $-4.657660$ & $6.848860$ & $-0.922657$ & $-0.408014$ & $-4.106396$ \\
        {\tt var\_kap\_fa05} & $-2.132856$ & $-2.904310$ & $4.546059$ & $-1.507022$ & $-2.168068$ & $-2.253347$ \\
        {\tt var\_kap\_opal} & $-2.059132$ & $-3.320437$ & $4.930581$ & $-1.475721$ & $-2.150590$ & $-2.271006$ \\
        {\tt var\_kap\_linear} & $-1.965837$ & $-3.360778$ & $4.866838$ & $-1.462410$ & $-2.494166$ & $-2.264241$ \\
        {\tt var\_atm\_phot} & $-2.413563$ & $-1.283567$ & $3.032487$ & $-2.910830$ & $-3.119726$ & $0.564136$ \\
        {\tt var\_mix\_noos} & $-2.039085$ & $-2.699709$ & $3.998840$ & $-1.378211$ & $-2.222036$ & $-2.258632$ \\
        {\tt var\_mix\_envos} & $-1.941593$ & $-8.940382$ & $10.926829$ & $2.806889$ & $4.609324$ & $-11.201532$ \\
    \hline
    \end{tabular}
\end{table*}

\bibliography{main1}{}

\begin{thebibliography}{}
\expandafter\ifx\csname natexlab\endcsname\relax\def\natexlab#1{#1}\fi
\providecommand{\url}[1]{\href{#1}{#1}}
\providecommand{\dodoi}[1]{doi:~\href{http://doi.org/#1}{\nolinkurl{#1}}}
\providecommand{\doeprint}[1]{\href{http://ascl.net/#1}{\nolinkurl{http://ascl.net/#1}}}
\providecommand{\doarXiv}[1]{\href{https://arxiv.org/abs/#1}{\nolinkurl{https://arxiv.org/abs/#1}}}

\bibitem[{{Abdurro'uf} {et~al.}(2022){Abdurro'uf}, {Accetta}, {Aerts}, {Silva Aguirre}, {Ahumada}, {Ajgaonkar}, {Filiz Ak}, {Alam}, {Allende Prieto}, {Almeida}, {Anders}, {Anderson}, {Andrews}, {Anguiano}, {Aquino-Ort{\'\i}z}, {Arag{\'o}n-Salamanca}, {Argudo-Fern{\'a}ndez}, {Ata}, {Aubert}, {Avila-Reese}, {Badenes}, {Barb{\'a}}, {Barger}, {Barrera-Ballesteros}, {Beaton}, {Beers}, {Belfiore}, {Bender}, {Bernardi}, {Bershady}, {Beutler}, {Bidin}, {Bird}, {Bizyaev}, {Blanc}, {Blanton}, {Boardman}, {Bolton}, {Boquien}, {Borissova}, {Bovy}, {Brandt}, {Brown}, {Brownstein}, {Brusa}, {Buchner}, {Bundy}, {Burchett}, {Bureau}, {Burgasser}, {Cabang}, {Campbell}, {Cappellari}, {Carlberg}, {Wanderley}, {Carrera}, {Cash}, {Chen}, {Chen}, {Cherinka}, {Chiappini}, {Choi}, {Chojnowski}, {Chung}, {Clerc}, {Cohen}, {Comerford}, {Comparat}, {da Costa}, {Covey}, {Crane}, {Cruz-Gonzalez}, {Culhane}, {Cunha}, {Dai}, {Damke}, {Darling}, {Davidson}, {Davies}, {Dawson}, {De Lee}, {Diamond-Stanic}, {Cano-D{\'\i}az}, {S{\'a}nchez},
  {Donor}, {Duckworth}, {Dwelly}, {Eisenstein}, {Elsworth}, {Emsellem}, {Eracleous}, {Escoffier}, {Fan}, {Farr}, {Feng}, {Fern{\'a}ndez-Trincado}, {Feuillet}, {Filipp}, {Fillingham}, {Frinchaboy}, {Fromenteau}, {Galbany}, {Garc{\'\i}a}, {Garc{\'\i}a-Hern{\'a}ndez}, {Ge}, {Geisler}, {Gelfand}, {G{\'e}ron}, {Gibson}, {Goddy}, {Godoy-Rivera}, {Grabowski}, {Green}, {Greener}, {Grier}, {Griffith}, {Guo}, {Guy}, {Hadjara}, {Harding}, {Hasselquist}, {Hayes}, {Hearty}, {Hern{\'a}ndez}, {Hill}, {Hogg}, {Holtzman}, {Horta}, {Hsieh}, {Hsu}, {Hsu}, {Huber}, {Huertas-Company}, {Hutchinson}, {Hwang}, {Ibarra-Medel}, {Chitham}, {Ilha}, {Imig}, {Jaekle}, {Jayasinghe}, {Ji}, {Johnson}, {Jones}, {J{\"o}nsson}, {Katkov}, {Khalatyan}, {Kinemuchi}, {Kisku}, {Knapen}, {Kneib}, {Kollmeier}, {Kong}, {Kounkel}, {Kreckel}, {Krishnarao}, {Lacerna}, {Lane}, {Langgin}, {Lavender}, {Law}, {Lazarz}, {Leung}, {Leung}, {Lewis}, {Li}, {Li}, {Lian}, {Liang}, {Lin}, {Lin}, {Lin}, {Lintott}, {Long}, {Longa-Pe{\~n}a}, {L{\'o}pez-Cob{\'a}}, {Lu},
  {Lundgren}, {Luo}, {Mackereth}, {de la Macorra}, {Mahadevan}, {Majewski}, {Manchado}, {Mandeville}, {Maraston}, {Margalef-Bentabol}, {Masseron}, {Masters}, {Mathur}, {McDermid}, {Mckay}, {Merloni}, {Merrifield}, {Meszaros}, {Miglio}, {Di Mille}, {Minniti}, {Minsley}, {Monachesi}, {Moon}, {Mosser}, {Mulchaey}, {Muna}, {Mu{\~n}oz}, {Myers}, {Myers}, {Nadathur}, {Nair}, {Nandra}, {Neumann}, {Newman}, {Nidever}, {Nikakhtar}, {Nitschelm}, {O'Connell}, {Garma-Oehmichen}, {Luan Souza de Oliveira}, {Olney}, {Oravetz}, {Ortigoza-Urdaneta}, {Osorio}, {Otter}, {Pace}, {Padilla}, {Pan}, {Pan}, {Parikh}, {Parker}, {Peirani}, {Pe{\~n}a Ram{\'\i}rez}, {Penny}, {Percival}, {Perez-Fournon}, {Pinsonneault}, {Poidevin}, {Poovelil}, {Price-Whelan}, {B{\'a}rbara de Andrade Queiroz}, {Raddick}, {Ray}, {Rembold}, {Riddle}, {Riffel}, {Riffel}, {Rix}, {Robin}, {Rodr{\'\i}guez-Puebla}, {Roman-Lopes}, {Rom{\'a}n-Z{\'u}{\~n}iga}, {Rose}, {Ross}, {Rossi}, {Rubin}, {Salvato}, {S{\'a}nchez}, {S{\'a}nchez-Gallego}, {Sanderson}, {Santana
  Rojas}, {Sarceno}, {Sarmiento}, {Sayres}, {Sazonova}, {Schaefer}, {Schiavon}, {Schlegel}, {Schneider}, {Schultheis}, {Schwope}, {Serenelli}, {Serna}, {Shao}, {Shapiro}, {Sharma}, {Shen}, {Shetrone}, {Shu}, {Simon}, {Skrutskie}, {Smethurst}, {Smith}, {Sobeck}, {Spoo}, {Sprague}, {Stark}, {Stassun}, {Steinmetz}, {Stello}, {Stone-Martinez}, {Storchi-Bergmann}, {Stringfellow}, {Stutz}, {Su}, {Taghizadeh-Popp}, {Talbot}, {Tayar}, {Telles}, {Teske}, {Thakar}, {Theissen}, {Tkachenko}, {Thomas}, {Tojeiro}, {Hernandez Toledo}, {Troup}, {Trump}, {Trussler}, {Turner}, {Tuttle}, {Unda-Sanzana}, {V{\'a}zquez-Mata}, {Valentini}, {Valenzuela}, {Vargas-Gonz{\'a}lez}, {Vargas-Maga{\~n}a}, {Alfaro}, {Villanova}, {Vincenzo}, {Wake}, {Warfield}, {Washington}, {Weaver}, {Weijmans}, {Weinberg}, {Weiss}, {Westfall}, {Wild}, {Wilde}, {Wilson}, {Wilson}, {Wilson}, {Wolf}, {Wood-Vasey}, {Yan}, {Zamora}, {Zasowski}, {Zhang}, {Zhao}, {Zheng}, {Zheng}, \& {Zhu}}]{Abdurrouf2022ApJS}
{Abdurro'uf}, {Accetta}, K., {Aerts}, C., {et~al.} 2022, \apjs, 259, 35, \dodoi{10.3847/1538-4365/ac4414}

\bibitem[{{Ahumada} {et~al.}(2020){Ahumada}, {Allende Prieto}, {Almeida}, {Anders}, {Anderson}, {Andrews}, {Anguiano}, {Arcodia}, {Armengaud}, {Aubert}, {Avila}, {Avila-Reese}, {Badenes}, {Balland}, {Barger}, {Barrera-Ballesteros}, {Basu}, {Bautista}, {Beaton}, {Beers}, {Benavides}, {Bender}, {Bernardi}, {Bershady}, {Beutler}, {Bidin}, {Bird}, {Bizyaev}, {Blanc}, {Blanton}, {Boquien}, {Borissova}, {Bovy}, {Brandt}, {Brinkmann}, {Brownstein}, {Bundy}, {Bureau}, {Burgasser}, {Burtin}, {Cano-D{\'\i}az}, {Capasso}, {Cappellari}, {Carrera}, {Chabanier}, {Chaplin}, {Chapman}, {Cherinka}, {Chiappini}, {Doohyun Choi}, {Chojnowski}, {Chung}, {Clerc}, {Coffey}, {Comerford}, {Comparat}, {da Costa}, {Cousinou}, {Covey}, {Crane}, {Cunha}, {Ilha}, {Dai}, {Damsted}, {Darling}, {Davidson}, {Davies}, {Dawson}, {De}, {de la Macorra}, {De Lee}, {Queiroz}, {Deconto Machado}, {de la Torre}, {Dell'Agli}, {du Mas des Bourboux}, {Diamond-Stanic}, {Dillon}, {Donor}, {Drory}, {Duckworth}, {Dwelly}, {Ebelke}, {Eftekharzadeh}, {Davis
  Eigenbrot}, {Elsworth}, {Eracleous}, {Erfanianfar}, {Escoffier}, {Fan}, {Farr}, {Fern{\'a}ndez-Trincado}, {Feuillet}, {Finoguenov}, {Fofie}, {Fraser-McKelvie}, {Frinchaboy}, {Fromenteau}, {Fu}, {Galbany}, {Garcia}, {Garc{\'\i}a-Hern{\'a}ndez}, {Garma Oehmichen}, {Ge}, {Geimba Maia}, {Geisler}, {Gelfand}, {Goddy}, {Gonzalez-Perez}, {Grabowski}, {Green}, {Grier}, {Guo}, {Guy}, {Harding}, {Hasselquist}, {Hawken}, {Hayes}, {Hearty}, {Hekker}, {Hogg}, {Holtzman}, {Horta}, {Hou}, {Hsieh}, {Huber}, {Hunt}, {Ider Chitham}, {Imig}, {Jaber}, {Jimenez Angel}, {Johnson}, {Jones}, {J{\"o}nsson}, {Jullo}, {Kim}, {Kinemuchi}, {Kirkpatrick}, {Kite}, {Klaene}, {Kneib}, {Kollmeier}, {Kong}, {Kounkel}, {Krishnarao}, {Lacerna}, {Lan}, {Lane}, {Law}, {Le Goff}, {Leung}, {Lewis}, {Li}, {Lian}, {Lin}, {Long}, {Longa-Pe{\~n}a}, {Lundgren}, {Lyke}, {Mackereth}, {MacLeod}, {Majewski}, {Manchado}, {Maraston}, {Martini}, {Masseron}, {Masters}, {Mathur}, {McDermid}, {Merloni}, {Merrifield}, {M{\'e}sz{\'a}ros}, {Miglio}, {Minniti},
  {Minsley}, {Miyaji}, {Mohammad}, {Mosser}, {Mueller}, {Muna}, {Mu{\~n}oz-Guti{\'e}rrez}, {Myers}, {Nadathur}, {Nair}, {Nandra}, {Correa do Nascimento}, {Nevin}, {Newman}, {Nidever}, {Nitschelm}, {Noterdaeme}, {O'Connell}, {Olmstead}, {Oravetz}, {Oravetz}, {Osorio}, {Pace}, {Padilla}, {Palanque-Delabrouille}, {Palicio}, {Pan}, {Pan}, {Parker}, {Paviot}, {Peirani}, {Ram{\'r}ez}, {Penny}, {Percival}, {Perez-Fournon}, {P{\'e}rez-R{\`a}fols}, {Petitjean}, {Pieri}, {Pinsonneault}, {Poovelil}, {Povick}, {Prakash}, {Price-Whelan}, {Raddick}, {Raichoor}, {Ray}, {Rembold}, {Rezaie}, {Riffel}, {Riffel}, {Rix}, {Robin}, {Roman-Lopes}, {Rom{\'a}n-Z{\'u}{\~n}iga}, {Rose}, {Ross}, {Rossi}, {Rowlands}, {Rubin}, {Salvato}, {S{\'a}nchez}, {S{\'a}nchez-Menguiano}, {S{\'a}nchez-Gallego}, {Sayres}, {Schaefer}, {Schiavon}, {Schimoia}, {Schlafly}, {Schlegel}, {Schneider}, {Schultheis}, {Schwope}, {Seo}, {Serenelli}, {Shafieloo}, {Shamsi}, {Shao}, {Shen}, {Shetrone}, {Shirley}, {Silva Aguirre}, {Simon}, {Skrutskie}, {Slosar},
  {Smethurst}, {Sobeck}, {Sodi}, {Souto}, {Stark}, {Stassun}, {Steinmetz}, {Stello}, {Stermer}, {Storchi-Bergmann}, {Streblyanska}, {Stringfellow}, {Stutz}, {Su{\'a}rez}, {Sun}, {Taghizadeh-Popp}, {Talbot}, {Tayar}, {Thakar}, {Theriault}, {Thomas}, {Thomas}, {Tinker}, {Tojeiro}, {Toledo}, {Tremonti}, {Troup}, {Tuttle}, {Unda-Sanzana}, {Valentini}, {Vargas-Gonz{\'a}lez}, {Vargas-Maga{\~n}a}, {V{\'a}zquez-Mata}, {Vivek}, {Wake}, {Wang}, {Weaver}, {Weijmans}, {Wild}, {Wilson}, {Wilson}, {Wolthuis}, {Wood-Vasey}, {Yan}, {Yang}, {Y{\`e}che}, {Zamora}, {Zarrouk}, {Zasowski}, {Zhang}, {Zhao}, {Zhao}, {Zheng}, {Zheng}, {Zhu}, \& {Zou}}]{Ahumada2020ApJS}
{Ahumada}, R., {Allende Prieto}, C., {Almeida}, A., {et~al.} 2020, \apjs, 249, 3, \dodoi{10.3847/1538-4365/ab929e}

\bibitem[{{Alongi} {et~al.}(1991){Alongi}, {Bertelli}, {Bressan}, \& {Chiosi}}]{Alongi1991A&A}
{Alongi}, M., {Bertelli}, G., {Bressan}, A., \& {Chiosi}, C. 1991, \aap, 244, 95

\bibitem[{{Anders} \& {Grevesse}(1989)}]{Anders1989GeCoA}
{Anders}, E., \& {Grevesse}, N. 1989, \gca, 53, 197, \dodoi{10.1016/0016-7037(89)90286-X}

\bibitem[{{Angulo} {et~al.}(1999){Angulo}, {Arnould}, {Rayet}, {Descouvemont}, {Baye}, {Leclercq-Willain}, {Coc}, {Barhoumi}, {Aguer}, {Rolfs}, {Kunz}, {Hammer}, {Mayer}, {Paradellis}, {Kossionides}, {Chronidou}, {Spyrou}, {degl'Innocenti}, {Fiorentini}, {Ricci}, {Zavatarelli}, {Providencia}, {Wolters}, {Soares}, {Grama}, {Rahighi}, {Shotter}, \& {Lamehi Rachti}}]{Angulo1999NuPhA}
{Angulo}, C., {Arnould}, M., {Rayet}, M., {et~al.} 1999, \nphysa, 656, 3, \dodoi{10.1016/S0375-9474(99)00030-5}

\bibitem[{{Asplund} {et~al.}(2021){Asplund}, {Amarsi}, \& {Grevesse}}]{Asplund2021A&A}
{Asplund}, M., {Amarsi}, A.~M., \& {Grevesse}, N. 2021, \aap, 653, A141, \dodoi{10.1051/0004-6361/202140445}

\bibitem[{{Asplund} {et~al.}(2009){Asplund}, {Grevesse}, {Sauval}, \& {Scott}}]{Asplund2009ARA&A}
{Asplund}, M., {Grevesse}, N., {Sauval}, A.~J., \& {Scott}, P. 2009, \araa, 47, 481, \dodoi{10.1146/annurev.astro.46.060407.145222}

\bibitem[{{Badnell} {et~al.}(2005){Badnell}, {Bautista}, {Butler}, {Delahaye}, {Mendoza}, {Palmeri}, {Zeippen}, \& {Seaton}}]{Badnell2005MNRAS}
{Badnell}, N.~R., {Bautista}, M.~A., {Butler}, K., {et~al.} 2005, \mnras, 360, 458, \dodoi{10.1111/j.1365-2966.2005.08991.x}

\bibitem[{{Baglin}(2003)}]{Baglin2003AdSpR}
{Baglin}, A. 2003, Advances in Space Research, 31, 345, \dodoi{10.1016/S0273-1177(02)00624-5}

\bibitem[{{Bahcall} \& {Ulrich}(1988)}]{Bahcall1988RvMP}
{Bahcall}, J.~N., \& {Ulrich}, R.~K. 1988, Reviews of Modern Physics, 60, 297, \dodoi{10.1103/RevModPhys.60.297}

\bibitem[{{Bedding} {et~al.}(2011){Bedding}, {Mosser}, {Huber}, {Montalb{\'a}n}, {Beck}, {Christensen-Dalsgaard}, {Elsworth}, {Garc{\'\i}a}, {Miglio}, {Stello}, {White}, {De Ridder}, {Hekker}, {Aerts}, {Barban}, {Belkacem}, {Broomhall}, {Brown}, {Buzasi}, {Carrier}, {Chaplin}, {di Mauro}, {Dupret}, {Frandsen}, {Gilliland}, {Goupil}, {Jenkins}, {Kallinger}, {Kawaler}, {Kjeldsen}, {Mathur}, {Noels}, {Silva Aguirre}, \& {Ventura}}]{Bedding2011Natur}
{Bedding}, T.~R., {Mosser}, B., {Huber}, D., {et~al.} 2011, \nat, 471, 608, \dodoi{10.1038/nature09935}

\bibitem[{{Blanton} {et~al.}(2017){Blanton}, {Bershady}, {Abolfathi}, {Albareti}, {Allende Prieto}, {Almeida}, {Alonso-Garc{\'\i}a}, {Anders}, {Anderson}, {Andrews}, {Aquino-Ort{\'\i}z}, {Arag{\'o}n-Salamanca}, {Argudo-Fern{\'a}ndez}, {Armengaud}, {Aubourg}, {Avila-Reese}, {Badenes}, {Bailey}, {Barger}, {Barrera-Ballesteros}, {Bartosz}, {Bates}, {Baumgarten}, {Bautista}, {Beaton}, {Beers}, {Belfiore}, {Bender}, {Berlind}, {Bernardi}, {Beutler}, {Bird}, {Bizyaev}, {Blanc}, {Blomqvist}, {Bolton}, {Boquien}, {Borissova}, {van den Bosch}, {Bovy}, {Brandt}, {Brinkmann}, {Brownstein}, {Bundy}, {Burgasser}, {Burtin}, {Busca}, {Cappellari}, {Delgado Carigi}, {Carlberg}, {Carnero Rosell}, {Carrera}, {Chanover}, {Cherinka}, {Cheung}, {G{\'o}mez Maqueo Chew}, {Chiappini}, {Choi}, {Chojnowski}, {Chuang}, {Chung}, {Cirolini}, {Clerc}, {Cohen}, {Comparat}, {da Costa}, {Cousinou}, {Covey}, {Crane}, {Croft}, {Cruz-Gonzalez}, {Garrido Cuadra}, {Cunha}, {Damke}, {Darling}, {Davies}, {Dawson}, {de la Macorra}, {Dell'Agli}, {De
  Lee}, {Delubac}, {Di Mille}, {Diamond-Stanic}, {Cano-D{\'\i}az}, {Donor}, {Downes}, {Drory}, {du Mas des Bourboux}, {Duckworth}, {Dwelly}, {Dyer}, {Ebelke}, {Eigenbrot}, {Eisenstein}, {Emsellem}, {Eracleous}, {Escoffier}, {Evans}, {Fan}, {Fern{\'a}ndez-Alvar}, {Fernandez-Trincado}, {Feuillet}, {Finoguenov}, {Fleming}, {Font-Ribera}, {Fredrickson}, {Freischlad}, {Frinchaboy}, {Fuentes}, {Galbany}, {Garcia-Dias}, {Garc{\'\i}a-Hern{\'a}ndez}, {Gaulme}, {Geisler}, {Gelfand}, {Gil-Mar{\'\i}n}, {Gillespie}, {Goddard}, {Gonzalez-Perez}, {Grabowski}, {Green}, {Grier}, {Gunn}, {Guo}, {Guy}, {Hagen}, {Hahn}, {Hall}, {Harding}, {Hasselquist}, {Hawley}, {Hearty}, {Gonzalez Hern{\'a}ndez}, {Ho}, {Hogg}, {Holley-Bockelmann}, {Holtzman}, {Holzer}, {Huehnerhoff}, {Hutchinson}, {Hwang}, {Ibarra-Medel}, {da Silva Ilha}, {Ivans}, {Ivory}, {Jackson}, {Jensen}, {Johnson}, {Jones}, {J{\"o}nsson}, {Jullo}, {Kamble}, {Kinemuchi}, {Kirkby}, {Kitaura}, {Klaene}, {Knapp}, {Kneib}, {Kollmeier}, {Lacerna}, {Lane}, {Lang}, {Law},
  {Lazarz}, {Lee}, {Le Goff}, {Liang}, {Li}, {Li}, {Lian}, {Lima}, {Lin}, {Lin}, {Bertran de Lis}, {Liu}, {de Icaza Lizaola}, {Long}, {Lucatello}, {Lundgren}, {MacDonald}, {Deconto Machado}, {MacLeod}, {Mahadevan}, {Geimba Maia}, {Maiolino}, {Majewski}, {Malanushenko}, {Malanushenko}, {Manchado}, {Mao}, {Maraston}, {Marques-Chaves}, {Masseron}, {Masters}, {McBride}, {McDermid}, {McGrath}, {McGreer}, {Medina Pe{\~n}a}, {Melendez}, {Merloni}, {Merrifield}, {Meszaros}, {Meza}, {Minchev}, {Minniti}, {Miyaji}, {More}, {Mulchaey}, {M{\"u}ller-S{\'a}nchez}, {Muna}, {Munoz}, {Myers}, {Nair}, {Nandra}, {Correa do Nascimento}, {Negrete}, {Ness}, {Newman}, {Nichol}, {Nidever}, {Nitschelm}, {Ntelis}, {O'Connell}, {Oelkers}, {Oravetz}, {Oravetz}, {Pace}, {Padilla}, {Palanque-Delabrouille}, {Alonso Palicio}, {Pan}, {Parejko}, {Parikh}, {P{\^a}ris}, {Park}, {Patten}, {Peirani}, {Pellejero-Ibanez}, {Penny}, {Percival}, {Perez-Fournon}, {Petitjean}, {Pieri}, {Pinsonneault}, {Pisani}, {Poleski}, {Prada}, {Prakash}, {Queiroz},
  {Raddick}, {Raichoor}, {Barboza Rembold}, {Richstein}, {Riffel}, {Riffel}, {Rix}, {Robin}, {Rockosi}, {Rodr{\'\i}guez-Torres}, {Roman-Lopes}, {Rom{\'a}n-Z{\'u}{\~n}iga}, {Rosado}, {Ross}, {Rossi}, {Ruan}, {Ruggeri}, {Rykoff}, {Salazar-Albornoz}, {Salvato}, {S{\'a}nchez}, {Aguado}, {S{\'a}nchez-Gallego}, {Santana}, {Santiago}, {Sayres}, {Schiavon}, {da Silva Schimoia}, {Schlafly}, {Schlegel}, {Schneider}, {Schultheis}, {Schuster}, {Schwope}, {Seo}, {Shao}, {Shen}, {Shetrone}, {Shull}, {Simon}, {Skinner}, {Skrutskie}, {Slosar}, {Smith}, {Sobeck}, {Sobreira}, {Somers}, {Souto}, {Stark}, {Stassun}, {Stauffer}, {Steinmetz}, {Storchi-Bergmann}, {Streblyanska}, {Stringfellow}, {Su{\'a}rez}, {Sun}, {Suzuki}, {Szigeti}, {Taghizadeh-Popp}, {Tang}, {Tao}, {Tayar}, {Tembe}, {Teske}, {Thakar}, {Thomas}, {Thompson}, {Tinker}, {Tissera}, {Tojeiro}, {Hernandez Toledo}, {de la Torre}, {Tremonti}, {Troup}, {Valenzuela}, {Martinez Valpuesta}, {Vargas-Gonz{\'a}lez}, {Vargas-Maga{\~n}a}, {Vazquez}, {Villanova}, {Vivek}, {Vogt},
  {Wake}, {Walterbos}, {Wang}, {Weaver}, {Weijmans}, {Weinberg}, {Westfall}, {Whelan}, {Wild}, {Wilson}, {Wood-Vasey}, {Wylezalek}, {Xiao}, {Yan}, {Yang}, {Ybarra}, {Y{\`e}che}, {Zakamska}, {Zamora}, {Zarrouk}, {Zasowski}, {Zhang}, {Zhao}, {Zheng}, {Zheng}, {Zhou}, {Zhou}, {Zhu}, {Zoccali}, \& {Zou}}]{Blanton2017AJ}
{Blanton}, M.~R., {Bershady}, M.~A., {Abolfathi}, B., {et~al.} 2017, \aj, 154, 28, \dodoi{10.3847/1538-3881/aa7567}

\bibitem[{{Bodenheimer}(1965)}]{Bodenheimer1965ApJ}
{Bodenheimer}, P. 1965, \apj, 142, 451, \dodoi{10.1086/148310}

\bibitem[{{Borucki} {et~al.}(2010){Borucki}, {Koch}, {Basri}, {Batalha}, {Brown}, {Caldwell}, {Caldwell}, {Christensen-Dalsgaard}, {Cochran}, {DeVore}, {Dunham}, {Dupree}, {Gautier}, {Geary}, {Gilliland}, {Gould}, {Howell}, {Jenkins}, {Kondo}, {Latham}, {Marcy}, {Meibom}, {Kjeldsen}, {Lissauer}, {Monet}, {Morrison}, {Sasselov}, {Tarter}, {Boss}, {Brownlee}, {Owen}, {Buzasi}, {Charbonneau}, {Doyle}, {Fortney}, {Ford}, {Holman}, {Seager}, {Steffen}, {Welsh}, {Rowe}, {Anderson}, {Buchhave}, {Ciardi}, {Walkowicz}, {Sherry}, {Horch}, {Isaacson}, {Everett}, {Fischer}, {Torres}, {Johnson}, {Endl}, {MacQueen}, {Bryson}, {Dotson}, {Haas}, {Kolodziejczak}, {Van Cleve}, {Chandrasekaran}, {Twicken}, {Quintana}, {Clarke}, {Allen}, {Li}, {Wu}, {Tenenbaum}, {Verner}, {Bruhweiler}, {Barnes}, \& {Prsa}}]{Borucki2010Sci}
{Borucki}, W.~J., {Koch}, D., {Basri}, G., {et~al.} 2010, Science, 327, 977, \dodoi{10.1126/science.1185402}

\bibitem[{{Bossini} {et~al.}(2015){Bossini}, {Miglio}, {Salaris}, {Pietrinferni}, {Montalb{\'a}n}, {Bressan}, {Noels}, {Cassisi}, {Girardi}, \& {Marigo}}]{Bossini2015MNRAS}
{Bossini}, D., {Miglio}, A., {Salaris}, M., {et~al.} 2015, \mnras, 453, 2290, \dodoi{10.1093/mnras/stv1738}

\bibitem[{{Buder} {et~al.}(2018){Buder}, {Asplund}, {Duong}, {Kos}, {Lind}, {Ness}, {Sharma}, {Bland-Hawthorn}, {Casey}, {de Silva}, {D'Orazi}, {Freeman}, {Lewis}, {Lin}, {Martell}, {Schlesinger}, {Simpson}, {Zucker}, {Zwitter}, {Amarsi}, {Anguiano}, {Carollo}, {Casagrande}, {{\v{C}}otar}, {Cottrell}, {da Costa}, {Gao}, {Hayden}, {Horner}, {Ireland}, {Kafle}, {Munari}, {Nataf}, {Nordlander}, {Stello}, {Ting}, {Traven}, {Watson}, {Wittenmyer}, {Wyse}, {Yong}, {Zinn}, {{\v{Z}}erjal}, \& {Galah Collaboration}}]{Buder2018MNRAS}
{Buder}, S., {Asplund}, M., {Duong}, L., {et~al.} 2018, \mnras, 478, 4513, \dodoi{10.1093/mnras/sty1281}

\bibitem[{{Buder} {et~al.}(2021){Buder}, {Sharma}, {Kos}, {Amarsi}, {Nordlander}, {Lind}, {Martell}, {Asplund}, {Bland-Hawthorn}, {Casey}, {de Silva}, {D'Orazi}, {Freeman}, {Hayden}, {Lewis}, {Lin}, {Schlesinger}, {Simpson}, {Stello}, {Zucker}, {Zwitter}, {Beeson}, {Buck}, {Casagrande}, {Clark}, {{\v{C}}otar}, {da Costa}, {de Grijs}, {Feuillet}, {Horner}, {Kafle}, {Khanna}, {Kobayashi}, {Liu}, {Montet}, {Nandakumar}, {Nataf}, {Ness}, {Spina}, {Tepper-Garc{\'\i}a}, {Ting}, {Traven}, {Vogrin{\v{c}}i{\v{c}}}, {Wittenmyer}, {Wyse}, {{\v{Z}}erjal}, \& {Galah Collaboration}}]{Buder2021MNRAS}
{Buder}, S., {Sharma}, S., {Kos}, J., {et~al.} 2021, \mnras, 506, 150, \dodoi{10.1093/mnras/stab1242}

\bibitem[{{Caffau} {et~al.}(2011){Caffau}, {Ludwig}, {Steffen}, {Freytag}, \& {Bonifacio}}]{Caffau2011SoPh}
{Caffau}, E., {Ludwig}, H.~G., {Steffen}, M., {Freytag}, B., \& {Bonifacio}, P. 2011, \solphys, 268, 255, \dodoi{10.1007/s11207-010-9541-4}

\bibitem[{{Caplan} {et~al.}(2022){Caplan}, {Bauer}, \& {Freeman}}]{Caplan2022MNRAS}
{Caplan}, M.~E., {Bauer}, E.~B., \& {Freeman}, I.~F. 2022, \mnras, 513, L52, \dodoi{10.1093/mnrasl/slac032}

\bibitem[{{Castelli} \& {Kurucz}(2003)}]{Castelli2003IAUS}
{Castelli}, F., \& {Kurucz}, R.~L. 2003, in Modelling of Stellar Atmospheres, ed. N.~{Piskunov}, W.~W. {Weiss}, \& D.~F. {Gray}, Vol. 210, A20, \dodoi{10.48550/arXiv.astro-ph/0405087}

\bibitem[{{Chanam{\'e}} {et~al.}(2022){Chanam{\'e}}, {Pinsonneault}, {Aguilera-G{\'o}mez}, \& {Zinn}}]{Chaname2022ApJ}
{Chanam{\'e}}, J., {Pinsonneault}, M.~H., {Aguilera-G{\'o}mez}, C., \& {Zinn}, J.~C. 2022, \apj, 933, 58, \dodoi{10.3847/1538-4357/ac70c8}

\bibitem[{{Chaplin} \& {Miglio}(2013)}]{Chaplin2013ARA&A}
{Chaplin}, W.~J., \& {Miglio}, A. 2013, \araa, 51, 353, \dodoi{10.1146/annurev-astro-082812-140938}

\bibitem[{{Choi} {et~al.}(2016){Choi}, {Dotter}, {Conroy}, {Cantiello}, {Paxton}, \& {Johnson}}]{Choi2016ApJ}
{Choi}, J., {Dotter}, A., {Conroy}, C., {et~al.} 2016, \apj, 823, 102, \dodoi{10.3847/0004-637X/823/2/102}

\bibitem[{{Christensen-Dalsgaard}(2015)}]{Christensen-Dalsgaard2015MNRAS}
{Christensen-Dalsgaard}, J. 2015, \mnras, 453, 666, \dodoi{10.1093/mnras/stv1656}

\bibitem[{{Christensen-Dalsgaard}(2021)}]{Christensen-Dalsgaard2021LRSP}
---. 2021, Living Reviews in Solar Physics, 18, 2, \dodoi{10.1007/s41116-020-00028-3}

\bibitem[{{Chugunov} {et~al.}(2007){Chugunov}, {Dewitt}, \& {Yakovlev}}]{Chugunov2007}
{Chugunov}, A.~I., {Dewitt}, H.~E., \& {Yakovlev}, D.~G. 2007, \prd, 76, 025028, \dodoi{10.1103/PhysRevD.76.025028}

\bibitem[{{Cox} \& {Giuli}(1968)}]{Cox1968pss}
{Cox}, J.~P., \& {Giuli}, R.~T. 1968, {Principles of stellar structure}

\bibitem[{{Cyburt} {et~al.}(2016){Cyburt}, {Fields}, {Olive}, \& {Yeh}}]{Cyburt2016RvMP}
{Cyburt}, R.~H., {Fields}, B.~D., {Olive}, K.~A., \& {Yeh}, T.-H. 2016, Reviews of Modern Physics, 88, 015004, \dodoi{10.1103/RevModPhys.88.015004}

\bibitem[{{Cyburt} {et~al.}(2010){Cyburt}, {Amthor}, {Ferguson}, {Meisel}, {Smith}, {Warren}, {Heger}, {Hoffman}, {Rauscher}, {Sakharuk}, {Schatz}, {Thielemann}, \& {Wiescher}}]{Cyburt2010ApJS}
{Cyburt}, R.~H., {Amthor}, A.~M., {Ferguson}, R., {et~al.} 2010, \apjs, 189, 240, \dodoi{10.1088/0067-0049/189/1/240}

\bibitem[{{De Silva} {et~al.}(2015){De Silva}, {Freeman}, {Bland-Hawthorn}, {Martell}, {de Boer}, {Asplund}, {Keller}, {Sharma}, {Zucker}, {Zwitter}, {Anguiano}, {Bacigalupo}, {Bayliss}, {Beavis}, {Bergemann}, {Campbell}, {Cannon}, {Carollo}, {Casagrande}, {Casey}, {Da Costa}, {D'Orazi}, {Dotter}, {Duong}, {Heger}, {Ireland}, {Kafle}, {Kos}, {Lattanzio}, {Lewis}, {Lin}, {Lind}, {Munari}, {Nataf}, {O'Toole}, {Parker}, {Reid}, {Schlesinger}, {Sheinis}, {Simpson}, {Stello}, {Ting}, {Traven}, {Watson}, {Wittenmyer}, {Yong}, \& {{\v{Z}}erjal}}]{DeSilva2015MNRAS}
{De Silva}, G.~M., {Freeman}, K.~C., {Bland-Hawthorn}, J., {et~al.} 2015, \mnras, 449, 2604, \dodoi{10.1093/mnras/stv327}

\bibitem[{{Demarque} {et~al.}(2004){Demarque}, {Woo}, {Kim}, \& {Yi}}]{Demarque2004ApJS}
{Demarque}, P., {Woo}, J.-H., {Kim}, Y.-C., \& {Yi}, S.~K. 2004, \apjs, 155, 667, \dodoi{10.1086/424966}

\bibitem[{{Deng} {et~al.}(2012){Deng}, {Newberg}, {Liu}, {Carlin}, {Beers}, {Chen}, {Chen}, {Christlieb}, {Grillmair}, {Guhathakurta}, {Han}, {Hou}, {Lee}, {L{\'e}pine}, {Li}, {Liu}, {Pan}, {Sellwood}, {Wang}, {Wang}, {Yang}, {Yanny}, {Zhang}, {Zhang}, {Zheng}, \& {Zhu}}]{Deng2012RAA}
{Deng}, L.-C., {Newberg}, H.~J., {Liu}, C., {et~al.} 2012, Research in Astronomy and Astrophysics, 12, 735, \dodoi{10.1088/1674-4527/12/7/003}

\bibitem[{{Desch} {et~al.}(2023){Desch}, {Dunlap}, {Dunham}, {Williams}, \& {Mane}}]{Desch2023Icar}
{Desch}, S.~J., {Dunlap}, D.~R., {Dunham}, E.~T., {Williams}, C.~D., \& {Mane}, P. 2023, \icarus, 402, 115607, \dodoi{10.1016/j.icarus.2023.115607}

\bibitem[{{Diaz Reeve} \& {Serenelli}(2023)}]{DiazReeve2023poster}
{Diaz Reeve}, P., \& {Serenelli}, A. 2023, in PLATO Stellar Science Conference 2023, 12, \dodoi{10.5281/zenodo.8108192}

\bibitem[{{D{\'\i}az Reeve} \& {Serenelli}(2023)}]{DiazReeve2023RNAAS}
{D{\'\i}az Reeve}, P., \& {Serenelli}, A. 2023, Research Notes of the American Astronomical Society, 7, 196, \dodoi{10.3847/2515-5172/acf9a2}

\bibitem[{{Eisenstein} {et~al.}(2011){Eisenstein}, {Weinberg}, {Agol}, {Aihara}, {Allende Prieto}, {Anderson}, {Arns}, {Aubourg}, {Bailey}, {Balbinot}, {Barkhouser}, {Beers}, {Berlind}, {Bickerton}, {Bizyaev}, {Blanton}, {Bochanski}, {Bolton}, {Bosman}, {Bovy}, {Brandt}, {Breslauer}, {Brewington}, {Brinkmann}, {Brown}, {Brownstein}, {Burger}, {Busca}, {Campbell}, {Cargile}, {Carithers}, {Carlberg}, {Carr}, {Chang}, {Chen}, {Chiappini}, {Comparat}, {Connolly}, {Cortes}, {Croft}, {Cunha}, {da Costa}, {Davenport}, {Dawson}, {De Lee}, {Porto de Mello}, {de Simoni}, {Dean}, {Dhital}, {Ealet}, {Ebelke}, {Edmondson}, {Eiting}, {Escoffier}, {Esposito}, {Evans}, {Fan}, {Femen{\'\i}a Castell{\'a}}, {Dutra Ferreira}, {Fitzgerald}, {Fleming}, {Font-Ribera}, {Ford}, {Frinchaboy}, {Garc{\'\i}a P{\'e}rez}, {Gaudi}, {Ge}, {Ghezzi}, {Gillespie}, {Gilmore}, {Girardi}, {Gott}, {Gould}, {Grebel}, {Gunn}, {Hamilton}, {Harding}, {Harris}, {Hawley}, {Hearty}, {Hennawi}, {Gonz{\'a}lez Hern{\'a}ndez}, {Ho}, {Hogg}, {Holtzman},
  {Honscheid}, {Inada}, {Ivans}, {Jiang}, {Jiang}, {Johnson}, {Jordan}, {Jordan}, {Kauffmann}, {Kazin}, {Kirkby}, {Klaene}, {Knapp}, {Kneib}, {Kochanek}, {Koesterke}, {Kollmeier}, {Kron}, {Lampeitl}, {Lang}, {Lawler}, {Le Goff}, {Lee}, {Lee}, {Leisenring}, {Lin}, {Liu}, {Long}, {Loomis}, {Lucatello}, {Lundgren}, {Lupton}, {Ma}, {Ma}, {MacDonald}, {Mack}, {Mahadevan}, {Maia}, {Majewski}, {Makler}, {Malanushenko}, {Malanushenko}, {Mandelbaum}, {Maraston}, {Margala}, {Maseman}, {Masters}, {McBride}, {McDonald}, {McGreer}, {McMahon}, {Mena Requejo}, {M{\'e}nard}, {Miralda-Escud{\'e}}, {Morrison}, {Mullally}, {Muna}, {Murayama}, {Myers}, {Naugle}, {Neto}, {Nguyen}, {Nichol}, {Nidever}, {O'Connell}, {Ogando}, {Olmstead}, {Oravetz}, {Padmanabhan}, {Paegert}, {Palanque-Delabrouille}, {Pan}, {Pandey}, {Parejko}, {P{\^a}ris}, {Pellegrini}, {Pepper}, {Percival}, {Petitjean}, {Pfaffenberger}, {Pforr}, {Phleps}, {Pichon}, {Pieri}, {Prada}, {Price-Whelan}, {Raddick}, {Ramos}, {Reid}, {Reyle}, {Rich}, {Richards}, {Rieke},
  {Rieke}, {Rix}, {Robin}, {Rocha-Pinto}, {Rockosi}, {Roe}, {Rollinde}, {Ross}, {Ross}, {Rossetto}, {S{\'a}nchez}, {Santiago}, {Sayres}, {Schiavon}, {Schlegel}, {Schlesinger}, {Schmidt}, {Schneider}, {Sellgren}, {Shelden}, {Sheldon}, {Shetrone}, {Shu}, {Silverman}, {Simmerer}, {Simmons}, {Sivarani}, {Skrutskie}, {Slosar}, {Smee}, {Smith}, {Snedden}, {Stassun}, {Steele}, {Steinmetz}, {Stockett}, {Stollberg}, {Strauss}, {Szalay}, {Tanaka}, {Thakar}, {Thomas}, {Tinker}, {Tofflemire}, {Tojeiro}, {Tremonti}, {Vargas Maga{\~n}a}, {Verde}, {Vogt}, {Wake}, {Wan}, {Wang}, {Weaver}, {White}, {White}, {Wilson}, {Wisniewski}, {Wood-Vasey}, {Yanny}, {Yasuda}, {Y{\`e}che}, {York}, {Young}, {Zasowski}, {Zehavi}, \& {Zhao}}]{Eisenstein2011AJ}
{Eisenstein}, D.~J., {Weinberg}, D.~H., {Agol}, E., {et~al.} 2011, \aj, 142, 72, \dodoi{10.1088/0004-6256/142/3/72}

\bibitem[{{Elsworth} {et~al.}(2019){Elsworth}, {Hekker}, {Johnson}, {Kallinger}, {Mosser}, {Pinsonneault}, {Hon}, {Kuszlewicz}, {Miglio}, {Serenelli}, {Stello}, {Tayar}, \& {Vrard}}]{Elsworth2019MNRAS}
{Elsworth}, Y., {Hekker}, S., {Johnson}, J.~A., {et~al.} 2019, \mnras, 489, 4641, \dodoi{10.1093/mnras/stz2356}

\bibitem[{{Farag} {et~al.}(2024){Farag}, {Fontes}, {Timmes}, {Bellinger}, {Guzik}, {Bauer}, {Wood}, {Mussack}, {Hakel}, {Colgan}, {Kilcrease}, {Sherrill}, {Raecke}, \& {Chidester}}]{Farag2024ApJ}
{Farag}, E., {Fontes}, C.~J., {Timmes}, F.~X., {et~al.} 2024, \apj, 968, 56, \dodoi{10.3847/1538-4357/ad4355}

\bibitem[{{Ferguson} {et~al.}(2005){Ferguson}, {Alexander}, {Allard}, {Barman}, {Bodnarik}, {Hauschildt}, {Heffner-Wong}, \& {Tamanai}}]{Ferguson2005ApJ}
{Ferguson}, J.~W., {Alexander}, D.~R., {Allard}, F., {et~al.} 2005, \apj, 623, 585, \dodoi{10.1086/428642}

\bibitem[{{Fu} {et~al.}(2015){Fu}, {Bressan}, {Molaro}, \& {Marigo}}]{Fu2015MNRAS}
{Fu}, X., {Bressan}, A., {Molaro}, P., \& {Marigo}, P. 2015, \mnras, 452, 3256, \dodoi{10.1093/mnras/stv1384}

\bibitem[{{Fuller} {et~al.}(1985){Fuller}, {Fowler}, \& {Newman}}]{Fuller1985ApJ}
{Fuller}, G.~M., {Fowler}, W.~A., \& {Newman}, M.~J. 1985, \apj, 293, 1, \dodoi{10.1086/163208}

\bibitem[{{Gaia Collaboration} {et~al.}(2016){Gaia Collaboration}, {Prusti}, {de Bruijne}, {Brown}, {Vallenari}, {Babusiaux}, {Bailer-Jones}, {Bastian}, {Biermann}, {Evans}, {Eyer}, {Jansen}, {Jordi}, {Klioner}, {Lammers}, {Lindegren}, {Luri}, {Mignard}, {Milligan}, {Panem}, {Poinsignon}, {Pourbaix}, {Randich}, {Sarri}, {Sartoretti}, {Siddiqui}, {Soubiran}, {Valette}, {van Leeuwen}, {Walton}, {Aerts}, {Arenou}, {Cropper}, {Drimmel}, {H{\o}g}, {Katz}, {Lattanzi}, {O'Mullane}, {Grebel}, {Holland}, {Huc}, {Passot}, {Bramante}, {Cacciari}, {Casta{\~n}eda}, {Chaoul}, {Cheek}, {De Angeli}, {Fabricius}, {Guerra}, {Hern{\'a}ndez}, {Jean-Antoine-Piccolo}, {Masana}, {Messineo}, {Mowlavi}, {Nienartowicz}, {Ord{\'o}{\~n}ez-Blanco}, {Panuzzo}, {Portell}, {Richards}, {Riello}, {Seabroke}, {Tanga}, {Th{\'e}venin}, {Torra}, {Els}, {Gracia-Abril}, {Comoretto}, {Garcia-Reinaldos}, {Lock}, {Mercier}, {Altmann}, {Andrae}, {Astraatmadja}, {Bellas-Velidis}, {Benson}, {Berthier}, {Blomme}, {Busso}, {Carry}, {Cellino}, {Clementini},
  {Cowell}, {Creevey}, {Cuypers}, {Davidson}, {De Ridder}, {de Torres}, {Delchambre}, {Dell'Oro}, {Ducourant}, {Fr{\'e}mat}, {Garc{\'\i}a-Torres}, {Gosset}, {Halbwachs}, {Hambly}, {Harrison}, {Hauser}, {Hestroffer}, {Hodgkin}, {Huckle}, {Hutton}, {Jasniewicz}, {Jordan}, {Kontizas}, {Korn}, {Lanzafame}, {Manteiga}, {Moitinho}, {Muinonen}, {Osinde}, {Pancino}, {Pauwels}, {Petit}, {Recio-Blanco}, {Robin}, {Sarro}, {Siopis}, {Smith}, {Smith}, {Sozzetti}, {Thuillot}, {van Reeven}, {Viala}, {Abbas}, {Abreu Aramburu}, {Accart}, {Aguado}, {Allan}, {Allasia}, {Altavilla}, {{\'A}lvarez}, {Alves}, {Anderson}, {Andrei}, {Anglada Varela}, {Antiche}, {Antoja}, {Ant{\'o}n}, {Arcay}, {Atzei}, {Ayache}, {Bach}, {Baker}, {Balaguer-N{\'u}{\~n}ez}, {Barache}, {Barata}, {Barbier}, {Barblan}, {Baroni}, {Barrado y Navascu{\'e}s}, {Barros}, {Barstow}, {Becciani}, {Bellazzini}, {Bellei}, {Bello Garc{\'\i}a}, {Belokurov}, {Bendjoya}, {Berihuete}, {Bianchi}, {Bienaym{\'e}}, {Billebaud}, {Blagorodnova}, {Blanco-Cuaresma}, {Boch},
  {Bombrun}, {Borrachero}, {Bouquillon}, {Bourda}, {Bouy}, {Bragaglia}, {Breddels}, {Brouillet}, {Br{\"u}semeister}, {Bucciarelli}, {Budnik}, {Burgess}, {Burgon}, {Burlacu}, {Busonero}, {Buzzi}, {Caffau}, {Cambras}, {Campbell}, {Cancelliere}, {Cantat-Gaudin}, {Carlucci}, {Carrasco}, {Castellani}, {Charlot}, {Charnas}, {Charvet}, {Chassat}, {Chiavassa}, {Clotet}, {Cocozza}, {Collins}, {Collins}, {Costigan}, {Crifo}, {Cross}, {Crosta}, {Crowley}, {Dafonte}, {Damerdji}, {Dapergolas}, {David}, {David}, {De Cat}, {de Felice}, {de Laverny}, {De Luise}, {De March}, {de Martino}, {de Souza}, {Debosscher}, {del Pozo}, {Delbo}, {Delgado}, {Delgado}, {di Marco}, {Di Matteo}, {Diakite}, {Distefano}, {Dolding}, {Dos Anjos}, {Drazinos}, {Dur{\'a}n}, {Dzigan}, {Ecale}, {Edvardsson}, {Enke}, {Erdmann}, {Escolar}, {Espina}, {Evans}, {Eynard Bontemps}, {Fabre}, {Fabrizio}, {Faigler}, {Falc{\~a}o}, {Farr{\`a}s Casas}, {Faye}, {Federici}, {Fedorets}, {Fern{\'a}ndez-Hern{\'a}ndez}, {Fernique}, {Fienga}, {Figueras}, {Filippi},
  {Findeisen}, {Fonti}, {Fouesneau}, {Fraile}, {Fraser}, {Fuchs}, {Furnell}, {Gai}, {Galleti}, {Galluccio}, {Garabato}, {Garc{\'\i}a-Sedano}, {Gar{\'e}}, {Garofalo}, {Garralda}, {Gavras}, {Gerssen}, {Geyer}, {Gilmore}, {Girona}, {Giuffrida}, {Gomes}, {Gonz{\'a}lez-Marcos}, {Gonz{\'a}lez-N{\'u}{\~n}ez}, {Gonz{\'a}lez-Vidal}, {Granvik}, {Guerrier}, {Guillout}, {Guiraud}, {G{\'u}rpide}, {Guti{\'e}rrez-S{\'a}nchez}, {Guy}, {Haigron}, {Hatzidimitriou}, {Haywood}, {Heiter}, {Helmi}, {Hobbs}, {Hofmann}, {Holl}, {Holland}, {Hunt}, {Hypki}, {Icardi}, {Irwin}, {Jevardat de Fombelle}, {Jofr{\'e}}, {Jonker}, {Jorissen}, {Julbe}, {Karampelas}, {Kochoska}, {Kohley}, {Kolenberg}, {Kontizas}, {Koposov}, {Kordopatis}, {Koubsky}, {Kowalczyk}, {Krone-Martins}, {Kudryashova}, {Kull}, {Bachchan}, {Lacoste-Seris}, {Lanza}, {Lavigne}, {Le Poncin-Lafitte}, {Lebreton}, {Lebzelter}, {Leccia}, {Leclerc}, {Lecoeur-Taibi}, {Lemaitre}, {Lenhardt}, {Leroux}, {Liao}, {Licata}, {Lindstr{\o}m}, {Lister}, {Livanou}, {Lobel}, {L{\"o}ffler},
  {L{\'o}pez}, {Lopez-Lozano}, {Lorenz}, {Loureiro}, {MacDonald}, {Magalh{\~a}es Fernandes}, {Managau}, {Mann}, {Mantelet}, {Marchal}, {Marchant}, {Marconi}, {Marie}, {Marinoni}, {Marrese}, {Marschalk{\'o}}, {Marshall}, {Mart{\'\i}n-Fleitas}, {Martino}, {Mary}, {Matijevi{\v{c}}}, {Mazeh}, {McMillan}, {Messina}, {Mestre}, {Michalik}, {Millar}, {Miranda}, {Molina}, {Molinaro}, {Molinaro}, {Moln{\'a}r}, {Moniez}, {Montegriffo}, {Monteiro}, {Mor}, {Mora}, {Morbidelli}, {Morel}, {Morgenthaler}, {Morley}, {Morris}, {Mulone}, {Muraveva}, {Musella}, {Narbonne}, {Nelemans}, {Nicastro}, {Noval}, {Ord{\'e}novic}, {Ordieres-Mer{\'e}}, {Osborne}, {Pagani}, {Pagano}, {Pailler}, {Palacin}, {Palaversa}, {Parsons}, {Paulsen}, {Pecoraro}, {Pedrosa}, {Pentik{\"a}inen}, {Pereira}, {Pichon}, {Piersimoni}, {Pineau}, {Plachy}, {Plum}, {Poujoulet}, {Pr{\v{s}}a}, {Pulone}, {Ragaini}, {Rago}, {Rambaux}, {Ramos-Lerate}, {Ranalli}, {Rauw}, {Read}, {Regibo}, {Renk}, {Reyl{\'e}}, {Ribeiro}, {Rimoldini}, {Ripepi}, {Riva}, {Rixon},
  {Roelens}, {Romero-G{\'o}mez}, {Rowell}, {Royer}, {Rudolph}, {Ruiz-Dern}, {Sadowski}, {Sagrist{\`a} Sell{\'e}s}, {Sahlmann}, {Salgado}, {Salguero}, {Sarasso}, {Savietto}, {Schnorhk}, {Schultheis}, {Sciacca}, {Segol}, {Segovia}, {Segransan}, {Serpell}, {Shih}, {Smareglia}, {Smart}, {Smith}, {Solano}, {Solitro}, {Sordo}, {Soria Nieto}, {Souchay}, {Spagna}, {Spoto}, {Stampa}, {Steele}, {Steidelm{\"u}ller}, {Stephenson}, {Stoev}, {Suess}, {S{\"u}veges}, {Surdej}, {Szabados}, {Szegedi-Elek}, {Tapiador}, {Taris}, {Tauran}, {Taylor}, {Teixeira}, {Terrett}, {Tingley}, {Trager}, {Turon}, {Ulla}, {Utrilla}, {Valentini}, {van Elteren}, {Van Hemelryck}, {van Leeuwen}, {Varadi}, {Vecchiato}, {Veljanoski}, {Via}, {Vicente}, {Vogt}, {Voss}, {Votruba}, {Voutsinas}, {Walmsley}, {Weiler}, {Weingrill}, {Werner}, {Wevers}, {Whitehead}, {Wyrzykowski}, {Yoldas}, {{\v{Z}}erjal}, {Zucker}, {Zurbach}, {Zwitter}, {Alecu}, {Allen}, {Allende Prieto}, {Amorim}, {Anglada-Escud{\'e}}, {Arsenijevic}, {Azaz}, {Balm}, {Beck}, {Bernstein},
  {Bigot}, {Bijaoui}, {Blasco}, {Bonfigli}, {Bono}, {Boudreault}, {Bressan}, {Brown}, {Brunet}, {Bunclark}, {Buonanno}, {Butkevich}, {Carret}, {Carrion}, {Chemin}, {Ch{\'e}reau}, {Corcione}, {Darmigny}, {de Boer}, {de Teodoro}, {de Zeeuw}, {Delle Luche}, {Domingues}, {Dubath}, {Fodor}, {Fr{\'e}zouls}, {Fries}, {Fustes}, {Fyfe}, {Gallardo}, {Gallegos}, {Gardiol}, {Gebran}, {Gomboc}, {G{\'o}mez}, {Grux}, {Gueguen}, {Heyrovsky}, {Hoar}, {Iannicola}, {Isasi Parache}, {Janotto}, {Joliet}, {Jonckheere}, {Keil}, {Kim}, {Klagyivik}, {Klar}, {Knude}, {Kochukhov}, {Kolka}, {Kos}, {Kutka}, {Lainey}, {LeBouquin}, {Liu}, {Loreggia}, {Makarov}, {Marseille}, {Martayan}, {Martinez-Rubi}, {Massart}, {Meynadier}, {Mignot}, {Munari}, {Nguyen}, {Nordlander}, {Ocvirk}, {O'Flaherty}, {Olias Sanz}, {Ortiz}, {Osorio}, {Oszkiewicz}, {Ouzounis}, {Palmer}, {Park}, {Pasquato}, {Peltzer}, {Peralta}, {P{\'e}turaud}, {Pieniluoma}, {Pigozzi}, {Poels}, {Prat}, {Prod'homme}, {Raison}, {Rebordao}, {Risquez}, {Rocca-Volmerange}, {Rosen},
  {Ruiz-Fuertes}, {Russo}, {Sembay}, {Serraller Vizcaino}, {Short}, {Siebert}, {Silva}, {Sinachopoulos}, {Slezak}, {Soffel}, {Sosnowska}, {Strai{\v{z}}ys}, {ter Linden}, {Terrell}, {Theil}, {Tiede}, {Troisi}, {Tsalmantza}, {Tur}, {Vaccari}, {Vachier}, {Valles}, {Van Hamme}, {Veltz}, {Virtanen}, {Wallut}, {Wichmann}, {Wilkinson}, {Ziaeepour}, \& {Zschocke}}]{Gaia2016A&A}
{Gaia Collaboration}, {Prusti}, T., {de Bruijne}, J.~H.~J., {et~al.} 2016, \aap, 595, A1, \dodoi{10.1051/0004-6361/201629272}

\bibitem[{{Gaia Collaboration} {et~al.}(2018){Gaia Collaboration}, {Brown}, {Vallenari}, {Prusti}, {de Bruijne}, {Babusiaux}, {Bailer-Jones}, {Biermann}, {Evans}, {Eyer}, {Jansen}, {Jordi}, {Klioner}, {Lammers}, {Lindegren}, {Luri}, {Mignard}, {Panem}, {Pourbaix}, {Randich}, {Sartoretti}, {Siddiqui}, {Soubiran}, {van Leeuwen}, {Walton}, {Arenou}, {Bastian}, {Cropper}, {Drimmel}, {Katz}, {Lattanzi}, {Bakker}, {Cacciari}, {Casta{\~n}eda}, {Chaoul}, {Cheek}, {De Angeli}, {Fabricius}, {Guerra}, {Holl}, {Masana}, {Messineo}, {Mowlavi}, {Nienartowicz}, {Panuzzo}, {Portell}, {Riello}, {Seabroke}, {Tanga}, {Th{\'e}venin}, {Gracia-Abril}, {Comoretto}, {Garcia-Reinaldos}, {Teyssier}, {Altmann}, {Andrae}, {Audard}, {Bellas-Velidis}, {Benson}, {Berthier}, {Blomme}, {Burgess}, {Busso}, {Carry}, {Cellino}, {Clementini}, {Clotet}, {Creevey}, {Davidson}, {De Ridder}, {Delchambre}, {Dell'Oro}, {Ducourant}, {Fern{\'a}ndez-Hern{\'a}ndez}, {Fouesneau}, {Fr{\'e}mat}, {Galluccio}, {Garc{\'\i}a-Torres},
  {Gonz{\'a}lez-N{\'u}{\~n}ez}, {Gonz{\'a}lez-Vidal}, {Gosset}, {Guy}, {Halbwachs}, {Hambly}, {Harrison}, {Hern{\'a}ndez}, {Hestroffer}, {Hodgkin}, {Hutton}, {Jasniewicz}, {Jean-Antoine-Piccolo}, {Jordan}, {Korn}, {Krone-Martins}, {Lanzafame}, {Lebzelter}, {L{\"o}ffler}, {Manteiga}, {Marrese}, {Mart{\'\i}n-Fleitas}, {Moitinho}, {Mora}, {Muinonen}, {Osinde}, {Pancino}, {Pauwels}, {Petit}, {Recio-Blanco}, {Richards}, {Rimoldini}, {Robin}, {Sarro}, {Siopis}, {Smith}, {Sozzetti}, {S{\"u}veges}, {Torra}, {van Reeven}, {Abbas}, {Abreu Aramburu}, {Accart}, {Aerts}, {Altavilla}, {{\'A}lvarez}, {Alvarez}, {Alves}, {Anderson}, {Andrei}, {Anglada Varela}, {Antiche}, {Antoja}, {Arcay}, {Astraatmadja}, {Bach}, {Baker}, {Balaguer-N{\'u}{\~n}ez}, {Balm}, {Barache}, {Barata}, {Barbato}, {Barblan}, {Barklem}, {Barrado}, {Barros}, {Barstow}, {Bartholom{\'e} Mu{\~n}oz}, {Bassilana}, {Becciani}, {Bellazzini}, {Berihuete}, {Bertone}, {Bianchi}, {Bienaym{\'e}}, {Blanco-Cuaresma}, {Boch}, {Boeche}, {Bombrun}, {Borrachero},
  {Bossini}, {Bouquillon}, {Bourda}, {Bragaglia}, {Bramante}, {Breddels}, {Bressan}, {Brouillet}, {Br{\"u}semeister}, {Brugaletta}, {Bucciarelli}, {Burlacu}, {Busonero}, {Butkevich}, {Buzzi}, {Caffau}, {Cancelliere}, {Cannizzaro}, {Cantat-Gaudin}, {Carballo}, {Carlucci}, {Carrasco}, {Casamiquela}, {Castellani}, {Castro-Ginard}, {Charlot}, {Chemin}, {Chiavassa}, {Cocozza}, {Costigan}, {Cowell}, {Crifo}, {Crosta}, {Crowley}, {Cuypers}, {Dafonte}, {Damerdji}, {Dapergolas}, {David}, {David}, {de Laverny}, {De Luise}, {De March}, {de Martino}, {de Souza}, {de Torres}, {Debosscher}, {del Pozo}, {Delbo}, {Delgado}, {Delgado}, {Di Matteo}, {Diakite}, {Diener}, {Distefano}, {Dolding}, {Drazinos}, {Dur{\'a}n}, {Edvardsson}, {Enke}, {Eriksson}, {Esquej}, {Eynard Bontemps}, {Fabre}, {Fabrizio}, {Faigler}, {Falc{\~a}o}, {Farr{\`a}s Casas}, {Federici}, {Fedorets}, {Fernique}, {Figueras}, {Filippi}, {Findeisen}, {Fonti}, {Fraile}, {Fraser}, {Fr{\'e}zouls}, {Gai}, {Galleti}, {Garabato}, {Garc{\'\i}a-Sedano}, {Garofalo},
  {Garralda}, {Gavel}, {Gavras}, {Gerssen}, {Geyer}, {Giacobbe}, {Gilmore}, {Girona}, {Giuffrida}, {Glass}, {Gomes}, {Granvik}, {Gueguen}, {Guerrier}, {Guiraud}, {Guti{\'e}rrez-S{\'a}nchez}, {Haigron}, {Hatzidimitriou}, {Hauser}, {Haywood}, {Heiter}, {Helmi}, {Heu}, {Hilger}, {Hobbs}, {Hofmann}, {Holland}, {Huckle}, {Hypki}, {Icardi}, {Jan{\ss}en}, {Jevardat de Fombelle}, {Jonker}, {Juh{\'a}sz}, {Julbe}, {Karampelas}, {Kewley}, {Klar}, {Kochoska}, {Kohley}, {Kolenberg}, {Kontizas}, {Kontizas}, {Koposov}, {Kordopatis}, {Kostrzewa-Rutkowska}, {Koubsky}, {Lambert}, {Lanza}, {Lasne}, {Lavigne}, {Le Fustec}, {Le Poncin-Lafitte}, {Lebreton}, {Leccia}, {Leclerc}, {Lecoeur-Taibi}, {Lenhardt}, {Leroux}, {Liao}, {Licata}, {Lindstr{\o}m}, {Lister}, {Livanou}, {Lobel}, {L{\'o}pez}, {Managau}, {Mann}, {Mantelet}, {Marchal}, {Marchant}, {Marconi}, {Marinoni}, {Marschalk{\'o}}, {Marshall}, {Martino}, {Marton}, {Mary}, {Massari}, {Matijevi{\v{c}}}, {Mazeh}, {McMillan}, {Messina}, {Michalik}, {Millar}, {Molina}, {Molinaro},
  {Moln{\'a}r}, {Montegriffo}, {Mor}, {Morbidelli}, {Morel}, {Morris}, {Mulone}, {Muraveva}, {Musella}, {Nelemans}, {Nicastro}, {Noval}, {O'Mullane}, {Ord{\'e}novic}, {Ord{\'o}{\~n}ez-Blanco}, {Osborne}, {Pagani}, {Pagano}, {Pailler}, {Palacin}, {Palaversa}, {Panahi}, {Pawlak}, {Piersimoni}, {Pineau}, {Plachy}, {Plum}, {Poggio}, {Poujoulet}, {Pr{\v{s}}a}, {Pulone}, {Racero}, {Ragaini}, {Rambaux}, {Ramos-Lerate}, {Regibo}, {Reyl{\'e}}, {Riclet}, {Ripepi}, {Riva}, {Rivard}, {Rixon}, {Roegiers}, {Roelens}, {Romero-G{\'o}mez}, {Rowell}, {Royer}, {Ruiz-Dern}, {Sadowski}, {Sagrist{\`a} Sell{\'e}s}, {Sahlmann}, {Salgado}, {Salguero}, {Sanna}, {Santana-Ros}, {Sarasso}, {Savietto}, {Schultheis}, {Sciacca}, {Segol}, {Segovia}, {S{\'e}gransan}, {Shih}, {Siltala}, {Silva}, {Smart}, {Smith}, {Solano}, {Solitro}, {Sordo}, {Soria Nieto}, {Souchay}, {Spagna}, {Spoto}, {Stampa}, {Steele}, {Steidelm{\"u}ller}, {Stephenson}, {Stoev}, {Suess}, {Surdej}, {Szabados}, {Szegedi-Elek}, {Tapiador}, {Taris}, {Tauran}, {Taylor},
  {Teixeira}, {Terrett}, {Teyssandier}, {Thuillot}, {Titarenko}, {Torra Clotet}, {Turon}, {Ulla}, {Utrilla}, {Uzzi}, {Vaillant}, {Valentini}, {Valette}, {van Elteren}, {Van Hemelryck}, {van Leeuwen}, {Vaschetto}, {Vecchiato}, {Veljanoski}, {Viala}, {Vicente}, {Vogt}, {von Essen}, {Voss}, {Votruba}, {Voutsinas}, {Walmsley}, {Weiler}, {Wertz}, {Wevers}, {Wyrzykowski}, {Yoldas}, {{\v{Z}}erjal}, {Ziaeepour}, {Zorec}, {Zschocke}, {Zucker}, {Zurbach}, \& {Zwitter}}]{Gaia2018A&A}
{Gaia Collaboration}, {Brown}, A.~G.~A., {Vallenari}, A., {et~al.} 2018, \aap, 616, A1, \dodoi{10.1051/0004-6361/201833051}

\bibitem[{{Gaia Collaboration} {et~al.}(2023){Gaia Collaboration}, {Vallenari}, {Brown}, {Prusti}, {de Bruijne}, {Arenou}, {Babusiaux}, {Biermann}, {Creevey}, {Ducourant}, {Evans}, {Eyer}, {Guerra}, {Hutton}, {Jordi}, {Klioner}, {Lammers}, {Lindegren}, {Luri}, {Mignard}, {Panem}, {Pourbaix}, {Randich}, {Sartoretti}, {Soubiran}, {Tanga}, {Walton}, {Bailer-Jones}, {Bastian}, {Drimmel}, {Jansen}, {Katz}, {Lattanzi}, {van Leeuwen}, {Bakker}, {Cacciari}, {Casta{\~n}eda}, {De Angeli}, {Fabricius}, {Fouesneau}, {Fr{\'e}mat}, {Galluccio}, {Guerrier}, {Heiter}, {Masana}, {Messineo}, {Mowlavi}, {Nicolas}, {Nienartowicz}, {Pailler}, {Panuzzo}, {Riclet}, {Roux}, {Seabroke}, {Sordo}, {Th{\'e}venin}, {Gracia-Abril}, {Portell}, {Teyssier}, {Altmann}, {Andrae}, {Audard}, {Bellas-Velidis}, {Benson}, {Berthier}, {Blomme}, {Burgess}, {Busonero}, {Busso}, {C{\'a}novas}, {Carry}, {Cellino}, {Cheek}, {Clementini}, {Damerdji}, {Davidson}, {de Teodoro}, {Nu{\~n}ez Campos}, {Delchambre}, {Dell'Oro}, {Esquej},
  {Fern{\'a}ndez-Hern{\'a}ndez}, {Fraile}, {Garabato}, {Garc{\'\i}a-Lario}, {Gosset}, {Haigron}, {Halbwachs}, {Hambly}, {Harrison}, {Hern{\'a}ndez}, {Hestroffer}, {Hodgkin}, {Holl}, {Jan{\ss}en}, {Jevardat de Fombelle}, {Jordan}, {Krone-Martins}, {Lanzafame}, {L{\"o}ffler}, {Marchal}, {Marrese}, {Moitinho}, {Muinonen}, {Osborne}, {Pancino}, {Pauwels}, {Recio-Blanco}, {Reyl{\'e}}, {Riello}, {Rimoldini}, {Roegiers}, {Rybizki}, {Sarro}, {Siopis}, {Smith}, {Sozzetti}, {Utrilla}, {van Leeuwen}, {Abbas}, {{\'A}brah{\'a}m}, {Abreu Aramburu}, {Aerts}, {Aguado}, {Ajaj}, {Aldea-Montero}, {Altavilla}, {{\'A}lvarez}, {Alves}, {Anders}, {Anderson}, {Anglada Varela}, {Antoja}, {Baines}, {Baker}, {Balaguer-N{\'u}{\~n}ez}, {Balbinot}, {Balog}, {Barache}, {Barbato}, {Barros}, {Barstow}, {Bartolom{\'e}}, {Bassilana}, {Bauchet}, {Becciani}, {Bellazzini}, {Berihuete}, {Bernet}, {Bertone}, {Bianchi}, {Binnenfeld}, {Blanco-Cuaresma}, {Blazere}, {Boch}, {Bombrun}, {Bossini}, {Bouquillon}, {Bragaglia}, {Bramante}, {Breedt},
  {Bressan}, {Brouillet}, {Brugaletta}, {Bucciarelli}, {Burlacu}, {Butkevich}, {Buzzi}, {Caffau}, {Cancelliere}, {Cantat-Gaudin}, {Carballo}, {Carlucci}, {Carnerero}, {Carrasco}, {Casamiquela}, {Castellani}, {Castro-Ginard}, {Chaoul}, {Charlot}, {Chemin}, {Chiaramida}, {Chiavassa}, {Chornay}, {Comoretto}, {Contursi}, {Cooper}, {Cornez}, {Cowell}, {Crifo}, {Cropper}, {Crosta}, {Crowley}, {Dafonte}, {Dapergolas}, {David}, {David}, {de Laverny}, {De Luise}, \& {De March}}]{Gaia2023A&A}
{Gaia Collaboration}, {Vallenari}, A., {Brown}, A.~G.~A., {et~al.} 2023, \aap, 674, A1, \dodoi{10.1051/0004-6361/202243940}

\bibitem[{{Gao} {et~al.}(2020){Gao}, {Lind}, {Amarsi}, {Buder}, {Bland-Hawthorn}, {Campbell}, {Asplund}, {Casey}, {de Silva}, {Freeman}, {Hayden}, {Lewis}, {Martell}, {Simpson}, {Sharma}, {Zucker}, {Zwitter}, {Horner}, {Munari}, {Nordlander}, {Stello}, {Ting}, {Traven}, {Wittenmyer}, \& {GALAH Collaboration}}]{Gao2020MNRAS}
{Gao}, X., {Lind}, K., {Amarsi}, A.~M., {et~al.} 2020, \mnras, 497, L30, \dodoi{10.1093/mnrasl/slaa109}

\bibitem[{{Garc{\'\i}a P{\'e}rez} {et~al.}(2016){Garc{\'\i}a P{\'e}rez}, {Allende Prieto}, {Holtzman}, {Shetrone}, {M{\'e}sz{\'a}ros}, {Bizyaev}, {Carrera}, {Cunha}, {Garc{\'\i}a-Hern{\'a}ndez}, {Johnson}, {Majewski}, {Nidever}, {Schiavon}, {Shane}, {Smith}, {Sobeck}, {Troup}, {Zamora}, {Weinberg}, {Bovy}, {Eisenstein}, {Feuillet}, {Frinchaboy}, {Hayden}, {Hearty}, {Nguyen}, {O'Connell}, {Pinsonneault}, {Wilson}, \& {Zasowski}}]{GarciaPerez2016AJ}
{Garc{\'\i}a P{\'e}rez}, A.~E., {Allende Prieto}, C., {Holtzman}, J.~A., {et~al.} 2016, \aj, 151, 144, \dodoi{10.3847/0004-6256/151/6/144}

\bibitem[{{Gilliland} {et~al.}(2010){Gilliland}, {Brown}, {Christensen-Dalsgaard}, {Kjeldsen}, {Aerts}, {Appourchaux}, {Basu}, {Bedding}, {Chaplin}, {Cunha}, {De Cat}, {De Ridder}, {Guzik}, {Handler}, {Kawaler}, {Kiss}, {Kolenberg}, {Kurtz}, {Metcalfe}, {Monteiro}, {Szab{\'o}}, {Arentoft}, {Balona}, {Debosscher}, {Elsworth}, {Quirion}, {Stello}, {Su{\'a}rez}, {Borucki}, {Jenkins}, {Koch}, {Kondo}, {Latham}, {Rowe}, \& {Steffen}}]{Gilliland2010PASP}
{Gilliland}, R.~L., {Brown}, T.~M., {Christensen-Dalsgaard}, J., {et~al.} 2010, \pasp, 122, 131, \dodoi{10.1086/650399}

\bibitem[{{Grevesse} {et~al.}(2007){Grevesse}, {Asplund}, \& {Sauval}}]{Grevesse2007SSRv}
{Grevesse}, N., {Asplund}, M., \& {Sauval}, A.~J. 2007, \ssr, 130, 105, \dodoi{10.1007/s11214-007-9173-7}

\bibitem[{{Grevesse} \& {Noels}(1993)}]{Grevesse1993oee}
{Grevesse}, N., \& {Noels}, A. 1993, in Origin and Evolution of the Elements, ed. N.~{Prantzos}, E.~{Vangioni-Flam}, \& M.~{Casse}, 15--25

\bibitem[{{Grevesse} \& {Sauval}(1998)}]{Grevesse1998SSRv}
{Grevesse}, N., \& {Sauval}, A.~J. 1998, \ssr, 85, 161, \dodoi{10.1023/A:1005161325181}

\bibitem[{{Harris} {et~al.}(2020){Harris}, {Millman}, {van der Walt}, {Gommers}, {Virtanen}, {Cournapeau}, {Wieser}, {Taylor}, {Berg}, {Smith}, {Kern}, {Picus}, {Hoyer}, {van Kerkwijk}, {Brett}, {Haldane}, {del R{\'\i}o}, {Wiebe}, {Peterson}, {G{\'e}rard-Marchant}, {Sheppard}, {Reddy}, {Weckesser}, {Abbasi}, {Gohlke}, \& {Oliphant}}]{Harris2020Natur}
{Harris}, C.~R., {Millman}, K.~J., {van der Walt}, S.~J., {et~al.} 2020, \nat, 585, 357, \dodoi{10.1038/s41586-020-2649-2}

\bibitem[{{Hauschildt} {et~al.}(1999{\natexlab{a}}){Hauschildt}, {Allard}, \& {Baron}}]{Hauschildt1999ApJa}
{Hauschildt}, P.~H., {Allard}, F., \& {Baron}, E. 1999{\natexlab{a}}, \apj, 512, 377, \dodoi{10.1086/306745}

\bibitem[{{Hauschildt} {et~al.}(1999{\natexlab{b}}){Hauschildt}, {Allard}, {Ferguson}, {Baron}, \& {Alexander}}]{Hauschildt1999ApJb}
{Hauschildt}, P.~H., {Allard}, F., {Ferguson}, J., {Baron}, E., \& {Alexander}, D.~R. 1999{\natexlab{b}}, \apj, 525, 871, \dodoi{10.1086/307954}

\bibitem[{{Hawkins} {et~al.}(2016){Hawkins}, {Masseron}, {Jofr{\'e}}, {Gilmore}, {Elsworth}, \& {Hekker}}]{Hawkins2016A&A}
{Hawkins}, K., {Masseron}, T., {Jofr{\'e}}, P., {et~al.} 2016, \aap, 594, A43, \dodoi{10.1051/0004-6361/201628812}

\bibitem[{{Hayden} {et~al.}(2014){Hayden}, {Holtzman}, {Bovy}, {Majewski}, {Johnson}, {Allende Prieto}, {Beers}, {Cunha}, {Frinchaboy}, {Garc{\'\i}a P{\'e}rez}, {Girardi}, {Hearty}, {Lee}, {Nidever}, {Schiavon}, {Schlesinger}, {Schneider}, {Schultheis}, {Shetrone}, {Smith}, {Zasowski}, {Bizyaev}, {Feuillet}, {Hasselquist}, {Kinemuchi}, {Malanushenko}, {Malanushenko}, {O'Connell}, {Pan}, \& {Stassun}}]{Hayden2014AJ}
{Hayden}, M.~R., {Holtzman}, J.~A., {Bovy}, J., {et~al.} 2014, \aj, 147, 116, \dodoi{10.1088/0004-6256/147/5/116}

\bibitem[{{Hayden} {et~al.}(2015){Hayden}, {Bovy}, {Holtzman}, {Nidever}, {Bird}, {Weinberg}, {Andrews}, {Majewski}, {Allende Prieto}, {Anders}, {Beers}, {Bizyaev}, {Chiappini}, {Cunha}, {Frinchaboy}, {Garc{\'\i}a-Her{\'n}andez}, {Garc{\'\i}a P{\'e}rez}, {Girardi}, {Harding}, {Hearty}, {Johnson}, {M{\'e}sz{\'a}ros}, {Minchev}, {O'Connell}, {Pan}, {Robin}, {Schiavon}, {Schneider}, {Schultheis}, {Shetrone}, {Skrutskie}, {Steinmetz}, {Smith}, {Wilson}, {Zamora}, \& {Zasowski}}]{Hayden2015ApJ}
{Hayden}, M.~R., {Bovy}, J., {Holtzman}, J.~A., {et~al.} 2015, \apj, 808, 132, \dodoi{10.1088/0004-637X/808/2/132}

\bibitem[{{Heged{\H{u}}s} {et~al.}(2023){Heged{\H{u}}s}, {M{\'e}sz{\'a}ros}, {Jofr{\'e}}, {Stringfellow}, {Feuillet}, {Garc{\'\i}a-Hern{\'a}ndez}, {Nitschelm}, \& {Zamora}}]{Hegedus2023A&A}
{Heged{\H{u}}s}, V., {M{\'e}sz{\'a}ros}, S., {Jofr{\'e}}, P., {et~al.} 2023, \aap, 670, A107, \dodoi{10.1051/0004-6361/202244813}

\bibitem[{{Hekker} {et~al.}(2020){Hekker}, {Angelou}, {Elsworth}, \& {Basu}}]{Hekker2020MNRAS}
{Hekker}, S., {Angelou}, G.~C., {Elsworth}, Y., \& {Basu}, S. 2020, \mnras, 492, 5940, \dodoi{10.1093/mnras/staa176}

\bibitem[{{Henyey} {et~al.}(1965){Henyey}, {Vardya}, \& {Bodenheimer}}]{Henyey1965ApJ}
{Henyey}, L., {Vardya}, M.~S., \& {Bodenheimer}, P. 1965, \apj, 142, 841, \dodoi{10.1086/148357}

\bibitem[{{Holweger}(2001)}]{Holweger2001AIPC}
{Holweger}, H. 2001, in American Institute of Physics Conference Series, Vol. 598, Joint SOHO/ACE workshop ``Solar and Galactic Composition'', ed. R.~F. {Wimmer-Schweingruber}, 23--30, \dodoi{10.1063/1.1433974}

\bibitem[{{Howell} {et~al.}(2014){Howell}, {Sobeck}, {Haas}, {Still}, {Barclay}, {Mullally}, {Troeltzsch}, {Aigrain}, {Bryson}, {Caldwell}, {Chaplin}, {Cochran}, {Huber}, {Marcy}, {Miglio}, {Najita}, {Smith}, {Twicken}, \& {Fortney}}]{Howell2014PASP}
{Howell}, S.~B., {Sobeck}, C., {Haas}, M., {et~al.} 2014, \pasp, 126, 398, \dodoi{10.1086/676406}

\bibitem[{{Huber} {et~al.}(2023){Huber}, {Pinsonneault}, {Beck}, {Bedding}, {Bland-Hawthorn}, {Breton}, {Bugnet}, {Chaplin}, {Garcia}, {Grunblatt}, {Guzik}, {Hekker}, {Kawaler}, {Mathis}, {Mathur}, {Metcalfe}, {Mosser}, {Ness}, {Piro}, {Serenelli}, {Sharma}, {Soderblom}, {Stassun}, {Stello}, {Tayar}, {van Belle}, \& {Zinn}}]{Huber2023arXiv}
{Huber}, D., {Pinsonneault}, M., {Beck}, P., {et~al.} 2023, arXiv e-prints, arXiv:2307.03237, \dodoi{10.48550/arXiv.2307.03237}

\bibitem[{{Hunter}(2007)}]{Hunter2007CSE}
{Hunter}, J.~D. 2007, Computing in Science and Engineering, 9, 90, \dodoi{10.1109/MCSE.2007.55}

\bibitem[{{Iben}(1965)}]{Iben1965ApJb}
{Iben}, Icko, J. 1965, \apj, 142, 1447, \dodoi{10.1086/148429}

\bibitem[{{Iglesias} \& {Rogers}(1993)}]{Iglesias1993ApJ}
{Iglesias}, C.~A., \& {Rogers}, F.~J. 1993, \apj, 412, 752, \dodoi{10.1086/172958}

\bibitem[{{Iglesias} \& {Rogers}(1996)}]{Iglesias1996ApJ}
---. 1996, \apj, 464, 943, \dodoi{10.1086/177381}

\bibitem[{{Irwin}(2004)}]{Irwin2004}
{Irwin}, A.~W. 2004, The FreeEOS Code for Calculating the Equation of State for Stellar Interiors.
\newblock \url{http://freeeos.sourceforge.net/}

\bibitem[{{Itoh} {et~al.}(1996){Itoh}, {Hayashi}, {Nishikawa}, \& {Kohyama}}]{Itoh1996}
{Itoh}, N., {Hayashi}, H., {Nishikawa}, A., \& {Kohyama}, Y. 1996, \apjs, 102, 411, \dodoi{10.1086/192264}

\bibitem[{{Jermyn} {et~al.}(2021){Jermyn}, {Schwab}, {Bauer}, {Timmes}, \& {Potekhin}}]{Jermyn2021ApJ}
{Jermyn}, A.~S., {Schwab}, J., {Bauer}, E., {Timmes}, F.~X., \& {Potekhin}, A.~Y. 2021, \apj, 913, 72, \dodoi{10.3847/1538-4357/abf48e}

\bibitem[{{Jermyn} {et~al.}(2023){Jermyn}, {Bauer}, {Schwab}, {Farmer}, {Ball}, {Bellinger}, {Dotter}, {Joyce}, {Marchant}, {Mombarg}, {Wolf}, {Sunny Wong}, {Cinquegrana}, {Farrell}, {Smolec}, {Thoul}, {Cantiello}, {Herwig}, {Toloza}, {Bildsten}, {Townsend}, \& {Timmes}}]{Jermyn2023ApJS}
{Jermyn}, A.~S., {Bauer}, E.~B., {Schwab}, J., {et~al.} 2023, \apjs, 265, 15, \dodoi{10.3847/1538-4365/acae8d}

\bibitem[{{Karakas} \& {Lattanzio}(2014)}]{Karakas2014PASA}
{Karakas}, A.~I., \& {Lattanzio}, J.~C. 2014, \pasa, 31, e030, \dodoi{10.1017/pasa.2014.21}

\bibitem[{{Khan} {et~al.}(2018){Khan}, {Hall}, {Miglio}, {Davies}, {Mosser}, {Girardi}, \& {Montalb{\'a}n}}]{Khan2018ApJ}
{Khan}, S., {Hall}, O.~J., {Miglio}, A., {et~al.} 2018, \apj, 859, 156, \dodoi{10.3847/1538-4357/aabf90}

\bibitem[{{Kjeldsen} {et~al.}(2010){Kjeldsen}, {Christensen-Dalsgaard}, {Handberg}, {Brown}, {Gilliland}, {Borucki}, \& {Koch}}]{Kjeldsen2010AN}
{Kjeldsen}, H., {Christensen-Dalsgaard}, J., {Handberg}, R., {et~al.} 2010, Astronomische Nachrichten, 331, 966, \dodoi{10.1002/asna.201011437}

\bibitem[{{Koch} {et~al.}(2010){Koch}, {Borucki}, {Basri}, {Batalha}, {Brown}, {Caldwell}, {Christensen-Dalsgaard}, {Cochran}, {DeVore}, {Dunham}, {Gautier}, {Geary}, {Gilliland}, {Gould}, {Jenkins}, {Kondo}, {Latham}, {Lissauer}, {Marcy}, {Monet}, {Sasselov}, {Boss}, {Brownlee}, {Caldwell}, {Dupree}, {Howell}, {Kjeldsen}, {Meibom}, {Morrison}, {Owen}, {Reitsema}, {Tarter}, {Bryson}, {Dotson}, {Gazis}, {Haas}, {Kolodziejczak}, {Rowe}, {Van Cleve}, {Allen}, {Chandrasekaran}, {Clarke}, {Li}, {Quintana}, {Tenenbaum}, {Twicken}, \& {Wu}}]{Koch2010ApJ}
{Koch}, D.~G., {Borucki}, W.~J., {Basri}, G., {et~al.} 2010, \apjl, 713, L79, \dodoi{10.1088/2041-8205/713/2/L79}

\bibitem[{{Kuhfuss}(1986)}]{Kuhfuss1986A&A}
{Kuhfuss}, R. 1986, \aap, 160, 116

\bibitem[{{Langanke} \& {Mart{\'\i}nez-Pinedo}(2000)}]{Langanke2000NuPhA}
{Langanke}, K., \& {Mart{\'\i}nez-Pinedo}, G. 2000, \nphysa, 673, 481, \dodoi{10.1016/S0375-9474(00)00131-7}

\bibitem[{{Lattanzio} {et~al.}(2015){Lattanzio}, {Siess}, {Church}, {Angelou}, {Stancliffe}, {Doherty}, {Stephen}, \& {Campbell}}]{Lattanzio2015MNRAS}
{Lattanzio}, J.~C., {Siess}, L., {Church}, R.~P., {et~al.} 2015, \mnras, 446, 2673, \dodoi{10.1093/mnras/stu2238}

\bibitem[{{Liagre} {et~al.}(2025){Liagre}, {Garc{\'\i}a}, {Mathur}, {Pinsonneault}, {Serenelli}, {Zinn}, {Cao}, {Godoy-Rivera}, {Tayar}, {Beck}, {Grossmann}, \& {Palakkatharappil}}]{Liagre2025A&A}
{Liagre}, B., {Garc{\'\i}a}, R.~A., {Mathur}, S., {et~al.} 2025, \aap, 702, A144, \dodoi{10.1051/0004-6361/202555167}

\bibitem[{{Lodders}(2003)}]{Lodders2003ApJ}
{Lodders}, K. 2003, \apj, 591, 1220, \dodoi{10.1086/375492}

\bibitem[{{Lodders} {et~al.}(2009){Lodders}, {Palme}, \& {Gail}}]{Lodders2009LanB}
{Lodders}, K., {Palme}, H., \& {Gail}, H.~P. 2009, Landolt B\&ouml;rnstein, 4B, 712, \dodoi{10.1007/978-3-540-88055-4_34}

\bibitem[{{Magg} {et~al.}(2022){Magg}, {Bergemann}, {Serenelli}, {Bautista}, {Plez}, {Heiter}, {Gerber}, {Ludwig}, {Basu}, {Ferguson}, {Gallego}, {Gamrath}, {Palmeri}, \& {Quinet}}]{Magg2022A&A}
{Magg}, E., {Bergemann}, M., {Serenelli}, A., {et~al.} 2022, \aap, 661, A140, \dodoi{10.1051/0004-6361/202142971}

\bibitem[{{Majewski} {et~al.}(2010){Majewski}, {Wilson}, {Hearty}, {Schiavon}, \& {Skrutskie}}]{Majewski2010IAUS}
{Majewski}, S.~R., {Wilson}, J.~C., {Hearty}, F., {Schiavon}, R.~R., \& {Skrutskie}, M.~F. 2010, in Chemical Abundances in the Universe: Connecting First Stars to Planets, ed. K.~{Cunha}, M.~{Spite}, \& B.~{Barbuy}, Vol. 265, 480--481, \dodoi{10.1017/S1743921310001298}

\bibitem[{{Majewski} {et~al.}(2017){Majewski}, {Schiavon}, {Frinchaboy}, {Allende Prieto}, {Barkhouser}, {Bizyaev}, {Blank}, {Brunner}, {Burton}, {Carrera}, {Chojnowski}, {Cunha}, {Epstein}, {Fitzgerald}, {Garc{\'\i}a P{\'e}rez}, {Hearty}, {Henderson}, {Holtzman}, {Johnson}, {Lam}, {Lawler}, {Maseman}, {M{\'e}sz{\'a}ros}, {Nelson}, {Nguyen}, {Nidever}, {Pinsonneault}, {Shetrone}, {Smee}, {Smith}, {Stolberg}, {Skrutskie}, {Walker}, {Wilson}, {Zasowski}, {Anders}, {Basu}, {Beland}, {Blanton}, {Bovy}, {Brownstein}, {Carlberg}, {Chaplin}, {Chiappini}, {Eisenstein}, {Elsworth}, {Feuillet}, {Fleming}, {Galbraith-Frew}, {Garc{\'\i}a}, {Garc{\'\i}a-Hern{\'a}ndez}, {Gillespie}, {Girardi}, {Gunn}, {Hasselquist}, {Hayden}, {Hekker}, {Ivans}, {Kinemuchi}, {Klaene}, {Mahadevan}, {Mathur}, {Mosser}, {Muna}, {Munn}, {Nichol}, {O'Connell}, {Parejko}, {Robin}, {Rocha-Pinto}, {Schultheis}, {Serenelli}, {Shane}, {Silva Aguirre}, {Sobeck}, {Thompson}, {Troup}, {Weinberg}, \& {Zamora}}]{Majewski2017AJ}
{Majewski}, S.~R., {Schiavon}, R.~P., {Frinchaboy}, P.~M., {et~al.} 2017, \aj, 154, 94, \dodoi{10.3847/1538-3881/aa784d}

\bibitem[{{Mamajek} {et~al.}(2015){Mamajek}, {Prsa}, {Torres}, {Harmanec}, {Asplund}, {Bennett}, {Capitaine}, {Christensen-Dalsgaard}, {Depagne}, {Folkner}, {Haberreiter}, {Hekker}, {Hilton}, {Kostov}, {Kurtz}, {Laskar}, {Mason}, {Milone}, {Montgomery}, {Richards}, {Schou}, \& {Stewart}}]{Mamajek2015arXiv}
{Mamajek}, E.~E., {Prsa}, A., {Torres}, G., {et~al.} 2015, arXiv e-prints, arXiv:1510.07674, \dodoi{10.48550/arXiv.1510.07674}

\bibitem[{{Marigo} \& {Aringer}(2009)}]{Marigo2009A&A}
{Marigo}, P., \& {Aringer}, B. 2009, \aap, 508, 1539, \dodoi{10.1051/0004-6361/200912598}

\bibitem[{{Marigo} {et~al.}(2022){Marigo}, {Aringer}, {Girardi}, \& {Bressan}}]{Marigo2022ApJ}
{Marigo}, P., {Aringer}, B., {Girardi}, L., \& {Bressan}, A. 2022, \apj, 940, 129, \dodoi{10.3847/1538-4357/ac9b40}

\bibitem[{{Martell} {et~al.}(2021){Martell}, {Simpson}, {Balasubramaniam}, {Buder}, {Sharma}, {Hon}, {Stello}, {Ting}, {Asplund}, {Bland-Hawthorn}, {De Silva}, {Freeman}, {Hayden}, {Kos}, {Lewis}, {Lind}, {Zucker}, {Zwitter}, {Campbell}, {{\v{C}}otar}, {Horner}, {Montet}, \& {Wittenmyer}}]{Martell2021MNRAS}
{Martell}, S.~L., {Simpson}, J.~D., {Balasubramaniam}, A.~G., {et~al.} 2021, \mnras, 505, 5340, \dodoi{10.1093/mnras/stab1356}

\bibitem[{{Martig} {et~al.}(2015){Martig}, {Rix}, {Silva Aguirre}, {Hekker}, {Mosser}, {Elsworth}, {Bovy}, {Stello}, {Anders}, {Garc{\'\i}a}, {Tayar}, {Rodrigues}, {Basu}, {Carrera}, {Ceillier}, {Chaplin}, {Chiappini}, {Frinchaboy}, {Garc{\'\i}a-Hern{\'a}ndez}, {Hearty}, {Holtzman}, {Johnson}, {Majewski}, {Mathur}, {M{\'e}sz{\'a}ros}, {Miglio}, {Nidever}, {Pan}, {Pinsonneault}, {Schiavon}, {Schneider}, {Serenelli}, {Shetrone}, \& {Zamora}}]{Martig2015MNRAS}
{Martig}, M., {Rix}, H.-W., {Silva Aguirre}, V., {et~al.} 2015, \mnras, 451, 2230, \dodoi{10.1093/mnras/stv1071}

\bibitem[{{Martig} {et~al.}(2016){Martig}, {Fouesneau}, {Rix}, {Ness}, {M{\'e}sz{\'a}ros}, {Garc{\'\i}a-Hern{\'a}ndez}, {Pinsonneault}, {Serenelli}, {Silva Aguirre}, \& {Zamora}}]{Martig2016MNRAS}
{Martig}, M., {Fouesneau}, M., {Rix}, H.-W., {et~al.} 2016, \mnras, 456, 3655, \dodoi{10.1093/mnras/stv2830}

\bibitem[{{Miglio} {et~al.}(2021){Miglio}, {Chiappini}, {Mackereth}, {Davies}, {Brogaard}, {Casagrande}, {Chaplin}, {Girardi}, {Kawata}, {Khan}, {Izzard}, {Montalb{\'a}n}, {Mosser}, {Vincenzo}, {Bossini}, {Noels}, {Rodrigues}, {Valentini}, \& {Mandel}}]{Miglio2021A&A}
{Miglio}, A., {Chiappini}, C., {Mackereth}, J.~T., {et~al.} 2021, \aap, 645, A85, \dodoi{10.1051/0004-6361/202038307}

\bibitem[{{Nataf} {et~al.}(2013){Nataf}, {Gould}, {Pinsonneault}, \& {Udalski}}]{Nataf2013ApJ}
{Nataf}, D.~M., {Gould}, A.~P., {Pinsonneault}, M.~H., \& {Udalski}, A. 2013, \apj, 766, 77, \dodoi{10.1088/0004-637X/766/2/77}

\bibitem[{{Ness} {et~al.}(2016){Ness}, {Hogg}, {Rix}, {Martig}, {Pinsonneault}, \& {Ho}}]{Ness2016ApJ}
{Ness}, M., {Hogg}, D.~W., {Rix}, H.~W., {et~al.} 2016, \apj, 823, 114, \dodoi{10.3847/0004-637X/823/2/114}

\bibitem[{{Nguyen} {et~al.}(2025){Nguyen}, {Bressan}, {Korn}, {Cescutti}, {Costa}, {Addari}, {Girardi}, {Fu}, {Chen}, \& {Marigo}}]{Nguyen2025A&A}
{Nguyen}, C.~T., {Bressan}, A., {Korn}, A.~J., {et~al.} 2025, \aap, 696, A136, \dodoi{10.1051/0004-6361/202451847}

\bibitem[{{Oda} {et~al.}(1994){Oda}, {Hino}, {Muto}, {Takahara}, \& {Sato}}]{Oda1994ADNDT}
{Oda}, T., {Hino}, M., {Muto}, K., {Takahara}, M., \& {Sato}, K. 1994, Atomic Data and Nuclear Data Tables, 56, 231, \dodoi{10.1006/adnd.1994.1007}

\bibitem[{{Paxton} {et~al.}(2011){Paxton}, {Bildsten}, {Dotter}, {Herwig}, {Lesaffre}, \& {Timmes}}]{Paxton2011}
{Paxton}, B., {Bildsten}, L., {Dotter}, A., {et~al.} 2011, \apjs, 192, 3, \dodoi{10.1088/0067-0049/192/1/3}

\bibitem[{{Paxton} {et~al.}(2013){Paxton}, {Cantiello}, {Arras}, {Bildsten}, {Brown}, {Dotter}, {Mankovich}, {Montgomery}, {Stello}, {Timmes}, \& {Townsend}}]{Paxton2013}
{Paxton}, B., {Cantiello}, M., {Arras}, P., {et~al.} 2013, \apjs, 208, 4, \dodoi{10.1088/0067-0049/208/1/4}

\bibitem[{{Paxton} {et~al.}(2015){Paxton}, {Marchant}, {Schwab}, {Bauer}, {Bildsten}, {Cantiello}, {Dessart}, {Farmer}, {Hu}, {Langer}, {Townsend}, {Townsley}, \& {Timmes}}]{Paxton2015}
{Paxton}, B., {Marchant}, P., {Schwab}, J., {et~al.} 2015, \apjs, 220, 15, \dodoi{10.1088/0067-0049/220/1/15}

\bibitem[{{Paxton} {et~al.}(2018){Paxton}, {Schwab}, {Bauer}, {Bildsten}, {Blinnikov}, {Duffell}, {Farmer}, {Goldberg}, {Marchant}, {Sorokina}, {Thoul}, {Townsend}, \& {Timmes}}]{Paxton2018}
{Paxton}, B., {Schwab}, J., {Bauer}, E.~B., {et~al.} 2018, \apjs, 234, 34, \dodoi{10.3847/1538-4365/aaa5a8}

\bibitem[{{Paxton} {et~al.}(2019){Paxton}, {Smolec}, {Schwab}, {Gautschy}, {Bildsten}, {Cantiello}, {Dotter}, {Farmer}, {Goldberg}, {Jermyn}, {Kanbur}, {Marchant}, {Thoul}, {Townsend}, {Wolf}, {Zhang}, \& {Timmes}}]{Paxton2019}
{Paxton}, B., {Smolec}, R., {Schwab}, J., {et~al.} 2019, \apjs, 243, 10, \dodoi{10.3847/1538-4365/ab2241}

\bibitem[{{Pietrinferni} {et~al.}(2004){Pietrinferni}, {Cassisi}, {Salaris}, \& {Castelli}}]{Pietrinferni2004ApJ}
{Pietrinferni}, A., {Cassisi}, S., {Salaris}, M., \& {Castelli}, F. 2004, \apj, 612, 168, \dodoi{10.1086/422498}

\bibitem[{{Pinsonneault}(1997)}]{Pinsonneault1997ARA&A}
{Pinsonneault}, M. 1997, \araa, 35, 557, \dodoi{10.1146/annurev.astro.35.1.557}

\bibitem[{{Pinsonneault} {et~al.}(1989){Pinsonneault}, {Kawaler}, {Sofia}, \& {Demarque}}]{Pinsonneault1989ApJ}
{Pinsonneault}, M.~H., {Kawaler}, S.~D., {Sofia}, S., \& {Demarque}, P. 1989, \apj, 338, 424, \dodoi{10.1086/167210}

\bibitem[{{Pinsonneault} {et~al.}(2014){Pinsonneault}, {Elsworth}, {Epstein}, {Hekker}, {M{\'e}sz{\'a}ros}, {Chaplin}, {Johnson}, {Garc{\'\i}a}, {Holtzman}, {Mathur}, {Garc{\'\i}a P{\'e}rez}, {Silva Aguirre}, {Girardi}, {Basu}, {Shetrone}, {Stello}, {Allende Prieto}, {An}, {Beck}, {Beers}, {Bizyaev}, {Bloemen}, {Bovy}, {Cunha}, {De Ridder}, {Frinchaboy}, {Garc{\'\i}a-Hern{\'a}ndez}, {Gilliland}, {Harding}, {Hearty}, {Huber}, {Ivans}, {Kallinger}, {Majewski}, {Metcalfe}, {Miglio}, {Mosser}, {Muna}, {Nidever}, {Schneider}, {Serenelli}, {Smith}, {Tayar}, {Zamora}, \& {Zasowski}}]{Pinsonneault2014ApJS}
{Pinsonneault}, M.~H., {Elsworth}, Y., {Epstein}, C., {et~al.} 2014, \apjs, 215, 19, \dodoi{10.1088/0067-0049/215/2/19}

\bibitem[{{Pinsonneault} {et~al.}(2018){Pinsonneault}, {Elsworth}, {Tayar}, {Serenelli}, {Stello}, {Zinn}, {Mathur}, {Garc{\'\i}a}, {Johnson}, {Hekker}, {Huber}, {Kallinger}, {M{\'e}sz{\'a}ros}, {Mosser}, {Stassun}, {Girardi}, {Rodrigues}, {Silva Aguirre}, {An}, {Basu}, {Chaplin}, {Corsaro}, {Cunha}, {Garc{\'\i}a-Hern{\'a}ndez}, {Holtzman}, {J{\"o}nsson}, {Shetrone}, {Smith}, {Sobeck}, {Stringfellow}, {Zamora}, {Beers}, {Fern{\'a}ndez-Trincado}, {Frinchaboy}, {Hearty}, \& {Nitschelm}}]{Pinsonneault2018ApJS}
{Pinsonneault}, M.~H., {Elsworth}, Y.~P., {Tayar}, J., {et~al.} 2018, \apjs, 239, 32, \dodoi{10.3847/1538-4365/aaebfd}

\bibitem[{{Pinsonneault} {et~al.}(2025){Pinsonneault}, {Zinn}, {Tayar}, {Serenelli}, {Garc{\'\i}a}, {Mathur}, {Vrard}, {Elsworth}, {Mosser}, {Stello}, {Bell}, {Bugnet}, {Corsaro}, {Gaulme}, {Hekker}, {Hon}, {Huber}, {Kallinger}, {Cao}, {Johnson}, {Liagre}, {Patton}, {Santos}, {Basu}, {Beck}, {Beers}, {Chaplin}, {Cunha}, {Frinchaboy}, {Girardi}, {Godoy-Rivera}, {Holtzman}, {J{\"o}nsson}, {M{\'e}sz{\'a}ros}, {Reyes}, {Rix}, {Shetrone}, {Smith}, {Spoo}, {Stassun}, \& {Wang}}]{Pinsonneault2025ApJS}
{Pinsonneault}, M.~H., {Zinn}, J.~C., {Tayar}, J., {et~al.} 2025, \apjs, 276, 69, \dodoi{10.3847/1538-4365/ad9fef}

\bibitem[{{Potekhin} \& {Chabrier}(2010)}]{Potekhin2010CoPP}
{Potekhin}, A.~Y., \& {Chabrier}, G. 2010, Contributions to Plasma Physics, 50, 82, \dodoi{10.1002/ctpp.201010017}

\bibitem[{{Rauer} {et~al.}(2014){Rauer}, {Catala}, {Aerts}, {Appourchaux}, {Benz}, {Brandeker}, {Christensen-Dalsgaard}, {Deleuil}, {Gizon}, {Goupil}, {G{\"u}del}, {Janot-Pacheco}, {Mas-Hesse}, {Pagano}, {Piotto}, {Pollacco}, {Santos}, {Smith}, {Su{\'a}rez}, {Szab{\'o}}, {Udry}, {Adibekyan}, {Alibert}, {Almenara}, {Amaro-Seoane}, {Eiff}, {Asplund}, {Antonello}, {Barnes}, {Baudin}, {Belkacem}, {Bergemann}, {Bihain}, {Birch}, {Bonfils}, {Boisse}, {Bonomo}, {Borsa}, {Brand{\~a}o}, {Brocato}, {Brun}, {Burleigh}, {Burston}, {Cabrera}, {Cassisi}, {Chaplin}, {Charpinet}, {Chiappini}, {Church}, {Csizmadia}, {Cunha}, {Damasso}, {Davies}, {Deeg}, {D{\'\i}az}, {Dreizler}, {Dreyer}, {Eggenberger}, {Ehrenreich}, {Eigm{\"u}ller}, {Erikson}, {Farmer}, {Feltzing}, {de Oliveira Fialho}, {Figueira}, {Forveille}, {Fridlund}, {Garc{\'\i}a}, {Giommi}, {Giuffrida}, {Godolt}, {Gomes da Silva}, {Granzer}, {Grenfell}, {Grotsch-Noels}, {G{\"u}nther}, {Haswell}, {Hatzes}, {H{\'e}brard}, {Hekker}, {Helled}, {Heng}, {Jenkins},
  {Johansen}, {Khodachenko}, {Kislyakova}, {Kley}, {Kolb}, {Krivova}, {Kupka}, {Lammer}, {Lanza}, {Lebreton}, {Magrin}, {Marcos-Arenal}, {Marrese}, {Marques}, {Martins}, {Mathis}, {Mathur}, {Messina}, {Miglio}, {Montalban}, {Montalto}, {Monteiro}, {Moradi}, {Moravveji}, {Mordasini}, {Morel}, {Mortier}, {Nascimbeni}, {Nelson}, {Nielsen}, {Noack}, {Norton}, {Ofir}, {Oshagh}, {Ouazzani}, {P{\'a}pics}, {Parro}, {Petit}, {Plez}, {Poretti}, {Quirrenbach}, {Ragazzoni}, {Raimondo}, {Rainer}, {Reese}, {Redmer}, {Reffert}, {Rojas-Ayala}, {Roxburgh}, {Salmon}, {Santerne}, {Schneider}, {Schou}, {Schuh}, {Schunker}, {Silva-Valio}, {Silvotti}, {Skillen}, {Snellen}, {Sohl}, {Sousa}, {Sozzetti}, {Stello}, {Strassmeier}, {{\v{S}}vanda}, {Szab{\'o}}, {Tkachenko}, {Valencia}, {Van Grootel}, {Vauclair}, {Ventura}, {Wagner}, {Walton}, {Weingrill}, {Werner}, {Wheatley}, \& {Zwintz}}]{Rauer2014ExA}
{Rauer}, H., {Catala}, C., {Aerts}, C., {et~al.} 2014, Experimental Astronomy, 38, 249, \dodoi{10.1007/s10686-014-9383-4}

\bibitem[{{Reback} {et~al.}(2022){Reback}, {jbrockmendel}, {McKinney}, {Van den Bossche}, {Augspurger}, {Roeschke}, {Hawkins}, {Cloud}, {gfyoung}, {Sinhrks}, {Hoefler}, {Klein}, {Petersen}, {Tratner}, {She}, {Ayd}, {Naveh}, {Darbyshire}, {Garcia}, {Shadrach}, {Schendel}, {Hayden}, {Saxton}, {Gorelli}, {Li}, {Zeitlin}, {Jancauskas}, {McMaster}, {W{\"o}rtwein}, \& {Battiston}}]{Reback2022zndo}
{Reback}, J., {jbrockmendel}, {McKinney}, W., {et~al.} 2022, {pandas-dev/pandas: Pandas 1.4.2}, v1.4.2, Zenodo,  Zenodo, \dodoi{10.5281/zenodo.3509134}

\bibitem[{{Refsdal} \& {Weigert}(1970)}]{Refsdal1970A&A}
{Refsdal}, S., \& {Weigert}, A. 1970, \aap, 6, 426

\bibitem[{{Ricker} {et~al.}(2015){Ricker}, {Winn}, {Vanderspek}, {Latham}, {Bakos}, {Bean}, {Berta-Thompson}, {Brown}, {Buchhave}, {Butler}, {Butler}, {Chaplin}, {Charbonneau}, {Christensen-Dalsgaard}, {Clampin}, {Deming}, {Doty}, {De Lee}, {Dressing}, {Dunham}, {Endl}, {Fressin}, {Ge}, {Henning}, {Holman}, {Howard}, {Ida}, {Jenkins}, {Jernigan}, {Johnson}, {Kaltenegger}, {Kawai}, {Kjeldsen}, {Laughlin}, {Levine}, {Lin}, {Lissauer}, {MacQueen}, {Marcy}, {McCullough}, {Morton}, {Narita}, {Paegert}, {Palle}, {Pepe}, {Pepper}, {Quirrenbach}, {Rinehart}, {Sasselov}, {Sato}, {Seager}, {Sozzetti}, {Stassun}, {Sullivan}, {Szentgyorgyi}, {Torres}, {Udry}, \& {Villasenor}}]{Ricker2015JATIS}
{Ricker}, G.~R., {Winn}, J.~N., {Vanderspek}, R., {et~al.} 2015, Journal of Astronomical Telescopes, Instruments, and Systems, 1, 014003, \dodoi{10.1117/1.JATIS.1.1.014003}

\bibitem[{{Roberts} {et~al.}(2024){Roberts}, {Pinsonneault}, {Johnson}, {Zinn}, {Weinberg}, {Vrard}, {Tayar}, {Stello}, {Mosser}, {Johnson}, {Cao}, {Stassun}, {Stringfellow}, {Serenelli}, {Mathur}, {Hekker}, {Garc{\'\i}a}, {Elsworth}, \& {Corsaro}}]{Roberts2024MNRAS}
{Roberts}, J.~D., {Pinsonneault}, M.~H., {Johnson}, J.~A., {et~al.} 2024, \mnras, 530, 149, \dodoi{10.1093/mnras/stae820}

\bibitem[{{Rodrigues} {et~al.}(2014){Rodrigues}, {Girardi}, {Miglio}, {Bossini}, {Bovy}, {Epstein}, {Pinsonneault}, {Stello}, {Zasowski}, {Allende Prieto}, {Chaplin}, {Hekker}, {Johnson}, {M{\'e}sz{\'a}ros}, {Mosser}, {Anders}, {Basu}, {Beers}, {Chiappini}, {da Costa}, {Elsworth}, {Garc{\'\i}a}, {Garc{\'\i}a P{\'e}rez}, {Hearty}, {Maia}, {Majewski}, {Mathur}, {Montalb{\'a}n}, {Nidever}, {Santiago}, {Schultheis}, {Serenelli}, \& {Shetrone}}]{Rodrigues2014MNRAS}
{Rodrigues}, T.~S., {Girardi}, L., {Miglio}, A., {et~al.} 2014, \mnras, 445, 2758, \dodoi{10.1093/mnras/stu1907}

\bibitem[{{Rogers} \& {Nayfonov}(2002)}]{Rogers2002ApJ}
{Rogers}, F.~J., \& {Nayfonov}, A. 2002, \apj, 576, 1064, \dodoi{10.1086/341894}

\bibitem[{{Salaris} {et~al.}(2018){Salaris}, {Cassisi}, {Schiavon}, \& {Pietrinferni}}]{Salaris2018A&A}
{Salaris}, M., {Cassisi}, S., {Schiavon}, R.~P., \& {Pietrinferni}, A. 2018, \aap, 612, A68, \dodoi{10.1051/0004-6361/201732340}

\bibitem[{{Salaris} {et~al.}(1993){Salaris}, {Chieffi}, \& {Straniero}}]{Salaris1993ApJ}
{Salaris}, M., {Chieffi}, A., \& {Straniero}, O. 1993, \apj, 414, 580, \dodoi{10.1086/173105}

\bibitem[{{Salaris} {et~al.}(2015){Salaris}, {Pietrinferni}, {Piersimoni}, \& {Cassisi}}]{Salaris2015A&A}
{Salaris}, M., {Pietrinferni}, A., {Piersimoni}, A.~M., \& {Cassisi}, S. 2015, \aap, 583, A87, \dodoi{10.1051/0004-6361/201526951}

\bibitem[{{Saumon} {et~al.}(1995){Saumon}, {Chabrier}, \& {van Horn}}]{Saumon1995ApJS}
{Saumon}, D., {Chabrier}, G., \& {van Horn}, H.~M. 1995, \apjs, 99, 713, \dodoi{10.1086/192204}

\bibitem[{{Schonhut-Stasik} {et~al.}(2024){Schonhut-Stasik}, {Zinn}, {Stassun}, {Pinsonneault}, {Johnson}, {Warfield}, {Stello}, {Elsworth}, {Garc{\'\i}a}, {Mathur}, {Mosser}, {Hon}, {Tayar}, {Stringfellow}, {Beaton}, {J{\"o}nsson}, \& {Minniti}}]{Schonhut-Stasik2024AJ}
{Schonhut-Stasik}, J., {Zinn}, J.~C., {Stassun}, K.~G., {et~al.} 2024, \aj, 167, 50, \dodoi{10.3847/1538-3881/ad0b13}

\bibitem[{{Sestito} \& {Randich}(2005)}]{Sestito2005A&A}
{Sestito}, P., \& {Randich}, S. 2005, \aap, 442, 615, \dodoi{10.1051/0004-6361:20053482}

\bibitem[{{Shetrone} {et~al.}(2019){Shetrone}, {Tayar}, {Johnson}, {Somers}, {Pinsonneault}, {Holtzman}, {Hasselquist}, {Masseron}, {M{\'e}sz{\'a}ros}, {J{\"o}nsson}, {Hawkins}, {Sobeck}, {Zamora}, \& {Garc{\'\i}a-Hern{\'a}ndez}}]{Shetrone2019ApJ}
{Shetrone}, M., {Tayar}, J., {Johnson}, J.~A., {et~al.} 2019, \apj, 872, 137, \dodoi{10.3847/1538-4357/aaff66}

\bibitem[{{Silva Aguirre} {et~al.}(2020){Silva Aguirre}, {Christensen-Dalsgaard}, {Cassisi}, {Miller Bertolami}, {Serenelli}, {Stello}, {Weiss}, {Angelou}, {Jiang}, {Lebreton}, {Spada}, {Bellinger}, {Deheuvels}, {Ouazzani}, {Pietrinferni}, {Mosumgaard}, {Townsend}, {Battich}, {Bossini}, {Constantino}, {Eggenberger}, {Hekker}, {Mazumdar}, {Miglio}, {Nielsen}, \& {Salaris}}]{SilvaAguirre2020A&A}
{Silva Aguirre}, V., {Christensen-Dalsgaard}, J., {Cassisi}, S., {et~al.} 2020, \aap, 635, A164, \dodoi{10.1051/0004-6361/201935843}

\bibitem[{{Soderblom}(2010)}]{Soderblom2010ARA&A}
{Soderblom}, D.~R. 2010, \araa, 48, 581, \dodoi{10.1146/annurev-astro-081309-130806}

\bibitem[{{Somers} {et~al.}(2020){Somers}, {Cao}, \& {Pinsonneault}}]{Somers2020ApJ}
{Somers}, G., {Cao}, L., \& {Pinsonneault}, M.~H. 2020, \apj, 891, 29, \dodoi{10.3847/1538-4357/ab722e}

\bibitem[{{Souto} {et~al.}(2019){Souto}, {Allende Prieto}, {Cunha}, {Pinsonneault}, {Smith}, {Garcia-Dias}, {Bovy}, {Garc{\'\i}a-Hern{\'a}ndez}, {Holtzman}, {Johnson}, {J{\"o}nsson}, {Majewski}, {Shetrone}, {Sobeck}, {Zamora}, {Pan}, \& {Nitschelm}}]{Souto2019ApJ}
{Souto}, D., {Allende Prieto}, C., {Cunha}, K., {et~al.} 2019, \apj, 874, 97, \dodoi{10.3847/1538-4357/ab0b43}

\bibitem[{{Tayar} {et~al.}(2015){Tayar}, {Ceillier}, {Garc{\'\i}a-Hern{\'a}ndez}, {Troup}, {Mathur}, {Garc{\'\i}a}, {Zamora}, {Johnson}, {Pinsonneault}, {M{\'e}sz{\'a}ros}, {Allende Prieto}, {Chaplin}, {Elsworth}, {Hekker}, {Nidever}, {Salabert}, {Schneider}, {Serenelli}, {Shetrone}, \& {Stello}}]{Tayar2015ApJ}
{Tayar}, J., {Ceillier}, T., {Garc{\'\i}a-Hern{\'a}ndez}, D.~A., {et~al.} 2015, \apj, 807, 82, \dodoi{10.1088/0004-637X/807/1/82}

\bibitem[{{Tayar} {et~al.}(2017){Tayar}, {Somers}, {Pinsonneault}, {Stello}, {Mints}, {Johnson}, {Zamora}, {Garc{\'\i}a-Hern{\'a}ndez}, {Maraston}, {Serenelli}, {Allende Prieto}, {Bastien}, {Basu}, {Bird}, {Cohen}, {Cunha}, {Elsworth}, {Garc{\'\i}a}, {Girardi}, {Hekker}, {Holtzman}, {Huber}, {Mathur}, {M{\'e}sz{\'a}ros}, {Mosser}, {Shetrone}, {Silva Aguirre}, {Stassun}, {Stringfellow}, {Zasowski}, \& {Roman-Lopes}}]{Tayar2017ApJ}
{Tayar}, J., {Somers}, G., {Pinsonneault}, M.~H., {et~al.} 2017, \apj, 840, 17, \dodoi{10.3847/1538-4357/aa6a1e}

\bibitem[{{Timmes} \& {Swesty}(2000)}]{Timmes2000ApJS}
{Timmes}, F.~X., \& {Swesty}, F.~D. 2000, \apjs, 126, 501, \dodoi{10.1086/313304}

\bibitem[{{Townsend}(2021)}]{Townsend2021zndo}
{Townsend}, R. 2021, {MESA SDK for Linux}, 21.4.1, Zenodo,  Zenodo, \dodoi{10.5281/zenodo.2603136}

\bibitem[{{van Saders} \& {Pinsonneault}(2012)}]{vanSaders2012ApJ}
{van Saders}, J.~L., \& {Pinsonneault}, M.~H. 2012, \apj, 746, 16, \dodoi{10.1088/0004-637X/746/1/16}

\bibitem[{{VandenBerg} {et~al.}(2006){VandenBerg}, {Bergbusch}, \& {Dowler}}]{VandenBerg2006ApJS}
{VandenBerg}, D.~A., {Bergbusch}, P.~A., \& {Dowler}, P.~D. 2006, \apjs, 162, 375, \dodoi{10.1086/498451}

\bibitem[{{Virtanen} {et~al.}(2020){Virtanen}, {Gommers}, {Oliphant}, {Haberland}, {Reddy}, {Cournapeau}, {Burovski}, {Peterson}, {Weckesser}, {Bright}, {van der Walt}, {Brett}, {Wilson}, {Millman}, {Mayorov}, {Nelson}, {Jones}, {Kern}, {Larson}, {Carey}, {Polat}, {Feng}, {Moore}, {VanderPlas}, {Laxalde}, {Perktold}, {Cimrman}, {Henriksen}, {Quintero}, {Harris}, {Archibald}, {Ribeiro}, {Pedregosa}, {van Mulbregt}, \& {SciPy 1. 0 Contributors}}]{Virtanen2020NatMe}
{Virtanen}, P., {Gommers}, R., {Oliphant}, T.~E., {et~al.} 2020, Nature Methods, 17, 261, \dodoi{10.1038/s41592-019-0686-2}

\bibitem[{{Vrard} {et~al.}(2025){Vrard}, {Pinsonneault}, {Elsworth}, {Hon}, {Kallinger}, {Kuszlewicz}, {Mosser}, {Garc{\'\i}a}, {Tayar}, {Bennett}, {Cao}, {Hekker}, {Loyer}, {Mathur}, \& {Stello}}]{Vrard2024A&A}
{Vrard}, M., {Pinsonneault}, M.~H., {Elsworth}, Y., {et~al.} 2025, \aap, 697, A165, \dodoi{10.1051/0004-6361/202452635}

\bibitem[{{Warfield} {et~al.}(2024){Warfield}, {Zinn}, {Schonhut-Stasik}, {Johnson}, {Pinsonneault}, {Johnson}, {Stello}, {Beaton}, {Elsworth}, {Garc{\'\i}a}, {Mathur}, {Mosser}, {Serenelli}, \& {Tayar}}]{Warfield2024AJ}
{Warfield}, J.~T., {Zinn}, J.~C., {Schonhut-Stasik}, J., {et~al.} 2024, \aj, 167, 208, \dodoi{10.3847/1538-3881/ad33bb}

\bibitem[{{Weinberg} {et~al.}(2022){Weinberg}, {Holtzman}, {Johnson}, {Hayes}, {Hasselquist}, {Shetrone}, {Ting}, {Beaton}, {Beers}, {Bird}, {Bizyaev}, {Blanton}, {Cunha}, {Fern{\'a}ndez-Trincado}, {Frinchaboy}, {Garc{\'\i}a-Hern{\'a}ndez}, {Griffith}, {Johnson}, {J{\"o}nsson}, {Lane}, {Leung}, {Mackereth}, {Majewski}, {M{\'e}sz{\'a}ros}, {Nitschelm}, {Pan}, {Schiavon}, {Schneider}, {Schultheis}, {Smith}, {Sobeck}, {Stassun}, {Stringfellow}, {Vincenzo}, {Wilson}, \& {Zasowski}}]{Weinberg2022ApJS}
{Weinberg}, D.~H., {Holtzman}, J.~A., {Johnson}, J.~A., {et~al.} 2022, \apjs, 260, 32, \dodoi{10.3847/1538-4365/ac6028}

\bibitem[{{Weiss} {et~al.}(2025){Weiss}, {Downing}, {Pinsonneault}, {Zinn}, {Stello}, {Bedding}, {Cao}, {Hon}, {Reyes}, {Gaudi}, {Wilson}, {Huber}, \& {Sharma}}]{Weiss2025ApJ}
{Weiss}, T.~J., {Downing}, N.~J., {Pinsonneault}, M.~H., {et~al.} 2025, \apj, 987, 181, \dodoi{10.3847/1538-4357/adde5b}

\bibitem[{{Wolf} \& {Schwab}(2017)}]{Wolf2017zndo}
{Wolf}, B., \& {Schwab}, J. 2017, {Wmwolf/Py\_Mesa\_Reader: Interact With Mesa Output}, 0.3.0, Zenodo,  Zenodo, \dodoi{10.5281/zenodo.826958}

\bibitem[{{Xu} {et~al.}(2013){Xu}, {Takahashi}, {Goriely}, {Arnould}, {Ohta}, \& {Utsunomiya}}]{Xu2013NuPhA}
{Xu}, Y., {Takahashi}, K., {Goriely}, S., {et~al.} 2013, \nphysa, 918, 61, \dodoi{10.1016/j.nuclphysa.2013.09.007}

\bibitem[{{York} {et~al.}(2000){York}, {Adelman}, {Anderson}, {Anderson}, {Annis}, {Bahcall}, {Bakken}, {Barkhouser}, {Bastian}, {Berman}, {Boroski}, {Bracker}, {Briegel}, {Briggs}, {Brinkmann}, {Brunner}, {Burles}, {Carey}, {Carr}, {Castander}, {Chen}, {Colestock}, {Connolly}, {Crocker}, {Csabai}, {Czarapata}, {Davis}, {Doi}, {Dombeck}, {Eisenstein}, {Ellman}, {Elms}, {Evans}, {Fan}, {Federwitz}, {Fiscelli}, {Friedman}, {Frieman}, {Fukugita}, {Gillespie}, {Gunn}, {Gurbani}, {de Haas}, {Haldeman}, {Harris}, {Hayes}, {Heckman}, {Hennessy}, {Hindsley}, {Holm}, {Holmgren}, {Huang}, {Hull}, {Husby}, {Ichikawa}, {Ichikawa}, {Ivezi{\'c}}, {Kent}, {Kim}, {Kinney}, {Klaene}, {Kleinman}, {Kleinman}, {Knapp}, {Korienek}, {Kron}, {Kunszt}, {Lamb}, {Lee}, {Leger}, {Limmongkol}, {Lindenmeyer}, {Long}, {Loomis}, {Loveday}, {Lucinio}, {Lupton}, {MacKinnon}, {Mannery}, {Mantsch}, {Margon}, {McGehee}, {McKay}, {Meiksin}, {Merelli}, {Monet}, {Munn}, {Narayanan}, {Nash}, {Neilsen}, {Neswold}, {Newberg}, {Nichol}, {Nicinski},
  {Nonino}, {Okada}, {Okamura}, {Ostriker}, {Owen}, {Pauls}, {Peoples}, {Peterson}, {Petravick}, {Pier}, {Pope}, {Pordes}, {Prosapio}, {Rechenmacher}, {Quinn}, {Richards}, {Richmond}, {Rivetta}, {Rockosi}, {Ruthmansdorfer}, {Sandford}, {Schlegel}, {Schneider}, {Sekiguchi}, {Sergey}, {Shimasaku}, {Siegmund}, {Smee}, {Smith}, {Snedden}, {Stone}, {Stoughton}, {Strauss}, {Stubbs}, {SubbaRao}, {Szalay}, {Szapudi}, {Szokoly}, {Thakar}, {Tremonti}, {Tucker}, {Uomoto}, {Vanden Berk}, {Vogeley}, {Waddell}, {Wang}, {Watanabe}, {Weinberg}, {Yanny}, {Yasuda}, \& {SDSS Collaboration}}]{York2000AJ}
{York}, D.~G., {Adelman}, J., {Anderson}, John~E., J., {et~al.} 2000, \aj, 120, 1579, \dodoi{10.1086/301513}

\bibitem[{{Zhao} {et~al.}(2012){Zhao}, {Zhao}, {Chu}, {Jing}, \& {Deng}}]{Zhao2012RAA}
{Zhao}, G., {Zhao}, Y.-H., {Chu}, Y.-Q., {Jing}, Y.-P., \& {Deng}, L.-C. 2012, Research in Astronomy and Astrophysics, 12, 723, \dodoi{10.1088/1674-4527/12/7/002}

\bibitem[{{Zinn} {et~al.}(2019){Zinn}, {Pinsonneault}, {Huber}, \& {Stello}}]{Zinn2019ApJ}
{Zinn}, J.~C., {Pinsonneault}, M.~H., {Huber}, D., \& {Stello}, D. 2019, \apj, 878, 136, \dodoi{10.3847/1538-4357/ab1f66}

\bibitem[{{Zinn} {et~al.}(2022){Zinn}, {Stello}, {Elsworth}, {Garc{\'\i}a}, {Kallinger}, {Mathur}, {Mosser}, {Hon}, {Bugnet}, {Jones}, {Reyes}, {Sharma}, {Sch{\"o}nrich}, {Warfield}, {Luger}, {Vanderburg}, {Kobayashi}, {Pinsonneault}, {Johnson}, {Huber}, {Buder}, {Joyce}, {Bland-Hawthorn}, {Casagrande}, {Lewis}, {Miglio}, {Nordlander}, {Davies}, {De Silva}, {Chaplin}, \& {Silva Aguirre}}]{Zinn2022ApJ}
{Zinn}, J.~C., {Stello}, D., {Elsworth}, Y., {et~al.} 2022, \apj, 926, 191, \dodoi{10.3847/1538-4357/ac2c83}

\end{thebibliography}
\bibliographystyle{aasjournal}

\end{document}